\documentclass[fleqn,usenatbib]{mnras}

\usepackage[T1]{fontenc}
\usepackage{ae,aecompl}

\usepackage{lmodern}	%
\usepackage{graphicx}	
\usepackage{amsmath}	
\usepackage{amssymb}
\usepackage{scalerel}
\usepackage{tikz}
\usepackage{mathtools}
\usepackage{ulem}
\usepackage{comment}

\usetikzlibrary{svg.path}

\definecolor{orcidlogocol}{HTML}{A6CE39}
\tikzset{
  orcidlogo/.pic={
    \fill[orcidlogocol] svg{M256,128c0,70.7-57.3,128-128,128C57.3,256,0,198.7,0,128C0,57.3,57.3,0,128,0C198.7,0,256,57.3,256,128z};
    \fill[white] svg{M86.3,186.2H70.9V79.1h15.4v48.4V186.2z}
                 svg{M108.9,79.1h41.6c39.6,0,57,28.3,57,53.6c0,27.5-21.5,53.6-56.8,53.6h-41.8V79.1z M124.3,172.4h24.5c34.9,0,42.9-26.5,42.9-39.7c0-21.5-13.7-39.7-43.7-39.7h-23.7V172.4z}
                 svg{M88.7,56.8c0,5.5-4.5,10.1-10.1,10.1c-5.6,0-10.1-4.6-10.1-10.1c0-5.6,4.5-10.1,10.1-10.1C84.2,46.7,88.7,51.3,88.7,56.8z};
  }
}

\newcommand\orcidicon[1]{\href{https://orcid.org/#1}{\mbox{\scalerel*{
\begin{tikzpicture}[yscale=-1,transform shape]
\pic{orcidlogo};
\end{tikzpicture}
}{|}}}}

\usepackage{xcolor}
\definecolor{light-gray}{gray}{0.95}
\definecolor{light-red}{rgb}{1. , 0.906, 0.298}

\title[Tree-based solvers for AMR code FLASH - III]{Tree-based solvers for adaptive mesh refinement code FLASH - III:
a novel scheme for radiation pressure on dust and gas and radiative transfer from diffuse sources.}

\author[Klepitko et al.]{A. Klepitko$^{1}$\thanks{E-mail: klepitko@ph1.uni-koeln.de} \orcidicon{0000-0002-5570-1184},
S. Walch$^{1,2}$\thanks{E-mail: walch@ph1.uni-koeln.de} \orcidicon{0000-0001-6941-7638},
R. W\"unsch$^{3}$\orcidicon{0000-0003-1848-8967},
D. Seifried$^{1,2}$\orcidicon{0000-0002-0368-9160}, \newauthor
F. Dinnbier$^{4}$\orcidicon{0000-0001-5532-4211},
S. Haid$^{1}$
\\
$^{1}$I. Physikalisches Insitut, Universit{\"a}t zu K{\"o}ln, Z\"ulpicher Str. 77, 50937 K\"oln, Germany\\
$^{2}$Center for data and simulation science, University of Cologne, www.cds.uni-koeln.de\\
$^{3}$Astronomical Institute, Academy of Sciences of the Czech Republic, Bocni II 1401, 141 31 Prague, Czech Republic\\
$^{4}$Astronomical Institute, Faculty of Mathematics and Physics, Charles University, V Hole\v{s}ovi\v{c}k\'{a}ch 2, 180 00 Praha 8, Czech Republic
}

\date{Accepted 2023 January 30. Received 2023 January 29; in original form 2022 April 19}

\pubyear{2023}

\begin{document}
\label{firstpage}
\pagerange{\pageref{firstpage}--\pageref{lastpage}}
\maketitle

\begin{abstract}
  Radiation is an important contributor to the energetics of the interstellar medium, yet its transport is difficult to solve numerically.
  We present a novel approach towards solving radiative transfer of diffuse sources via backwards ray tracing.
  Here we focus on the radiative transfer of infrared radiation and the radiation pressure on dust.
  The new module, \textsc{TreeRay/RadPressure}, is an extension to the novel radiative transfer method \textsc{TreeRay} implemented in the grid-based MHD code {\sc Flash}. In \textsc{TreeRay/RadPressure}, every cell and every star particle is a source of infrared radiation.
  We also describe how gas, dust and radiation are coupled via a chemical network. This allows us to compute the local dust temperature in thermal equilibrium, leading to a significantly improvement over the classical grey approximation.
  In several tests, we demonstrate that the scheme produces the correct radiative intensities as well as the correct momentum input by radiation pressure. 
  Subsequently, we apply our new scheme to model massive star formation from a collapsing, turbulent core of 150 ${\rm M}_\odot$. We include the effects of both, ionizing and infrared radiation on the dynamics of the core.
  We find that the newborn massive  star prevents fragmentation in its proximity due to radiative heating. Over time, dust and radiation temperature equalize, while the gas temperature can be either warmer due to shock heating or colder due to insufficient dust-gas coupling. Compared to gravity, the effects of radiation pressure are insignificant for the stellar mass on the simulated time scale in this work.
\end{abstract}




\section{Introduction}

Radiation provides a channel through which energy and momentum may be transported independently from the flow of mass. This makes it an integral part of the physical processes of the interstellar medium (ISM). The impact of radiation ranges from peaceful processes such as line emission cooling, continuum radiation cooling and heating -- to violently driven HII regions \citep{spitzer1978} and radiation pressure (RP) driven outflows. This enables radiative processes to regulate the star formation efficiency both ways -- increasing it through cooling processes, enabling gas to collapse, and decreasing it through explosive expansion resulting in the removal of gas. Especially in recent works, RP is discussed as a mechanism to limit star formation. \cite{Matzner2002} argues that giant molecular clouds are supported by the feedback of their most massive stars. Sites of massive star formation are those with column densities greater than 0.1 $\mathrm{g\, cm^{-2}}$ \citep{massiveReview2014}. \cite{thompson2005radiation} discuss the importance of RP on galactic scales and argue that RP from dust acts as an anti-catalyst for the star formation rate in ultra-luminous infrared galaxies. 

The combination of photoionization and RP is discussed by \cite{kuiper2018} in the context of massive star formation, where they find that RP plays a role at later stages and may limit the final mass of a massive star. \cite{rosen2019} find that RP may lead to cavities, which only allow for disk accretion onto the star at later stages. Eventually the mass flow onto the disk is cut off resulting in starvation of the disk and limiting the total accretion onto the star.
 
Solving radiative transfer (RT) numerically is difficult to accomplish, because RT can be considered a non-local problem to solve in the context of hydrodynamical simulations. This makes solving RT to a certain degree similar to the problem of solving gravity, where its numerical solution requires a lot of communication. However, unlike gravity, radiation may be shielded or reprocessed, rendering the problem even more difficult than solving gravity numerically. Recent improvements in computer hardware, ever-growing supercomputers, and the development of numerical methods enabled RT to be treated on the fly in state-of-the-art numerical simulations. \cite{bisbas2015} show benchmarks of 11 independent approaches towards solving the D-type expansion of an HII region driven by a single source. This highlights the effort that is put into solving RT in current numerical codes.

The pioneering work of \cite{levermore1981flux} formulates a flux-limited diffusion (FLD) theory, of which core concepts are still used in state-of-the-art numerical simulations and codes today \citep[e.g.][and many others]{krumholz2007, kuiper2018}. Here, radiation is treated in the form of an energy density and evolved according to the diffusion equation. This formalism transforms RT into a local problem, which is beneficial from a computational perspective as very little communication is required during the computation. As a trade-off the timestep is limited through the speed of light. This limitation can be partially overcome in a reduced speed of light approximation.

M1 closure builds on the FLD method and introduces more accuracy to the solution by taking into account one more moment to advect its radiation energy density. This leads to M1 closure capturing shadows more accurate than FLD \citep{levermore1984, rosdahl2015scheme, kannan2019}. However, M1 closure may result in colliding flows of radiation if mulitiple sources are present \citep{menon2022}. The work of \cite{menon2022} employs a variable Eddington tensor method which greatly improves the result of the radiation field. Their algorithm, \textsc{VETTAM}, uses ray-tracing to compute the variable Eddington tensor in a hybrid characteristics approach.

A different approach to solving RT is ray tracing, where multiple rays are cast to probe the environment along different lines of sight. Ray tracing yields more accurate results in terms of shadow casting than moment methods. This is especially the case if the radiation originates from many pointlike sources, with the trade-off of being computationally more demanding compared to moment methods or FLD. Examples of ray tracing can be found in the works of \cite{fervent2015} and also \cite{kim2018} where they solve the forward propagation of ionizing radiation based on the methods of \cite{abel2002}. Forward ray tracing (FRT) may quickly become infeasible if the number of sources grows, limiting the applicability of FRT. For this reason FRT is not suited for modelling reprocessed radiation, because of the sheer number of sources involved i.e. each gaseous or dusty cell can be a source of radiation. \cite{rosen2017} and \cite{kuiper2020} use a hybrid solution where radiation originating from pointlike sources is modelled with FRT while the reprocessed radiation is modelled with FLD. The early work of \cite{kessel-deynet2000} shows a different approach to ray tracing, where they use backwards ray tracing to solve RT of ionizing radiation. Similarly, the work of \cite{Altay2013} shows an application of backwards ray tracing to treat ionizing radiation in the context of smoothed particle hydrodynamics. Another example can be found in the work of \cite{grond2019} where a single ray from each point in space is cast to every source of radiation along which RT is solved. Their approach, named \textsc{TREVR}, merges sources and adaptively refines on extincting material to save on computational effort. Over all, \textsc{TREVR} scales with an $(N \, \log^2(N))$ relation. Following this approach calculations on taking into account emission from dust may become infeasible.

\cite{2018wunsch} solve gravity in adaptive mesh refinement (AMR) simulations with a tree-solver operating on an octal-spatial tree (Octtree). In a next step, \cite{wunsch2021} expand the method to treat ionizing radiation from pointlike sources in a backwards ray tracing approach, called {\sc TreeRay/OnTheSpot}. This work will build on both these papers and will expand the applicability of {\sc TreeRay} to treat reprocessed radiation from dust. Or more general, this work will explore solving radiative transfer from macroscopic sources via backwards ray tracing.

This work discusses three types of radiation: ionising radiation from point
sources (e.g. stars), non-ionising radiation from point sources, and
non-ionising radiation from dust. The transport of the first one is treated by
the {\sc TreeRay/OnTheSpot} module described in \cite{wunsch2021}, calculating its absorption by gas
through the case B recombination and the gas heating. Here, we add the
interaction of ionising radiation with dust, specifically, its radiation
pressure on dust and gas. The absorption of ionising radiation by
dust is neglected. The {\sc TreeRay/RadPressure} module
described here calculates the transport of the non-ionising radiation from point
sources and from dust. Both types of the non-ionising radiation interact
directly only with the dust. In the latter we assume that
dust and gas are always dynamically coupled and the momentum inserted by the
radiation pressure to the dust is immediately transferred to the gas.

This manuscript is structured as follows. In \S\ref{s:theory} we recall basics and simplifications on radiative transfer relevant to this work. In \S\ref{s: Methods} we summarize the numerical methods used in this work followed by a detailed explanation of the novel radiative transfer scheme given in \S\ref{ss:rad_transfer}. We show tests verifying the correctness of the scheme in \S\ref{s:benchmarks}. We follow this up with two more sophisticated setups covering the expansion of an HII region in \S\ref{s:hii_region} and a star forming setup in \S\ref{s:star_forming_setup}.

\section{Simple, Dusty Radiative Transfer} \label{s:theory}
In this section we briefly define the essentials of radiative transfer while focussing on dust. We employ simplifications to aid the speedup of the algorithm.

The radiative transfer equation for the frequency-dependent radiation intensity, $I_\nu$, propagating along a unit vector, $\hat{\mathbf{n}}$, reads
\begin{eqnarray}
  \left( \frac{1}{c} \frac{\partial}{\partial t} + \hat{\mathbf{n}} \cdot \nabla \right) I_\nu = - a_\nu I_\nu + j_\nu, \label{eq:rt}
\end{eqnarray}
where $t$, $\nu$, $a_\nu$ and $j_\nu$ are the unit time, the frequency of the light and the extinction and emission coefficient, respectively \citep{mihalas2013foundations}. In the following we will assume $c$ to approach infinity in eq. \ref{eq:rt}. As a consequence, contributions from the time derivative vanish, such that one arrives at an equilibrium solution. By doing this, we can not model the propagation of light within a small time window $\mathrm{d}t$ accurately. However, given that the sound crossing time is much larger than the light crossing time on astrophysical scales, this is a reasonable approximation. Further we drop all frequency dependencies by introducing a Planck weighting.\\
By doing so we arrive at a formulation of eq.~\ref{eq:rt} that is greatly simplified, namely
\begin{eqnarray}
  \hat{\mathbf{n}} \cdot \nabla I = - a I + j\, , \label{eq:rtsimple}
\end{eqnarray}
where $a$ and $j$ are the Planck-weighted mean extinction and emission coefficient, respectively, and $I$ is the total radiation intensity. We summarize
\begin{eqnarray}
    I &=& \int_0^\infty I_\nu d\nu \, , \\   
    a &=& \kappa_{\rm P}\rho \, , \label{eq:alpha}\\
    j &=& \frac{\sigma}{\pi}\kappa_{\rm P}\rho T^4\, , \\
    \kappa_{\rm P} &=& \frac{\int_0^\infty \kappa_\nu B_\nu \mathrm{d}\nu }{ \int_0^\infty B_\nu \mathrm{d}\nu}\, ,
\end{eqnarray}
where $\kappa_{\rm P}$, $\sigma$ are the Planck mean opacity and Stefan-Boltzmann constant, respectively. Throughout this work, we employ the Planck mean dust opacity based on \cite{semenov2003rosseland}. The model description we choose is the same as stated by \cite{krumholz2012direct}:
\begin{eqnarray}
  \kappa_\mathrm{P}(T) = 10^{-1}  \frac{\mathrm{cm^2}}{100 \,\mathrm{K^2\, g}} \times 
  \begin{cases}
    T^2 & {\rm for}\; T<150\,\mathrm{K} \\
    (150\,\mathrm{K})^2 & \mathrm{else}
  \end{cases}\,.
\end{eqnarray}
In the following we drop the index P such that $\kappa \coloneqq \kappa_{\rm P}$. 

We define the radiative flux, $\mathbf{F}$, and the mean radiative intensity, $\bar{J}$, in the following way
\begin{eqnarray}
        \mathbf{F} &=& - \oint_{\mathcal{S}}I \, \hat{\mathbf{n}}_\mathrm{rad} \, \mathrm{d}\Omega\, ,\label{eq:rad_flux}\\
    \bar{J} &=& \frac{1}{4\pi} \oint_{\mathcal{S}} I \, \mathrm{d}\Omega\,.
\end{eqnarray}
Here, $\hat{\mathbf{n}}_\mathrm{rad}$ is the radial unit vector pointing from the origin towards the surface of a unit sphere, $\mathcal{S}$, and $\mathrm{d}\Omega$ is the solid angle around $\hat{\mathbf{n}}_\mathrm{rad}$. The minus sign in Eq. \ref{eq:rad_flux} defines the flux in the opposite direction of $\hat{\mathbf{n}}_\mathrm{rad}$.

Note that our approximation for the radiative transfer scheme does not account for scattering. Hence, also the momentum of RP is caused only by first absorption and reemission by thermal photons but not by scattering. However, in the infrared regime the scattering albedo is very small \citep[e.g. for the dust models of ][]{weingartner2001} and hence absorption and reemission events are significantly more frequent than scattering events.

\subsection{Effective area of an infinitesimal volume of dust} \label{ss: eff area}

\begin{figure}
    \centering
    \includegraphics[width=0.5\columnwidth]{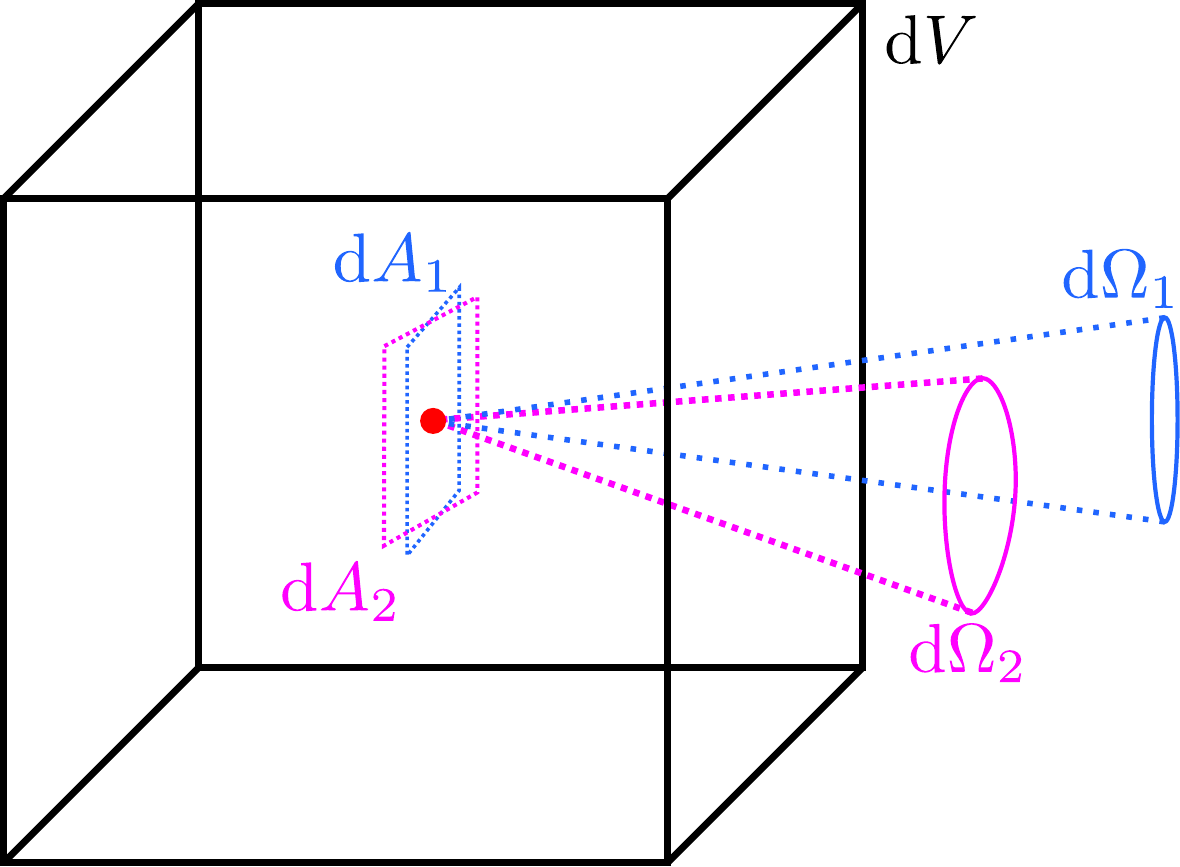}
    \caption{Schematic picture of an infinitesimal volume, $\mathrm{d}V$, from which radiation is emitted into solid angles, $\mathrm{d}\Omega_1$ and $\mathrm{d}\Omega_2$. The associated areas of the solid angles, $\mathrm{d}\Omega_1$ and $\mathrm{d}\Omega_2$, are labeled $\mathrm{d}\mathbf{A}_1$ and $\mathrm{d}\mathbf{A}_2$, respectively. The relation $| \mathrm{d}\mathbf{A}_1 |  = | \mathrm{d}\mathbf{A}_2 | $ holds.
    The source radiates with $L^0_{\Omega} = \frac{\sigma}{\pi} T^4$ into every solid angle (see Eq. \ref{eq: thermal rad}). \label{fig:vol_to_area}}
\end{figure}

In this section we discuss the emission properties of dust in the context of three-dimensional emission. In the following, we argue that the effective area, $\mathrm{d}A$, from which dust inside an infinitesimal volume, $\mathrm{d}V$ of mass density $\rho$ emits radiation into a solid angle $\mathrm{d}\Omega$, is given by
\begin{eqnarray}
    \cos \vartheta \, \mathrm{d}A \mathrm{d}\Omega= \kappa\rho \, \mathrm{d}V \mathrm{d}\Omega \label{eq: area_to_vol}\, ,
\end{eqnarray} 

 where $\vartheta$ is the angle between the surface normal of $\mathrm{d}A$ and solid angle $\mathrm{d}\Omega$.

Following Eq. \ref{eq:rtsimple}, the decrease in radiation intensity as a function of distance from the radiation source is governed by the extinction coefficient. Both the effects of scattering and absorption contribute to $a$. If we neglect scattering, we can express $a$ as the product of the mass density and the dust opacity (Eq. \ref{eq:alpha}). Following its dimensionality, $L^2 M^{-1}$, the dust opacity can be understood as an associated area per unit mass. It is the effective area through which dust interacts with radiation.

For the cubic, infinitesimal volume, $\mathrm{d}V=(\mathrm{d}r)^3$, filled with material of mass density $\rho$, we can express the contained mass as $\mathrm{d}m = \rho \, \mathrm{d}V$. A one-dimensional ray penetrating $\mathrm{d}V$ and traveling through its length, $\mathrm{d}r$, would act on an area $\mathrm{d}A = \kappa \, \mathrm{d}m = \kappa \rho \, \mathrm{d}V = (\mathrm{d}r)^2 (\kappa \rho \, \mathrm{d}r)$. The last term 
\begin{eqnarray}
    \tau = \kappa \rho \, \mathrm{d}r
\end{eqnarray} represents the optical depth,  $\tau$, through the volume element. Here, in this instance, we only consider volumes which are so small that they are certainly optically thin (in the numerical implementation we separately treat optically thin and thick volumes; see \S\ref{ss:rad_transfer}). 

At the same time, $\mathrm{d}A$ is the effective area from which thermal radiation is emitted by the volume into the path of the ray. Note, that $\mathrm{d}A$ is invariant under rotation, since the spatial distribution of dust within $\mathrm{d}V$ is assumed to be homogeneous (i.e. $|\mathrm{d}{\bf A}_1| = |\mathrm{d}{\bf A}_2| = : \mathrm{d}A$ in Fig. \ref{fig:vol_to_area}). For this reason, $\mathrm{d}V$ emits and absorbs radiation isotropically. Because of the isotropic emission behaviour of such a volume of dust, we can neglect the $\cos \vartheta$ factor as there always exists an area $\mathrm{d}\mathbf{A}$ which is normal to a choosen $\mathrm{d}\Omega$. This yields $\vartheta=0$ and therefore $\cos \vartheta = 1$.\

\subsection{Thermal emission of an infinitesimal volume}
We model the radiation of dust to be of thermal nature.
\noindent Let $L^0_{\Omega} \left( T \right) \cos \vartheta \, \mathrm{d}A \, \mathrm{d}\Omega$ measure the luminosity of thermal radiation emitted from an effective surface element, $\mathrm{d}A$, at temperature $T$ into a solid angle $\mathrm{d}\Omega$. The following equation holds \citep{mihalas2013foundations}:
\begin{eqnarray}
    L^0_{\Omega} \left( T \right) \cos \vartheta \, \mathrm{d}A \, \mathrm{d}\Omega &=& \frac{\sigma}{\pi} T^4 \cos \vartheta \, \mathrm{d}A \, \mathrm{d}\Omega\, . \label{eq: thermal rad}
\end{eqnarray}
Combining eq. \ref{eq: area_to_vol} and eq. \ref{eq: thermal rad} yields
\begin{eqnarray}
    L^0_{\Omega} \left( T \right) \kappa\rho \, \mathrm{d}V \, \mathrm{d}\Omega &=& \frac{\sigma}{\pi} \kappa\rho \, T^4 \, \mathrm{d}V \, \mathrm{d}\Omega \, , 
    \label{eq:thermal_lum}
\end{eqnarray}
where $\sigma$ is the Stefan-Boltzmann constant. Fig. \ref{fig:vol_to_area} schematically shows how an area $\mathrm{d}\mathbf{A}_1$ represents a fraction of the surface area of $\mathrm{d}V$.\\
Since we aim to model radiative transfer from dust and we assume the dust to be distributed smoothly within $\mathrm{d}V$, we can use $\mathrm{d}A$ to model the emission from a small volume $\mathrm{d}V$.

\begin{figure}
    \includegraphics[width=\columnwidth]{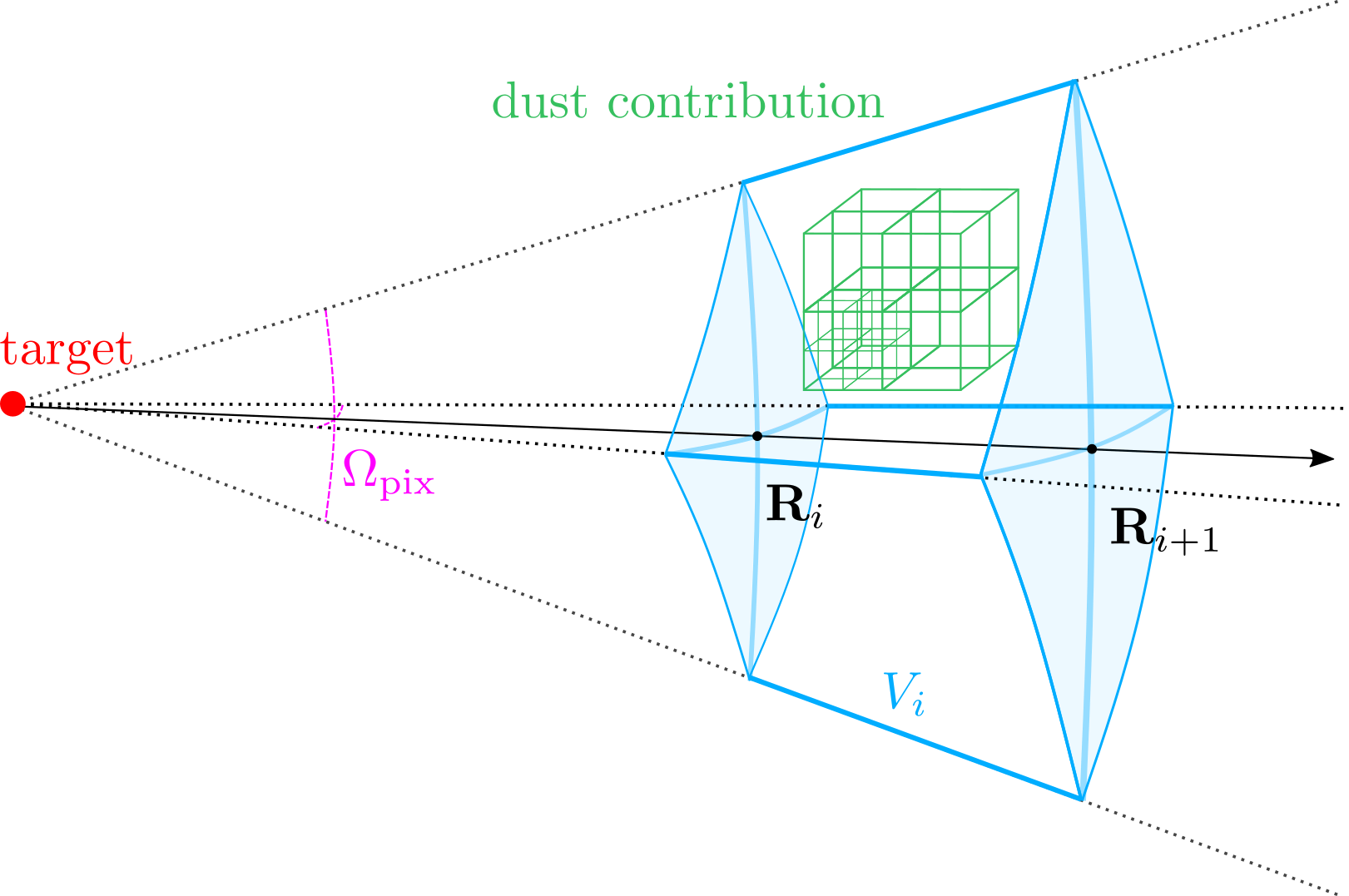}
    \caption{Graphical representation of a single {\sc HealPix} ray. The ray is of angular size $\Omega_\mathrm{pix}$ (pink) and has multiple evaluation points along its radial direction. One such evaluation point is labeled with $i$ and has an associated distance $\mathbf{R}_i$ from the target point (black dot). The volume $V_i$ (blue) associated with evaluation point $i$ is bound by $\mathbf{R}_i$ and $\mathbf{R}_{i+1}$ in radial direction. The dust contributions (green) exist on an octree structure and can be mapped to rays adaptively.\label{fig: ray bin}}
\end{figure}
\section{Numerical Methods}\label{s: Methods}
We use the highly scalable code {\sc flash 4.3} \citep{fryxell2000flash} to solve the ideal hydrodynamics equations. The equations are solved with a directionally split scheme based on the 5-wave Bouchut solver HLL5R \citep{bouchut2007,bouchut2010}. Gravity is accounted for using the tree-based solver described in the work of \cite{2018wunsch}. Additionally, we include the treatment of ionizing \citep{wunsch2021} and non-ionizing radiation in this work. For the first time we present results from our backwards ray tracing infrared radiative transfer method, {\sc TreeRay/RadPressure}, implemented in {\sc flash4.3}\footnote{The method has been transferred to {\sc flash4.6} as well.}. These numerical methods form the baseline upon which we will benchmark \textsc{TreeRay/RadPressure}.

{\sc TreeRay/RadPressure} allows to compute radiation fields from both pointlike (e.g. stars represented by sink particles) and macroscopic sources (e.g. radiation from dust) taking into account absorption and emission on the fly. The latter is accomplished by allowing for every cell to be a source. With the radiation fields at hand we can treat the effects of radiation pressure and radiative heating and cooling through dust. Here, {\sc TreeRay/RadPressure} is coupled to a chemical non-equilibrium network \citep{nelson1997,walch2015silcc}. We explain the details of {\sc TreeRay/RadPressure} throughout section \ref{ss:rad_transfer} and how we treat heating and cooling of gas through dust in section \ref{ss:heating_and_cooling_of_dust}. 

\section{Radiative transfer algorithm}\label{ss:rad_transfer}
The algorithm presented in this work builds on {\sc TreeRay} \citep{wunsch2021} as a framework. The general approach uses backwards ray tracing to solve radiative transfer. For the algorithm to work efficiently, it is necessary that the information of the computational domain is stored in a volume dividing tree. To do this, we employ an octree in the current implementation. From each point of interest we cast rays aligned with the pixel defined by the {\sc HEALPix} algorithm \citep{HEALPix}. Fig. \ref{fig: ray bin} illustrates the shape and components of a ray originating from a point of interest (target position, red dot). For each ray we map contributions to the ray based on their geometric intersection. The contributions are mapped on a node basis, as they naturally occur in an oct-tree. If the nodes appear larger than an angle, $\theta_\mathrm{lim}$, while being mapped to a ray, they are rejected and opened instead. This refinement process is repeated until we eventually reach leaves of the tree, in which case they are guaranteed to be accepted. In a final step we integrate the rays to obtain useful quantities. In our case we focus on dust, a macroscopic source potentially permeating the entire computational domain. Applications of {\sc TreeRay} solving radiative transfer for ionizing radiation can be found in the works of \cite{haid18}, \cite{haid19} and \cite{dinnbier2020}.

In the general spirit of {\sc TreeRay}, we will solve radiative transfer for infrared radiation with backwards ray tracing. The method consists of multiple conceptual parts: (1) we save necessary quantities on the tree (see \S\ref{sss: tree build}); (2) we cast rays from our target point in order to probe the surroundings, tree-nodes are mapped to the rays (see \S\ref{sss: ray mapping}); (3) we integrate individual rays (see \S\ref{sss: ray integration}); (4) perform final calculations after knowing results from all rays (see \S\ref{sss: ray finalization}). In \S\ref{sss: momentum from flux} we compute the momentum from our flux. The additional features presented in this work recover the same computational strong and weak scaling as presented in the work of \cite{wunsch2021}. A derivation of the performance was carried out by \cite{grond2019} where they conclude a scaling of $N_\mathrm{sink}\log(N)$, where $N_\mathrm{sink}$ and $N$ are the number of sink particles and cells, respectively. In our scheme, the scaling is the same as in the work of \cite{grond2019} except that $N_\mathrm{sink}$ is substituted by the number of cells, $N$.

The main goals of this module are to calculate both the acceleration caused by RP, $\mathbf{a}_\mathrm{RP}$, projected onto the cartesian coordinates $x,\,y,\,\mathrm{and}\,z$ as well as the radiation intensity, $J$. The accelerations are stored in 3 fields within \textsc{FLASH} named \texttt{RPAX}, \texttt{RPAY} and \texttt{RPAZ} and have units cm s$^{-2}$ each. We explain the calculation of $\mathbf{a}_\mathrm{RP}$ starting in \S\ref{sss: tree build} and arrive at the final expression in \S\ref{sss: momentum from flux}. The calculation of the mean radiative intensity, $J$, which is stored in the field variable \texttt{IRXE} of units erg s$^{-1}$ cm$^{-2}$ sr$^{-1}$, is finalized in section \S\ref{sss: ray finalization}. The directly deposited radiation pressure due to first absorption within a cell, $P_\mathrm{src}$, is stored in the field \texttt{PRAD} and has the units dyn cm$^{-2}$. We use this quantity to account for RP due to direct absorption of radiation from a point source (like a star) within a cell. The dust temperature, $T_\mathrm{dust}$, is stored in the field \texttt{TDUS} with units K.

Due to the difference of radiation being emitted from an optically thin volume compared to that of an optically thick volume, we need to treat either case inherently separate. While optically thin material can radiate from its entire volume, optically thick material can only radiate from its surface. Since we use backwards ray tracing that adaptively refines the grid, we need to make simplifications, as the exact geometry of the material is unknown. For our simplifications we assume a complete mixing for the optically thin material and assume a compact geometry for optically thick material at given radial distances for the entire solid angle of the rays.

\subsection{Tree build} \label{sss: tree build}
Information about regions in space is stored in an oct-tree structure. We use the term \textit{node} whenever we talk about data stored inside the tree. The nodes that represent the AMR grid at the highest local refinement are called \textit{bottom nodes} ($:=$ \textit{b-node}). \textit{Higher level nodes} ($:=$ \textit{h-node}) are all nodes that are not bottom nodes. By our definition, bottom nodes correspond to leaf nodes of a tree such that they have the highest resolution meaning that they resolve the smallest structures. On the other hand, higher level nodes correspond to branches of a tree structure, therefore offering less resolution.
\subsubsection{Bottom nodes}
We store different quantities on a b-node depending on the b-node being optically thin or thick. A node is labeled optically thin if its optical depth, $\tau_\mathrm{b-node}$, is less than 1 and optically thick otherwise. We calculate $\tau_\mathrm{b-node}$ for our nodes of cubic shape with volume $\mathrm{d}V = \mathrm{d}r^3$ as such:
\begin{eqnarray}
    \tau_\mathrm{b-node} = \rho \kappa \mathrm{d}r\, .
\end{eqnarray}

For an optically thin node we store:
\begin{eqnarray}
  L_\mathrm{b-node,\,thin} &=& 4\sigma \rho \kappa (T_\mathrm{dust})^4 \mathrm{d}V, \label{eq: lthinbnode}\\
  \widetilde{A}_\mathrm{b-node} &=& \rho \kappa\mathrm{d}V, \label{eq: athinbnode}\\
  S_\mathrm{b-node} &=& L_\mathrm{src,\, IR} \times e^{-\tau_\mathrm{b-node}/2},
\end{eqnarray}
where $L_\mathrm{b-node,\,thin}$ is the optically thin luminosity of the node and $\widetilde{A}$ is the associated area, $S_\mathrm{b-node}$ are luminosity contributions from pointlike sources like sink particles. $L_\mathrm{src}$ is the non ionizing luminosity of pointlike particles present in a bottom node (see  eq. \ref{eq:LsrcIR}). The complement of the pointlike sources contributions is used to heat the node and added as a trapped radiation pressure. $\widetilde{A}$ can be interpreted as an associated area for a volume to interact with radiation analogue to the concept discussed in \S\ref{ss: eff area}. Fig. \ref{fig:vol_to_area} shows the associated area of an optically thin volume. Later on $\widetilde{A}$ will be used to compute extinction along the ray.

For an optically thick node we store:
\begin{eqnarray}
  \underline{L}_\mathrm{b-node,\,thick} &=& \frac{\sigma}{\pi} (T_\mathrm{dust})^4, \label{eq:lthickbnode}
\end{eqnarray}
 $\widetilde{A}_\mathrm{b-node}$ and $S_\mathrm{b-node}$, where $\underline{L}_\mathrm{b-node,\,thick}$ is the luminosity per area emitted from a solid black body into a solid angle. Note that we indicate the different units between the luminosities stored on optically thin and thick nodes by an underline (compare Eq. \ref{eq: lthinbnode} and Eq. \ref{eq:lthickbnode}). In our implementation we toggle the bit responsible for storing the $\pm$-sign of $L_\mathrm{b-node}$ in the memory to indicate whether a node is optically thin or thick on the bottom layer. By doing this we keep the memory footprint of the bottom nodes small by containing only 3 floats. This enhancement is doable because a negative luminosity is not meaningful in our scheme.

\subsubsection{Higher level nodes}
Higher level nodes are created by joining 8 subvolumes together into a bigger volume. We refer to a subvolume of a higher level node as a \textit{sub-node}. The edge length of a node is twice that of its sub-node in our implementation.

We propagate the information from the bottom of the tree upwards to each h-node by summation over all sub-nodes. The details can be found in \S\ref{s:fillin_tree_from_bottom_to_top} of the appendix. The quantities are stored separately for optically thin and thick material in h-nodes, so that we can trace the optically thin and thick content of $L$, $\widetilde{A}$ and $V$ inside each h-node.

Each of the h-nodes holds $\mathbf{r}_\mathrm{COL}$ pointing from the geometric center of the h-node to its center of luminosity (COL). We find that taking into account $\mathbf{r}_\mathrm{COL}$ while mapping nodes helps improving errors introduced by coarsening at greater distances. Again the details can be found in \S\ref{s:fillin_tree_from_bottom_to_top} of the appendix. 

\subsection{Ray mapping} \label{sss: ray mapping}
Each target is equipped with a set of rays defined by the pixels generated by the \textsc{HEALPix} algorithm \citep{HEALPix}. We will distinguish the rays with the index $h$. The total number of rays is given by $N_\mathrm{pix}$. Note that each cell in our computation is considered to be a target point.

A single ray has multiple evaluation points along its direction, which will be labeled with the index $i$. Their spacing from the target, $\mathbf{R}_i$, grows quadratically. We choose this spacing because it is on par with the angular resolution criterion, where we satisfy that nodes are opened until they appear to be smaller than a certain angle. A more detailed discussion can be found in the work of \cite{wunsch2021}. The volume, $V_i$, associated with evaluation point $i$ is the space between evaluation point $i$ and $i+1$. Fig. \ref{fig: ray bin} illustrates the shape of a ray and all the quantities just mentioned.

Two kernels in the form of lookup tables are used in order to map nodes to rays. First, the angular overlapping fraction, $\alpha_\mathrm{node}(h)$, of a node with a ray, $h$, is looked up in the corresponding kernel. The kernel holds montecarlo precalculated values for nodes of different node sizes at discrete angles ,$\vartheta_{\mathrm{fine}, j}$ and $\varphi_{\mathrm{fine}, j}$, and discrete distances, $R_{\mathrm{fine},\, j}$. These discrete coordinates are chosen such that they closely represent the true position of the node (i.e. by taking the closest matching discrete value for either value). The second kernel treats radial mapping. It ensures that quantities affected by the inverse square law stay conserved, when they are mapped from their true position, $\mathbf{r}_\mathrm{node}$, to an evaluation point at distance $R_i$. The kernel holds pre-calculated correction values for nodes at discrete distances $R_{\mathrm{fine},\, j}$. By choosing a large set of $R_{\mathrm{fine},\, j}$ we approximate the true correction factor, $\beta^*_\mathrm{node}(i)$, with, $\beta_\mathrm{node}(i)$, such that
\begin{eqnarray}
    \beta_\mathrm{node}(i) = \frac{R_{\mathrm{fine},\, j}^2}{R_i^2} \approx \frac{r_\mathrm{node}^2}{R_i^2} = \beta^*_\mathrm{node}(i).
\end{eqnarray}
For a given node at distance $r_\mathrm{node}$, $R_{\mathrm{fine},\, j}$ is choosen so that 
\begin{eqnarray}
    (R_{\mathrm{fine},\, j} < r_\mathrm{node}) \land (R_{\mathrm{fine},\, j+1} > r_\mathrm{node})
\end{eqnarray}
hold. We allocate the radial kernel inside shared memory, because of its large size. This allows us to keep the memory footprint small while allowing for great precision.

We implement periodic boundary conditions during ray mapping by considering periodic copies of nodes. Out of all periodic copies, only the closest ones are considered. This treatment is analogous to that described by \cite{wunsch2021}.

\subsubsection{Optically thin nodes} \label{sss: opt thin}
We correct inverse square law sensitive quantities in the following way:
\begin{eqnarray}
    \hat{L}_\mathrm{node} &=& \beta_\mathrm{node} \, L_\mathrm{node}, \\
    \hat{A}_\mathrm{node} &=& \beta_\mathrm{node} \, \widetilde{A}_\mathrm{node}.
\end{eqnarray}
This way the flux seen from the luminosity $\hat{L}_\mathrm{node}$ at distance $R_i$ approximates the flux seen from $L_\mathrm{node}$ at distance $r_\mathrm{node}$. The same holds for $\hat{A}_\mathrm{node}$.

Each evaluation point, $i$, of a ray, $h$, holds
\begin{eqnarray}
  \hat{L}_{\mathrm{thin},\, i}^h &=& \sum^{\mathrm{nodes}(i)} \alpha_\mathrm{node}(h) \hat{L}_\mathrm{node,\,thin} ,\label{eq: Lthinmap}\\
    \hat{A}_{\mathrm{thin},\, i}^h &=& \sum^{\mathrm{nodes}(i)} \alpha_\mathrm{node}(h) \, \hat{A}_\mathrm{node} , \label{eq: Athinmap}\\
    \hat{V}_{\mathrm{thin},\, i}^h &=& \sum^{\mathrm{nodes}(i)} \alpha_\mathrm{node}(h) \, V_\mathrm{node} ,\label{eq: Vthinmap}
\end{eqnarray}
where nodes$(i)$ describes the set of nodes mapped to evaluation point $i$. $\hat{L}^h_i$ can be understood as the total unextincted luminosity of segment $i$ of ray $h$. In section \ref{sss: ray integration} we will use $\hat{A}^h_i$ to compute the corrected luminosity of our segment $i$. $V_\mathrm{node}$ is the volume of the node  volume (which is known).

\subsubsection{Optically thick nodes} \label{sss: opt thick}
For optically thick nodes we map the following quantities to rays:
\begin{eqnarray}
  \hat{L}_{\mathrm{thick},\, i}^h &=& \sum^{\mathrm{nodes}(i)} \alpha_\mathrm{node}(h) \,\underline{L}_\mathrm{node,\,thick} V_\mathrm{node} ,\\
    \hat{A}_{\mathrm{thick},\, i}^h &=& \sum^{\mathrm{nodes}(i)} A_\mathrm{node} , \\
    \hat{V}_{\mathrm{thick},\, i}^h &=& \sum^{\mathrm{nodes}(i)} \alpha_\mathrm{node}(h) \, V_\mathrm{node} , \\
    \hat{Q}_{\mathrm{thick},\, i}^h &=& \sum^{\mathrm{nodes}(i)} \alpha_\mathrm{node}(h) \, \beta_\mathrm{node}(i)\, V_\mathrm{node} .
\end{eqnarray}
Similar to eq. \ref{eq: Lthinmap}, \ref{eq: Athinmap} and \ref{eq: Vthinmap} in the optically thin case we map the luminosity, area and volume in the optically thick case as well. A fourth quantity is mapped, $\hat{Q}_{\mathrm{thick},\, i}^h$, which will later on be used to compute the angular size of the optically thick material. $\hat{Q}_{\mathrm{thick},\, i}^h$ is the only quantity that holds information about the inverse square law corretions concerning the treatment of optically thick nodes.

\subsection{Ray integration} \label{sss: ray integration}
In the following we will benefit from eq. \ref{eq:rtsimple} being 1 dimensional and aligned with our radial direction. We will apply eq. \ref{eq:rtsimple} to integrate each ray individually from the inside out. For each ray segment, $i$, along a pixel, $h$, we will compute extinction (see section \ref{sss: optdepth}) and emission (see section \ref{sss: emission}) from thin and thick material. Finally we will obtain the radiative flux, $\mathbf{F}$, and mean intensity, $\bar{J}$, in section \ref{sss: ray finalization} below.

To simplfify, we focus on one ray and drop the labeling of individual pixels, $h$, throughout this section.

If optically thick material of a node is detected to be spread across 2 adjacent segments in radial direction, then those segments are merged to one segment instead. To account for errors introduced by the inverse square law, we multiply by the factor $T(i,j) = R_i^2/R_j^2$ before adding the inverse square law sensitive content of segment $j$ into segment $i$. These quantities are $\hat{L}_{\mathrm{thin}}$, $\hat{A}_{\mathrm{thin}}$ and $\hat{Q}_{\mathrm{thin}}$. All other quantities are not affected by the inverse square law and can simply be added together during merging. This step is required, since we assume the optically thick material to take the shape of a compact object.

\subsubsection{Optical Depth and Extinction}\label{sss: optdepth}
We assume that the optically thin matter is evenly mixed within one segment and spread over the whole volume of that segment. Together with the angular size of a ray, $\omega = {4\pi}/{N_\mathrm{pix}}$, we can compute the optical depth of a ray segment $i$ accounting for optically thin matter, $\tau_{\mathrm{thin}, i}$, with the following equation:
\begin{eqnarray}
    \tau_{\mathrm{thin}, i} &=& \begin{cases} 0 \, , & \mathrm{for} \quad i=0 \\
    \frac{\hat{A}_{\mathrm{thin},\,i}}{\omega R_i^2}  \, , & \mathrm{for} \quad i>0 \end{cases}. \label{eq:tauthin}
\end{eqnarray}

Concerning the optically thick matter, we assume that the volume forms a compact sphere. Assuming a perfect mixing for optically thick material is overestimating the luminosity output at greater distances drastically.\footnote{Changing the shape of an optically thick body also changes its luminosity, as $L \propto A$ if a body is optically thick.} To be self-consistent with respect to previous assumptions, we compute the optical depth of the optically thick material, $\tau_{\mathrm{thick}, i}$, of segment $i$ by
\begin{eqnarray}
  \omega_{\mathrm{thick}, i} &=& \min \left(\frac{\hat{Q}_{\mathrm{thick},i}}{\hat{V}_{\mathrm{thick}, i}}\frac{(\hat{V}_{\mathrm{thick}, i})^{2/3}}{R_i^2},\omega \right)\, ,\label{eq: omegaThick} \\
  \tau_{\mathrm{thick}, i} &=& \begin{cases} 0 \, , & \mathrm{for} \quad i=0 \\
  \frac{\hat{A}_{\mathrm{thick},\,i}}{\omega_{\mathrm{thick}, i} R_i^2}  \, , & \mathrm{for} \quad i>0 \end{cases}\, . \label{eq: opt thick}
\end{eqnarray}
Eq. \ref{eq: omegaThick} computes the angular size, $\omega_{\mathrm{thick}, i}$, of our compact volume limited to be less or equal to the angular size of a ray, $\omega$.

The ray is integrated starting from the target position (see Fig. \ref{fig: ray bin}) going radially outwards. For the optically thin material we compute the extinction in the following way
\begin{eqnarray}
  \tau_{\mathrm{thin},\, 0\rightarrow i} &=& \sum^{i-1}_{j=0} \tau_{\mathrm{thin}, j}, \\
  \epsilon_{\mathrm{thin}, \, 0 \rightarrow i} &=& \exp(-\tau_{\mathrm{thin},\, 0\rightarrow i})\, .
\end{eqnarray}
Concerning the extinction caused by the  optically thick material we take into account its angular size. First we compute the fraction of a ray occupied by optically thick matter, $f_{\mathrm{thick},\, i}$, and its complement, $\bar{f}_{\mathrm{thick},\, i}$. Secondly we compute the cumulative effects of extinction of all segments between the target and segment $i$, $\epsilon_{\mathrm{thick},  \, 0 \rightarrow i}$.
\begin{eqnarray}
  f_{\mathrm{thick},\, i} &=& \frac{\omega_{\mathrm{thick}, i}}{\omega}, \\
  \bar{f}_{\mathrm{thick},\, i} &=& 1 - f_{\mathrm{thick},\, i}, \\
  \epsilon_{\mathrm{thick},  \, 0 \rightarrow i} &=& \prod^{i-1}_{j=0} \left(\bar{f}_{\mathrm{thick}, \, j} + f_{\mathrm{thick}, \, j} \, \exp(-\tau_{\mathrm{thick},\, j})\right)\,.
\end{eqnarray}
The combined extinction of optically thin and thick material working together is then straight forward computed by
\begin{eqnarray}
  \epsilon_{0\rightarrow i} &=& \epsilon_{\mathrm{thin},  \, 0 \rightarrow i} \cdot \epsilon_{\mathrm{thick},  \, 0 \rightarrow i}\, . \label{eq: ext inner}
\end{eqnarray}

\subsubsection{Emission} \label{sss: emission}
In this section we explain how we compute the flux at our target position arising from optically thin and thick material within each segment $i$. The equations stating the received flux (eq. \ref{eq: fthin}, eq. \ref{eq: fthick} and finally eq. \ref{eq: fcombined}) already account for the self extinction within segment $i$.

The flux received from optically thin material without extinction from inner material is given by
\begin{eqnarray}
  F_{\mathrm{thin}, i}^* = \frac{1-\exp(-\tau_{\mathrm{thin}, i})}{\tau_{\mathrm{thin}, i}} \times \frac{\hat{L}_{\mathrm{thin},i}}{4\pi R_i^2} \label{eq: fthin}\, .
\end{eqnarray}
Eq. \ref{eq: fthin} takes into account self extinction of the material inside segment $i$. We direct the reader to section \ref{s: limits} for discussion of the limits of eq. \ref{eq: fthin}.

The flux received from optically thick material without extinction from inner material is given by
\begin{eqnarray}
  F_{\mathrm{thick}, i}^* = \omega_{\mathrm{thick}, i} \frac{\hat{L}_{\mathrm{thick},i}}{\hat{V}_{\mathrm{thick}, i}}\,. \label{eq: fthick}
\end{eqnarray}

Finally we compute a linear combination of both source types based on the optical depth:
\begin{eqnarray}
  F_{i}^* = \frac{F_{\mathrm{thin}, i}^* \cdot \hat{A}_{\mathrm{thin},i}+ F_{\mathrm{thick}, i}^* \cdot f_{\mathrm{thick}, i} \hat{A}_{\mathrm{thick},i}}{\hat{A}_{\mathrm{thin}, i} + f_{\mathrm{thick}, i} \hat{A}_{\mathrm{thick},i}} \label{eq: fcombined}\, .
\end{eqnarray}
In order to account for the extinction of the inner part, we use $\epsilon_{0\rightarrow i}$ from eq. \ref{eq: ext inner} to compute
\begin{eqnarray}
  F_{i} =\epsilon_{0\rightarrow i}F_{i}^*\, .
\end{eqnarray}
and finally sum up over all evalutation points by taking the sum
\begin{eqnarray}
  F =\sum_i F_{i} \label{eq: finalpixelflux}\, .
\end{eqnarray}
$F$ is now the final flux we receive from our pixel.

\subsection{Ray finalization} \label{sss: ray finalization}
After section \ref{sss: ray integration} we know the flux received from each pixel, $F^h$ (eq. \ref{eq: finalpixelflux}). To compute the net flux, $\mathbf{F}$, at our target position we sum up all pixels taking into account the orientation of each ray, $\hat{\mathbf{n}}^h$. Note that $|| \hat{\mathbf{n}}^h ||= 1$. Since the radiation is antiparallel with the ray, we have to introduce a minus sign. Finally we compute
\begin{eqnarray}
    \mathbf{F} &=& \sum_{h=1}^{N_\mathrm{pix}} -F^h \, \hat{\mathbf{n}}^h\, .
\end{eqnarray}
Similar, the mean intensity, $\bar{J}$, can be computed by summation over all pixels and by including the local heating intensity generated by a source, $J_\mathrm{src}$. We summarize
\begin{eqnarray}
  J_\mathrm{src,\, IR} &=& \left(1 - e^{-\tau_\mathrm{b-node}/2} \right) \frac{L_\mathrm{src,\, IR}V_\mathrm{b-node}}{V_\mathrm{b-node}^{2/3}V_\mathrm{src}}, \label{eq:jsrc}\\
  \bar{J} &=& \frac{1}{4\pi} \sum_{h=0}^{N_\mathrm{pix}} F^h + J_\mathrm{src, \, IR},
\end{eqnarray}
where $\tau_\mathrm{b-node}=\rho\kappa \mathrm{d}x$ and $V_\mathrm{b-node}$ are the optical depth and volume of the bottom node containing the point source, respectively. Note that $J_\mathrm{src}=0$ if the bottom node does not contain a source. We use $\bar{J}$ in order to heat the dust with radiation coming from surrounding material and point sources as explained in section \ref{ss:heating_and_cooling_of_dust}. $\bar{J}$ is then stored in the field \texttt{IRXE} in units erg s$^{-1}$ cm$^{-2}$ sr$^{-1}$.

\subsection{Momentum from Flux}\label{sss: momentum from flux}
The momentum equation reads
\begin{eqnarray}
  \frac{\partial \rho \mathbf{v}}{\partial t} + \nabla \left( \rho \mathbf{v} \otimes \mathbf{v} + \left( P + P_\mathrm{src} \right) \right)= \frac{\kappa \rho}{c} \mathbf{F}\, ,  \label{eq:fullMomentum}
\end{eqnarray}
where we obtain the trapped radiation pressure, $P_\mathrm{src}$, as discussed below and the flux, $\mathbf{F}$, as discussed in section \ref{sss: ray finalization}. $\rho$, $\mathbf{v}$, $P$, $\kappa$ and $c$ are the density, velocity, thermal pressure and speed of light, respectively. $P_\mathrm{src}$ is the radiation pressure that is generated by the absorption of radiation within the hosting bottom node of the point source. $P_\mathrm{src}$ is computed in the following way
\begin{eqnarray}
  P_\mathrm{src} &=& \left(1 - e^{-\tau_\mathrm{b-node}/2} \right) \frac{L_\mathrm{src}}{c A_\mathrm{src}} \,, \label{eq: psrc}
\end{eqnarray}
where $L_\mathrm{src}$ and $A_\mathrm{src}$ are the luminosity and the area of the volume containing the source. Eq. \ref{eq: psrc} does not account for the multiscattering limit which is expected to boost RP by the optical depth for IR photons \citep[see eg.][]{hopkins2011}. Instead, Eq. \ref{eq:jsrc} generates additional boosting of RP by heating the dust locally around the source in a given b-node and therefore increases the received flux for surrounding cells.

In the following we will explain how we compute the rate of momentum, $\dot{\mathbf{P}}$, absorbed by a cell of volume, $\mathrm{d}V$, from an incoming flux, $\mathbf{F}$. We assume the cell to be cubic with an edge length $\mathrm{d}x$. Density $\rho$ and opacity of the cell are given by $\rho$ and $\kappa$, respectively.

If the cell's optical depth, $\tau = \rho \kappa \mathrm{d}x$, is low so that we can neglect extinction within the cell, we can assume the flux through the cell to be constant. In that particular case $\dot{\mathbf{P}}$ can be computed as follows 
\begin{eqnarray}
  \dot{\mathbf{P}}_{\tau << 1} = \frac{\mathbf{F}}{c}\,\kappa \rho \mathrm{d}V . \label{eq: mom on cell thin}
\end{eqnarray}
In the limit where the cell is optically thick, we must have
\begin{eqnarray}
  \dot{\mathbf{P}}_{\tau > 1} \le \frac{\mathbf{F}}{c}\, \mathrm{d}A, \label{eq: mom on cell thick}
\end{eqnarray}
where $\mathrm{d}A = (\mathrm{d}x)^{2}$ is the geometric surface of the cell.

One can compute $\dot{\mathbf{P}}$ in a scenario where $\mathbf{F}$ is not affected by the inverse square law and hits the cube face on. In this case we find $\dot{\mathbf{P}}$ can be expressed in the following way
\begin{eqnarray}
  \dot{\mathbf{P}} = \frac{\mathbf{F}}{c} \kappa \rho \frac{1-e^{-\tau}}{\tau} \mathrm{d}V\, . \label{eq: mom on cell}
\end{eqnarray}
Note that eq. \ref{eq: mom on cell} recovers booth limits mentioned by eq. \ref{eq: mom on cell thin} and eq. \ref{eq: mom on cell thick}.
Finally we store the acceleration, $\mathbf{a}_\mathrm{RP} = \dot{\mathbf{P}}/(\rho \mathrm{d}V)$, in $x,\, y,\, \mathrm{and}\, z$ direction in the fields \texttt{RPAX}, \texttt{RPAY} and \texttt{RPAZ}. Eq. \ref{eq: mom on cell} may be prone to under- and over-flows during its numerical evaluation. We approximate the fraction holding $\tau$ to be 1 if $\tau<10^{-4}$.

The deposition of momentum by ionizing radiation onto dust is treated in an analogous manner. We receive a flux of ionizing radiation at our target position provided by {\sc TreeRay/OnTheSpot} \citep{wunsch2021} and assume a constant kappa, $\sigma_\mathrm{d} = 1.5\times 10^{-21} \, \mathrm{cm^2}$ per hydrogen atom, for UV photons. We explain the exact details of how gas and dust receive their momenta from ionized radiation in \S\ref{ss:uvrp}.

\subsection{Heating and Cooling of Dust Grains}\label{ss:heating_and_cooling_of_dust}
For each cell we compute the equilibrium temperature of dust by taking into account heating and cooling rates of dimension $\left[ \mathrm{erg \, cm^{-3} \, s^{-1}} \right]$ for multiple physical processes. These processes are:
\begin{itemize}
  \item $\Gamma_\mathrm{dust-gas}$: dust grain and gas collisional interactions \citep[][given by eq. 3.25 of theirs]{Hollenbach1979},
  \item $\Gamma_\mathrm{ISRF}$: heating by an interstellar radiation field \citep[][given by eq. 42 of theirs]{bakes1994} which we set to 0 here,
  \item $\Gamma_\mathrm{H2}$: heating from H$_2$ formation on grain surface \citep[][ their eq. 45]{Hollenbach1979, glover2007},
  \item $\Lambda_\mathrm{BB-cool}$: cooling by thermal emission,
  \item $\Gamma_\mathrm{BB-heat}$: heating by thermal radiation.
\end{itemize}
We list the equations for $\Gamma_\mathrm{dust-gas}$, $\Gamma_\mathrm{ISRF}$ and $\Gamma_\mathrm{H2}$ in appendix \ref{s:eq_heat_cool}. During the dust temperature calculation we do not allow dust to cool beyond $T_\mathrm{dust,\,floor} = 2.7 \mathrm{K}$. 

The rates of the thermal radiative processes are modelled in the following way:
\begin{eqnarray}
  \Lambda_\mathrm{BB-cool} &=& \rho \xi 4\sigma\kappa T_\mathrm{dust}^4, \label{eq: dust-cool} \\
  \Gamma_\mathrm{BB-heat} &=& \rho \xi 4\pi\kappa\bar{J} ,\label{eq: dust-heat}
\end{eqnarray}
where $\rho$ and $\xi$ are the density and dust to gas ratio respectively. Eq. \ref{eq: dust-cool} and eq. \ref{eq: dust-heat} together model cooling and heating due to thermal radiation accurately. In the limit, where we consider dust being embedded inside an optically thick medium, the mean intensity, $J$, will resemble the mean temperature of the surroundings. This is the temperature the dust will attempt to converge to. However, if the medium is optically thin along one or more \textsc{HEALPix} generated surfaces, $\bar{J}$ will have contributions of $0 \, \mathrm{K}$ allowing the dust to cool and thus remove energy from the computational domain.

In combination with the effects of $\Gamma_\mathrm{dust-gas}$, dust may drain energy from the gas and radiate it away given the right circumstances. These circumstances are given if the gas is substantially hotter than the dust temperature according to the mean radiatiave intensity, $\bar{J}$. We can assign a temperature to the radiation field by solving for the equilibrium between $\Lambda_\mathrm{BB-cool}$ and $\Gamma_\mathrm{BB-heat}$, which yields
\begin{eqnarray}
  T_\mathrm{\bar{J}} = \sqrt[4]{\frac{\pi\bar{J}}{\sigma}}\,. \label{eq:radtemp}
\end{eqnarray}

\section{Benchmark of \textsc{TreeRay/RadPressure}}\label{s:benchmarks}
We will present results of \textsc{TreeRay/RadPressure} and verify them against analytically known solutions. The tests performed within this scope include a single radiating point source (see \S\ref{ss:point_source}), an optically thick blob of dust radiating thermal emission (see \S\ref{ss:rp_test_opt_thick}), dust chemistry (see \S\ref{ss:dust_chemistry}) and a radiation pressure driven expansion of a bubble to verify the correct momentum deposition (see \S\ref{ss:rp_bubble}). We perform all tests with \textsc{TreeRay} set to 48 rays. Note that we perform an additional test in \S\ref{ss:sf_discussion} with an increased number of rays using a more realistic setup of star formation.
\subsection{Point source inside a homogeneous medium} \label{ss:point_source}
\begin{figure}
    \center
    \includegraphics[width=\columnwidth]{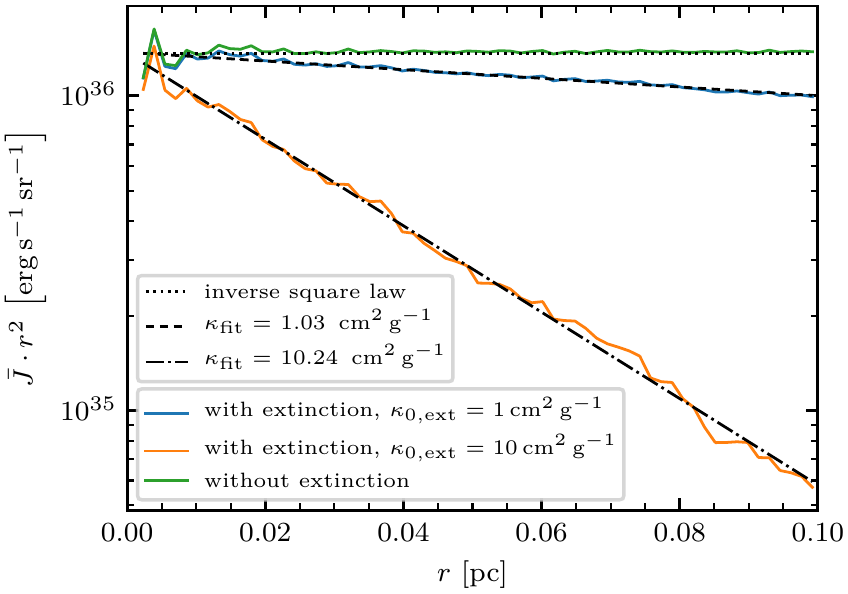}
    \caption{A single point source inside a homogeneous density medium. We compare three scenarios, two with extinction where $\kappa_{0,\,\mathrm{ext}}=1\,\mathrm{cm^2\,g^{-1}}$ (solid blue line) and $\kappa_{0,\,\mathrm{ext}}=10\,\mathrm{cm^2\,g^{-1}}$ (solid orange line) and one without extinction (solid green line). We show the mean intensity times the distance squared as a function of distance. Without extinction the inverse-square law is recovered. For the runs with extinction we fit an exponential to the simulation results}. We recover a dust opacity of $\kappa_\mathrm{fit} = 1.03\,\mathrm{cm^2\,g^{-1}}$ and $\kappa_\mathrm{fit} = 10.24\,\mathrm{cm^2\,g^{-1}}$ as shown by the dashed and dash-dotted lines.\label{fig:point_source}
\end{figure}
We consider a point source with a luminosity of $5.7\times10^4\, L_\odot$ inside a $(0.1 \, \mathrm{pc})^3$ computational domain of homogeneous density $\rho_0 = 10^{-18} \, \mathrm{g \, cm^{-3}}$. Background radiation is set to $0\,\mathrm{K}$ and the boundary conditions are isolated such that there is no effect from the ambient medium. We disable re-emission from dust and only consider radiation from the principal source so that we can validate the radiative transfer for a single point source in a simplified way. We perform three runs with a constant dust opacity of $\kappa_{0, \mathrm{ext}}= 1\, \mathrm{cm^2\,g^{-1}}$, $\kappa_{0, \mathrm{ext}}= 10\, \mathrm{cm^2\,g^{-1}}$ and $\kappa_{0, \mathrm{no \, ext}} = 0 \, \mathrm{cm^2\,g^{-1}}$. The two groups of runs are labeled with and without extinction, respectively.

We expect the radial profile of the radiation field to follow a combination of the inverse square law and the Beer-Lambert law, namely
\begin{eqnarray}
    \bar{J} = \bar{J}_0 \times \frac{e^{-\tau}}{4\pi r^{-2}}\,. 
\end{eqnarray} Fig. \ref{fig:point_source} shows the mean radiative intensity times the distance squared, $\bar{J}\cdot r^{2}$, effectively disposing the effects of the inverse square law. Here, we expect the inverse square law to be a horizontal line as shown by the black dotted line. Our run without extinction is in agreement with the inverse square law, while the runs with extinction decreases exponentially according to the Beer-Lambert law. By fitting an exponential to the runs with extinction we recover the dust opacities with a value of $\kappa_{\mathrm{fit}} = 1.03\,\mathrm{cm^2\,g^{-1}}$  and $\kappa_{\mathrm{fit}} = 10.24\,\mathrm{cm^2\,g^{-1}}$, respectively. The dashed and dash-dotted lines in Fig. \ref{fig:point_source} show the fitted curves. Note that the momentum input is given by Eq.~\ref{eq: mom on cell} which means that an accurate radiation field will yield the correct momentum input caused by radiation pressure in this test. This is because $\mathbf{F}$ and $4\pi\bar{J}$ can be considered to have the same magnitude, as the single point source is the only source of radiation. In our case we are overshooting the extinction by $\sim 3$\,\% and thus underestimating the momentum input from our point source. This behaviour results from inaccuracies introduced by not tracking the material in front of the source accurately. Rather, all of the material including the source appear to be mixed inside each radial volume along the ray.

\subsection{Optically thick blob of dust}\label{ss:rp_test_opt_thick}
In this section we want to show that the superposition of optically thin and thick contributions add up to the correct solution.

Let us consider a spherical, optically thick body with radius $r_\mathrm{c}$ and density $\rho_\mathrm{c} = 10^{-17}\,\mathrm{g\, cm^{-3}}$ at temperature $T_\mathrm{c}=30\,\mathrm{K}$. The blob is placed inside a low density ambient medium with temperature $T_\mathrm{a} = 10 \, \mathrm{K}$ and density $\rho_\mathrm{a} = 10^{-19}\, \mathrm{g\, cm^{-3}}$ and the background radiation temperature is set to zero. The density field of a blob with radial size $r_\mathrm{c}$ is given by the following equation
\begin{eqnarray}
  \rho\left(\mathbf{r}\right) = \begin{cases} \rho_\mathrm{c} \quad \mathrm{if}\, r \le r_\mathrm{c}\\ \rho_\mathrm{a} \quad \mathrm{else} \end{cases}\, ,
\end{eqnarray}
where $r$ is the distance measured from the centre of the blob. Such a scenario is difficult to handle for backwards ray tracing, as the geometric size of an optically thick object has to be tracked accurately, because the luminosity scales with area.

We realise 3 different blob sizes each inside a $(15\, \mathrm{pc})^3$ sized computational domain and label them S, M and L in an increasing order. The blobs S, M and L are of size 0.75 pc, 1.5 pc and 3 pc in diameter, respectively. For each blob size we perform a run taking into account emission and absorption from the surrounding ambient medium.

\begin{figure}
    \includegraphics[width=\columnwidth]{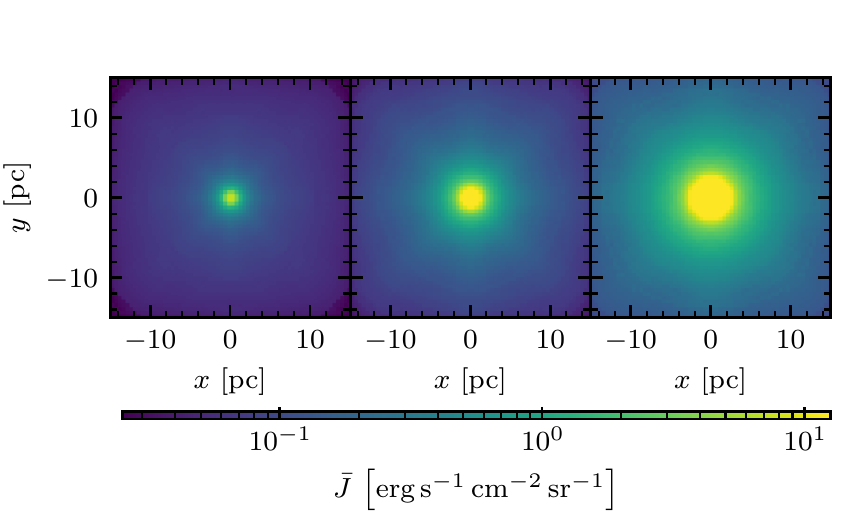}
    \caption{Slice through $z=0$ of small (S), medium (M) and large (L) sized blobs with central temperature $30 \, \mathrm{K}$ showing the mean radiation intensity, $\bar{J}$. The blobs are 2, 4 and 8 grid cells in diameter.\label{fig: slblob}}
\end{figure}

Fig. \ref{fig: slblob} shows $\bar{J}$ in a slice through $z=0$ for S, M and L. The blobs are 2, 4 and 8 grid cells wide. The radiation field produced by the blob of hot dust are spherically symmetric in all three cases.

We perform two additional runs for S, which we label S$^-$ and S$^+$. S$^-$ is run without emission and absorption from the ambient medium, so that $\bar{J}$ as a function of distance approximately follows an $r^{-2}$ profile, because the blob can be approximated by a point source at large distance. The run S$^+$ has an increased density of the ambient medium by a factor of 10. This causes the ambient medium to be optically thick on the length scale of the computational domain. We expect $\bar{J}(r)$ to converge to a value of 10 K, which is the equilibrium temperature, of greater distances. For the other runs, S, M and L, we expect the profile to drop below the $\bar{J}(T=10 \, \mathrm{K})$ value, because the outer regions cool radiatively since we set the background temperature to zero.

\begin{figure}
    \includegraphics[width=\columnwidth]{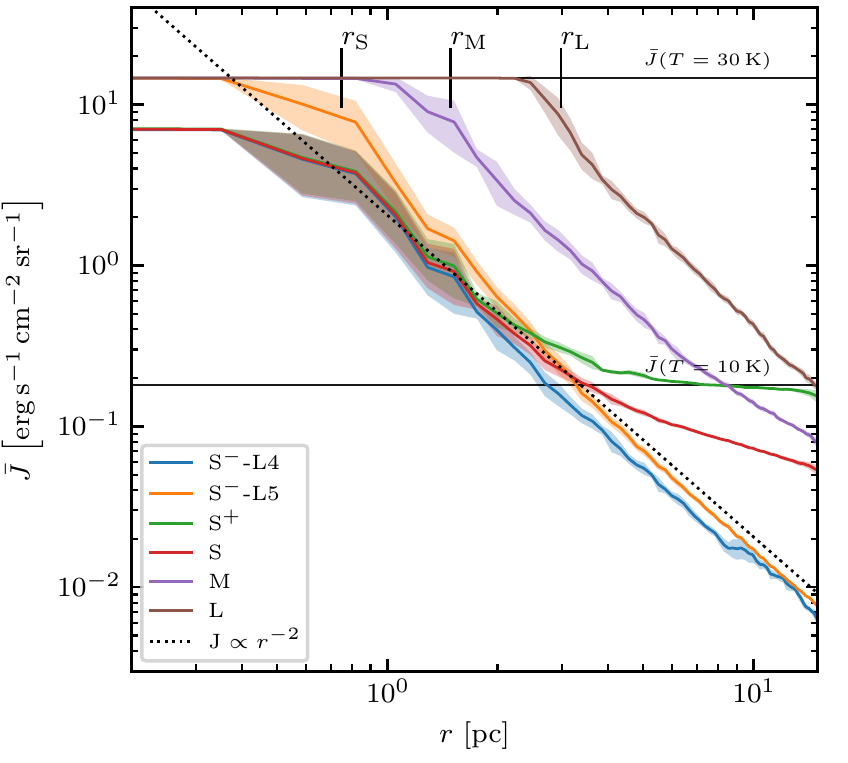}
    \caption{Radial average of the the mean intensity, $\bar{J}$, for different optically thick blobs of hot dust at $T_\mathrm{dust} = 30\,\mathrm{K}$ inside a 10 K cold optically thin medium. Small (S, red solid line), medium (M, purple solid line) and large (L, brown solid line) sized blobs with have a density of  $\rho_\mathrm{c}=10^{-17}\,\mathrm{g\, cm^{-3}}$. Three additional runs for the small blobs are shown, one where the background medium is turned optically thick (S$^+$, green solid line) and two where it is removed (S$^-$-L4, blue solid line, and S$^-$-L5, orange solid line). The run S$^-$-L4 is run at $(64)^3$ resolution and the run S$^-$-L5 at $(128)^3$. The shadows of the respective curves indicate the $1\, \sigma$ region. The black dotted line shows the Inverse-square law for an object of luminosity equal to that of model S. The black solid horizontal lines show the value of $\bar{J}$ corresponding to 30 K and 10 K, respectively. The vertical marks labeled $r_\mathrm{S}$, $r_\mathrm{M}$ and $r_\mathrm{L}$ mark the radius of the blobs S, M and L, respectively.\label{fig: blob}}
\end{figure}

Fig. \ref{fig: blob} shows the mean radiation intensity, $\bar{J}$, as a function of distance for our five runs. The black dotted line shows the analytic solution for a point particle with a luminosity corresponding to the central dense object in the run labelled S. The analytic expression we use to model the small blob of radius $R_\mathrm{S}$ is given by
\begin{eqnarray}
    \bar{J}_\mathrm{S}(r) = \frac{1}{4\pi}\sigma T^4\frac{R_\mathrm{S}^2}{r^2}\, .    
\end{eqnarray}
Additionally, black horizontal lines mark the values $\bar{J}(T=10 \, \mathrm{K})$ and $\bar{J}(T=30 \, \mathrm{K})$.We can see that the runs S, S$^+$ and S$^-$ agree with the $r^{-2}$ profile of the black dotted line for small distances. Note that S does not reach the full 30 K value of $\bar{J}$ because the core is only realised by $2\times 2 \times 2$ cubic cells. Evaluating the radiation field at one of those 8 cells will create HEALPix pixel which are not populated with 30 K material so that we do not reach the full $\bar{J}(T=30\,\mathrm{K})$ value. For greater distances only S$^-$ follows the $r^{-2}$ profile, while the ambient medium contributes additional radiation in the case of S and S$^+$ causing deviation. S$^+$ does enter the expected plateau phase in the ambient medium, similar to M and L which plateau in the optically dense core at $\bar{J}(T=30\,\mathrm{K})$. However outside we do not see a flattening caused by the ambient medium for M and L.

From this test we conclude that the method of splitting material in optically thin and thick contributions (see section \ref{ss:rad_transfer}) converges to the correct $\bar{J}$ in an optically thick embedded case. Also note that the interior of the core remains at constant temperature, hence there is acceleration caused by radiation pressure, because $\mathbf{F}= 0$ is the case inside. Additionally, there is no trapped radiation pressure inside, since a source of radiation does not exist. Outside in close proximity, there is a region where we can model the blob to be a point source, such that $\mathbf{F} \propto r^{-2} \mathbf{e}_r$ is roughly satisfied. Further outwards, we reach a state where the ambient medium shields the radiation from the blob and dominates in emission. Again we find $\mathbf{F} = 0$ in that region.
\subsection{Dust Chemistry}\label{ss:dust_chemistry}
We show results from our method discussed in section \ref{ss:heating_and_cooling_of_dust} in this section. We setup a homogeneous density field of density $\rho_0 = 10^{-16}\,\mathrm{g\,cm^{-3}}$ at a gas temperature of $T_\mathrm{gas}=2000\, \mathrm{K}$. Initially, the dust temperature is set to $T_\mathrm{dust} = 150\, \mathrm{K}$.  The entire domain is $(30\,\mathrm{pc})^3$ at a resolution of $(8)^3$ grid cells. The background radiation field is set to $T_{\mathrm{BG}} = 0\, \mathrm{K}$. We disable any hydrodynamical evolution and focus on the interactions of dust and the non-equilibrium chemical network.

Dust tends to cool to the background temperature, $T_{\mathrm{BG}} =0\, \mathrm{K}$, and is floored at $2.7 \, \mathrm{K}$. We use a dust to gas ratio of $\frac{1}{100}$. Heating of dust originates from dust-gas coupling given by $\Gamma_\mathrm{dust-gas}$ within this setup. In addition, emission of dust from elsewhere may contribute to the local $\bar{J}$ (see eq. \ref{eq: dust-heat}) and partially stall the cooling process of dust.

\begin{figure}
    \includegraphics[width=\columnwidth]{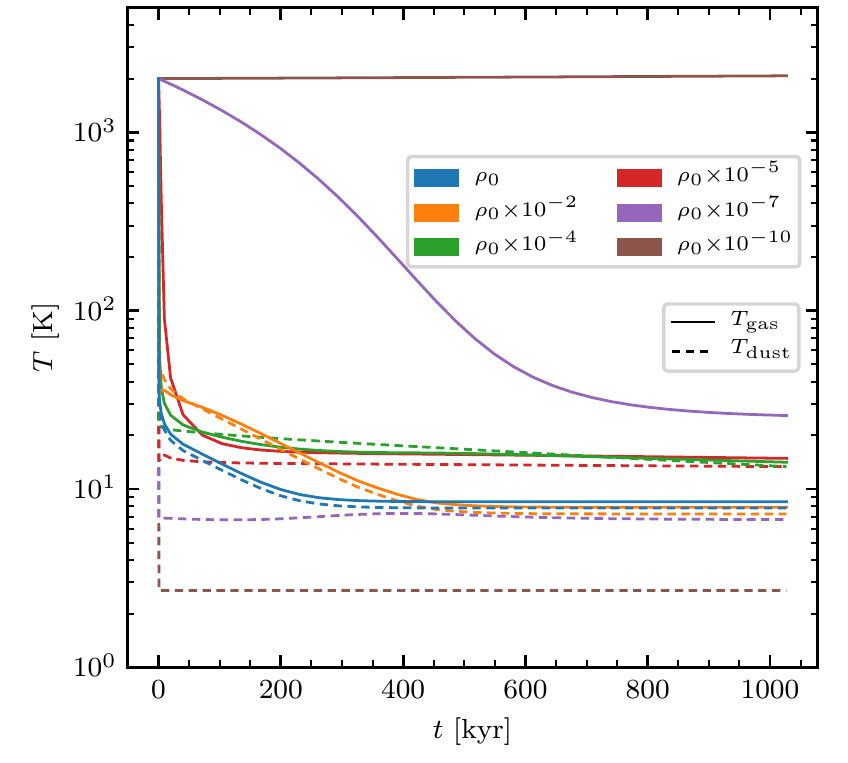}
    \caption{Dust temperature calculation showing the effects of dust cooling on the gas temperature. The initial gas temperature is set to 2000 K with $\rho_0 = 10^{-16} \, \mathrm{g\, cm^{-3}}$. The dust temperature is calculated according to \S\ref{ss:heating_and_cooling_of_dust}. Dust tends to cool to $T_\mathrm{dust, floor}=2.7\,\mathrm{K}$ radiatively but the gas dust coupling heats the dust, preventing the dust from reaching $T_\mathrm{dust, floor}$. Because $\Gamma_\mathrm{dust-gas}\propto  \rho^2 (T_\mathrm{gas} - T_\mathrm{dust})$ is the case, dust gas coupling is effective at high densities and high temperature differences. Thus in the early stage dust is able to impact the gas temperature more effectively due to larger differences in temperature. In the run $\rho_0\times10^{-10}$ the radiative cooling of dust is stronger than the gas-dust coupling and so that dust reaches $T_\mathrm{dust, floor}$. \label{fig: dtemp}}
\end{figure}

Fig. \ref{fig: dtemp} shows the temperature of dust and gas as a function of time for different densities, $f\times\rho_0$, where $f=10^{0},\,10^{-2},\,10^{-4},\,10^{-5},\,10^{-7},\,10^{-10}$. At low densities dust and gas remain uncoupled and the dust quickly cools down to $T_\mathrm{min}$ while the gas remains warm. As the density increases, dust begins to have an impact on the gas temperature through the collisional interaction terms and vice versa. Gas and dust are heated by additional $\mathrm{H}_2$ formation as the gas starts cooling down. As a function of density, the $\mathrm{H}_2$ formation process happens on different timescales. Dust and gas remain at a temperature greater than $10 \,\mathrm{K}$ as long as this process is ongoing in our test (see red, green and orange lines in fig. \ref{fig: dtemp}). Once this process has finished, the mixture of gas and dust is able to cool below $10 \, \mathrm{K}$ (see orange line in fig. \ref{fig: dtemp}). In our test the H$_2$ formation is still ongoing for densities smaller than $\rho_0 \times 10^{-4}$ at the end of the simulation time resulting in temperatures between 19 K and 16 K. At densities greater than $\rho_0 \times 10^{-4}$ the formation of H$_2$ is completed before the simulation ends and thus the temperature is able to drop below 10 K. We show the H$_\mathrm{2}$ fraction in \ref{s: add_dust_chemistry}.

We conclude that the non-equilibrium chemical modelling may influence the dust and gas temperatures depending on the simulated timescale and densities. For example, chemical reactions may heat the gas, which may in return influence the dust temperature through collisional coupling. Thus, dust may also be indirectly heated by chemical reactions.

\subsection{Radiation pressure driven bubble} \label{ss:rp_bubble}
\begin{figure}
    \includegraphics[width=\columnwidth]{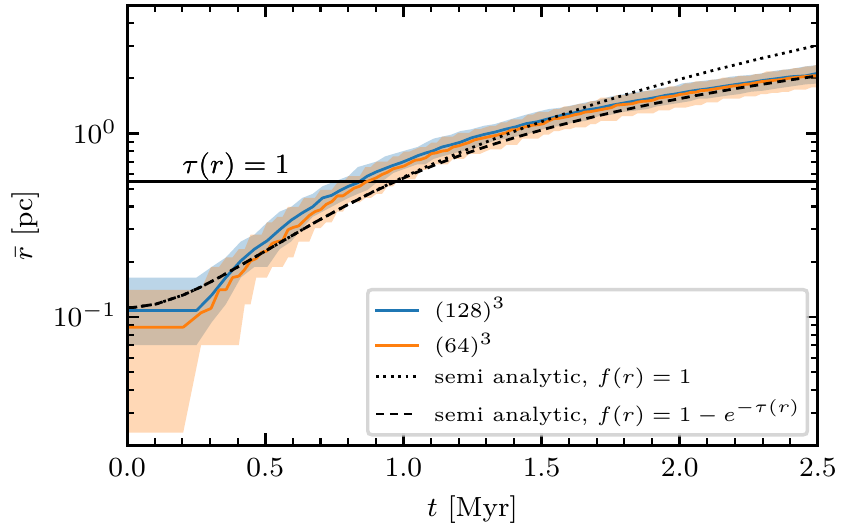}
     \caption{Mean radius of the shell, $\bar{r}$, vs. time of a single source driving a bubble through radiation pressure. The blue and orange solid lines show the numerical solutions obtained through {\sc TreeRay/RadPressure} running with $(128)^3$ and $(64)^3$ cells, respectively. The black dotted and dashed lines show semi-analytic solutions of eq. \ref{eq:EOMrpbub} with $f(r)=1$ and $f(r) = 1-e^{-\tau(r)}$, respectively. For $f(r) = 1-e^{-\tau(r)}$ (dashed line), radiation can escape the shell once it gets optically thin resulting in a momentum conserving phase above the $\tau(r)=1$ line. The other case, $f(r)=1$ (dotted line), does not account for leakage resulting in a greater $\bar{r}$ compared to $f(r) = 1-e^{-\tau(r)}$.} \label{fig:rp_bubble}
\end{figure}
In this section, we want to verify that the momentum injected through radiation pressure is correct. We consider a single point source of luminosity, $L = 1/8 \times 10^6 \, L_\odot$, embedded inside a small compact spherical core of radius, $r_\mathrm{core} = 0.15 \, \mathrm{pc}$, at density $\rho_\mathrm{c} = 1.08 \times 10^{-16}\, \mathrm{g\,cm^{-3}}$. The mass of the core is given as $M_\mathrm{c} = \frac{4}{3}\pi r_\mathrm{c}^3 \rho_\mathrm{c}$. The ambient density is $1.08\times10^{-28} \, \mathrm{g\,cm^{-3}}$. The entire setup is placed inside a $(3.0 \, \mathrm{pc})^3$ sized computational domain so that the centre of the cloud, as well as the source are positioned in one of the corners. We set the faces adjacent to the source to be reflecting (see \cite{wunsch2021} for details on the boundary condition). By doing so we only have to simulate 1/8 of the entire setup due to symmetry. The entire mass of the cloud is $M=2.2\times10^{4}\,M_\odot$, where we also take into account the mass that is not actually simulated. Faces that are not reflecting have contributions of $T_\mathrm{BG} = 0\,\mathrm{K}$.

The source is expected to inject radial momentum per unit time at a rate of $\dot{p} = L/c$ given that all of its radiation is absorbed. If the line of sight from the source outwards is optically thin, we expect the rate to decrease by a factor $f(r) = 1 - e^{- \tau(r)}$, where $\tau(r)$ measures the optical depth along the line of sight from the source through a thin shell at distance $r$. The optical depth can be calculated as the product of the surface density, $\Sigma(r) = M / 4\pi r^2$, and the dust opacity, $\kappa(T)$. We compute $\Sigma(r)$ by assuming that the cloud behaves like a thin shell, where all its mass is concentrated at a radial distance, $\bar{r}$, and distributed smoothly across the entire angular space. The temperature, $T$, for $\kappa(T)$ is computed by solving for the equilibrium between the two rates given by eq. \ref{eq: dust-cool} and eq. \ref{eq: dust-heat}.The equation of motion reads
\begin{eqnarray}
  \ddot{r} = f(r) \cdot \frac{L}{cM}, \label{eq:EOMrpbub}
\end{eqnarray}
where $M$ denotes the mass of the shell. 

Two black lines in Fig. \ref{fig:rp_bubble} show the radius of the thin shell, $\bar{r}$, as a function of time for two numerically integrated solutions of Eq. \ref{eq:EOMrpbub}. Both lines employ $f(r) = 1$ (black dotted line) and $f(r) = 1 - e^{- \tau(r)}$ (black dashed line), respectively. The two semi-analytic solutions deviate past the point, where $\tau(r) < 1$ (above black solid horizontal line). The initial radius of the thin shell has been set to match the center of mass along the radial direction for the cloud at $\frac{3}{4}r_\mathrm{core}$.

 Additionally, we show two runs performed with {\sc TreeRay/RadPressure}, where we use $(64)^3$ and $(128)^3$ cells for the simulated domain, respectively. For these runs, the radius is determined by taking the radial distance of maximum density along 48 rays. The rays are cast along different directions that originate from the source. We compute the mean among all the rays to determine the average radius, $\bar{r}$. Additionally, we estimate the error of the shell by also tracking the location of half the maximum density. The two locations in front and behind $\bar{r}$ are shown as shadows of both curves Comparing the analytical solutions, we find the solution taking into account leaking of radiation to match the computational results of \textsc{TreeRay/RadPressure.} With this, the radial momentum carried by the light may also escape, as very little is absorbed, since $\tau(r)$ tends towards 0 for large values of $r$. Here, the shell enters a momentum conserving phase as $f(r) = 1 - e^{- \tau(r)}$ approaches zero for decreasing $\tau$.

\section{Expanding HII region}\label{s:hii_region}
\begin{table*}
	\centering
	\caption{This table organizes the parameters used in \S\ref{s:hii_region} across the two different setups in rows. The symbols $n$, $L$ and $t_\mathrm{f}$ represent the number density, luminosity and normalization time, respectively. The variables $r_\mathrm{D}$ and $r_\mathrm{\mathrm{RP}}$ are the radii the ionisation front will expand to based on the effects of D-type and RP driven expansion, respectively. Both setups run with RP turned on and off. The normalisation of time, $t_\mathrm{f}$, is calculated as explained in eq. \ref{eq:timenormalisation}.}
	\label{tab:hiio}
	\begin{tabular}{lcccccr} 
		\hline
    Case & $n \, \left[\mathrm{cm^{-3}}\right]$ & $\dot{N}_\mathrm{LyC}\,\left[\mathrm{s}^{-1}\right]$ & $L \, \left[ L_\odot \right]$ &  $r_\mathrm{RP} \, \left[ \mathrm{pc} \right]$ & $r_\mathrm{D} \, \left[ \mathrm{pc} \right]$ & $t_\mathrm{f} \, \left[ \mathrm{Myr} \right]$\\
		\hline
        \hline
    D-type dominated & $10^3$ & $10^{49}$ & $5.64\times10^{4}$ &  $1.86$ & $1.64\times10^{1}$ & 9.2\\
    RP dominated & $10^9$ & $10^{52}$ & $5.64\times10^{7}$ &  $5.87\times10^{-2}$ & $1.64\times10^{-2}$ & 0.014\\

		\hline
	\end{tabular}
\end{table*}
\begin{figure}
    \includegraphics[width=\columnwidth]{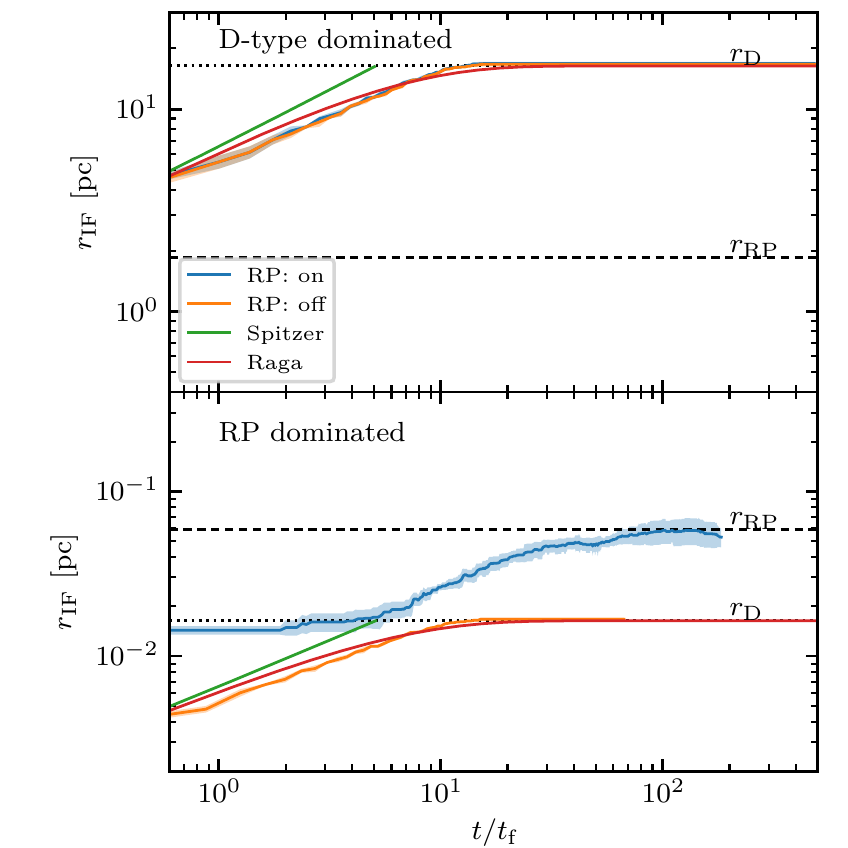}
    \caption{
    Radius of the ionisation front, $r_\mathrm{IF}$, measured from the central source vs normalised time. The blue and orange lines show the runs with and without radiation pressure, respectively.
    The green line shows the Spitzer solution. The black dashed and dotted horizontal lines show the expected radiation pressure and ionisation driven radii $r_{\mathrm{RP}}$ (eq. \ref{eq: RPstall}) and $r_{\mathrm{D}}$ (eq. \ref{eq: DTstall}), respectively.
    The red line is the numerical solution to eq. 8 of \protect\cite{raga2012}. The upper and lower panel show the case where D-type and RP are dominating, respectively.
    \label{fig:hii_region}
    }
\end{figure}
This section features a D-type expansion of an HII region that is radiation pressure assisted. We compare the importance of the thermal driving caused by ionisation to the radiation pressure driving for two cases. In one instance, radiation pressure will play a minuscule role for the dynamics of the HII region and in the other instance, radiation pressure will play the dominating part. Either case will be run with radiation pressure turned off and on, respectively, to highlight a direct comparison. We use similar setups to those of \cite{rosdahl2015scheme} in their \S3.

For the setup, we initiate a single source inside a computational domain of homogeneous number density, $n$. The source radiates at a luminosity of $L$ corresponding to a rate of $\dot{N}_\mathrm{LyC}=L/E_\mathrm{LyC}$ photons in the Lyman continuum, where $E_\mathrm{LyC}=13.6 \, \mathrm{eV}$. We place the ionizing source in one corner of the computational domain and set the boundary conditions of the cube to be reflecting on all faces connected to said corner. Table \ref{tab:hiio} summarizes the parameters we use for our simulations. Radiation pressure from non-ionizing radiation is neglected and turned off. In addition, we reduce the density of cells within a distance of 2 grid cells of the source by a factor of 100. This is to ensure that an HII region is spawned immediately, since \textsc{TreeRay/OnTheSpot} can not resolve the R-type expansion of an HII region. In particular, the Str\"omgren-radius must exceed one grid cell for an HII region to spawn. We set the number of rays within \textsc{TreeRay} to be 48 and $T_\mathrm{BG} = 0\, \mathrm{K}$.

The Str\"omgren radius \citep{stroemgren1939} is given by
\begin{eqnarray}
  R_\mathrm{s} = \left(\frac{3\dot{N}_\mathrm{LyC}m_\mathrm{p}^2}{4\pi \alpha_\mathrm{B}X^2\rho_0^2}\right)^{1/3}, \label{eq:stroemgren}
\end{eqnarray}
where $m_\mathrm{p}$, $\alpha_\mathrm{B} = 2.5\cdot10^{-13} \, \mathrm{cm^3\,s^{-1}}$ and $X$ are the mass of a proton, the case B recombination rate and the mass fraction of hydrogen in the neutral medium \citep{wunsch2021}, respectively. The equation of motion for an ionisation front expanding into a neutral homogeneous medium through time, $t$, is given by the Spitzer solution, namely
\begin{eqnarray}
    r_\mathrm{i} = R_\mathrm{s} \left(1 + \frac{7}{4} \frac{c_\mathrm{i} t}{R_\mathrm{s}}\right)^{4/7}\,,
\end{eqnarray}
where $r_\mathrm{i}$ measures the radius of the ionisation front and $c_\mathrm{i}$ the sound speed of the ionised medium.

One can express equilibrium radii at which the driving through radiation pressure and thermal expansion stalls, similar to the work of \cite{rosdahl2015scheme}. The radiation pressure is expected to be in pressure balance, if the thermal ambient medium pressure, $P_\mathrm{T_0}=n_0 k T_0$, is comparable to that of the radiation pressure at the bubble surface, $P_\mathrm{RP}=L/4\pi c r_\mathrm{RP}^2$. Here, $n_\mathrm{0}$, $k$, $T_0$ and $r_\mathrm{RP}$ are the number density and temperature of the neutral medium, the Boltzmann constant and the radius of the radiation pressure bubble, respectively. In that case we find the radius to be
\begin{eqnarray}
    r_\mathrm{RP} = \sqrt{\frac{L m_\mathrm{p} \mu_0}{4 \pi c \rho_0 k T_0}} \label{eq: RPstall}.
\end{eqnarray}
A similar argument can be used to determine the final radius of the HII region driven by its D-type expansion. We expect the bubble to expand until the pressure between the ionised and neutral medium are equal, $P_\mathrm{i} = P_\mathrm{0}$. This relation is given by
\begin{eqnarray}
    n_\mathrm{i} T_i =\frac{\rho_\mathrm{i}}{\mu_\mathrm{i} m_\mathrm{p}}T_\mathrm{i} = \frac{\rho_0}{\mu_0 m_\mathrm{p}}T_0 = n_0 T_0\, , \label{eq:peq}
\end{eqnarray}
$\rho = m_\mathrm{p}n \mu$, $\mu$ and T are the mass density, mean molecular mass and temperature, respectively, and we label the ionised and neutral medium with the subscript $\mathrm{i}$ and $0$. Our source is able to keep a specified amount of hydrogen atoms ionised, according to eq. \ref{eq:stroemgren}. Using eq. \ref{eq:stroemgren} and solving for the expected ionized number density, $n_\mathrm{i}= \rho_\mathrm{0}/(m_\mathrm{p}\mu_{\mathrm{i}})$, yields
\begin{eqnarray}
        n_\mathrm{i} = \frac{1}{X\mu_\mathrm{i}} \sqrt{r_\mathrm{D}^{-3} \, \frac{3\dot{N}_\mathrm{LyC}}{4\pi \alpha_\mathrm{B}}}\, . \label{eq:ni}
\end{eqnarray}
We can use this constraint to express $n_\mathrm{i}$ in terms of a corresponding radius, $r_\mathrm{D}$, for the equilibrium radius of the HII region driven by heating through ionization. By combining the pressure constraint given by Eq. \ref{eq:peq} and the ionized number density, $n_i$, from Eq. \ref{eq:ni}, we arrive at
\begin{eqnarray}
    r_\mathrm{D} &=& \left( \frac{T_\mathrm{i}}{T_0} \frac{\mu_0}{\mu_\mathrm{i}}\right)^{2/3} R_\mathrm{s}\rvert_{\rho_0}\, .  \label{eq: DTstall}
\end{eqnarray}
We find a slightly altered description taking into account the mean molecular weight of the neutral and ionised species in our calculations (compare their eq. 65 to our eq. \ref{eq: DTstall}).

We construct two cases corresponding to $r_\mathrm{D}>r_\mathrm{RP}$ and $r_\mathrm{RP} < r_\mathrm{D}$ where we turn RP on and off for each instance. We expect the presence of RP to have no effect on the run where $r_\mathrm{D}>r_\mathrm{RP}$ holds. On the other hand, we expect RP to drastically change the outcome of the runs where $r_\mathrm{RP}>r_\mathrm{D}$.

Fig. \ref{fig:hii_region} shows the radial distance of the ionisation front, $r_\mathrm{IF}$, vs. time for our different setups using {\sc TreeRay/RadPressure}. We determine $r_\mathrm{IF}$ as the mean distance along multiple rays aligned with the \textsc{HEALPix} algorithm. The time is normalised by
\begin{eqnarray}
    t_\mathrm{f} = \frac{r_\mathrm{D}}{c_\mathrm{i}},  \label{eq:timenormalisation}  
\end{eqnarray}
where $c_\mathrm{i}$ is the sound speed of the ionised medium. The top panel shows the runs where the expansion is thermal pressure dominated (i.e. $r_\mathrm{D} > r_\mathrm{RP}$). Both runs, regardless of RP being present, show almost identical dynamics of the ionisation front in the upper panel. This is because, the dominating effect is the D-type expansion itself. Their solution follows the solution of \cite{raga2012} and converges to the limit given by $r_\mathrm{D}$. The Spitzer solution is capped at $r_\mathrm{D}$ (dotted line) which corresponds to the D-type expansion radius from the semi-analytical solution by \cite{raga2012} (red line). The lower panel shows the RP dominated setup. The run with RP expands more quickly and further than the run without RP. The run without RP stalls at a size in agreement with the solution of \cite{raga2012}. The run with RP in the lower panel reaches a final radius matching $r_{\mathrm{RP}}$ given by Eq. \ref{eq: RPstall} and has clearly surpassed the limits provided by $r_\mathrm{D}$ and the solution of \cite{raga2012}.

\begin{figure*}
    \includegraphics[width=\textwidth]{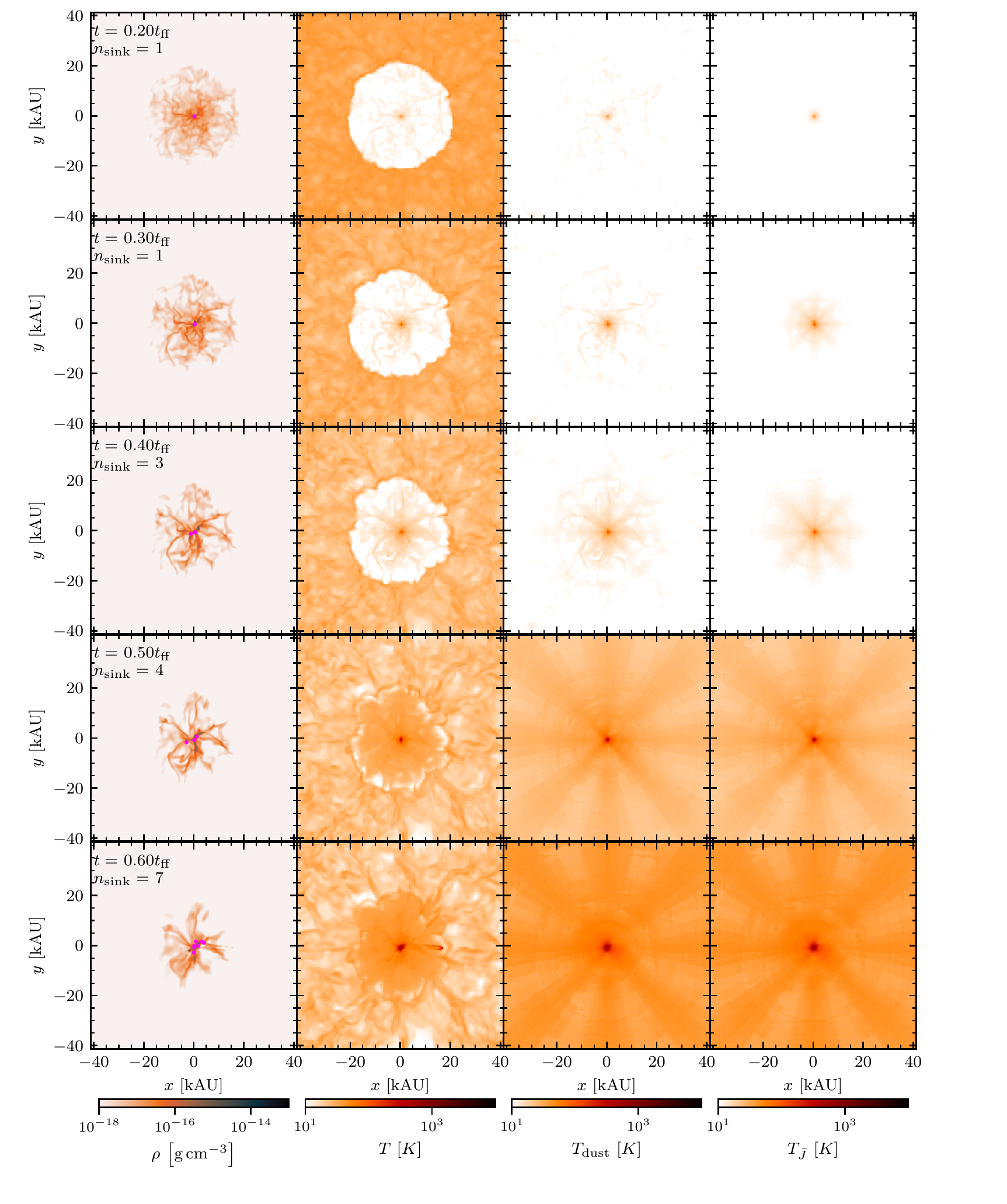}
    \caption{Slices through $z=0$. Organized in rows we show from left to right density, gas temperature, dust temperature and radiation temperature across different times. The time, $t$, is measured in free-fall times, $t_\mathrm{ff} = 42.3\, \mathrm{kyr}$. The number of sink particles present is labeled $n_\mathrm{sink}$. This figure pertains to the SF core setup of \S\ref{s:star_forming_setup}.\label{fig: rosen_all}}
\end{figure*}

\section{Star Forming Setup}\label{s:star_forming_setup}
\begin{figure*}
    \includegraphics[width=\textwidth]{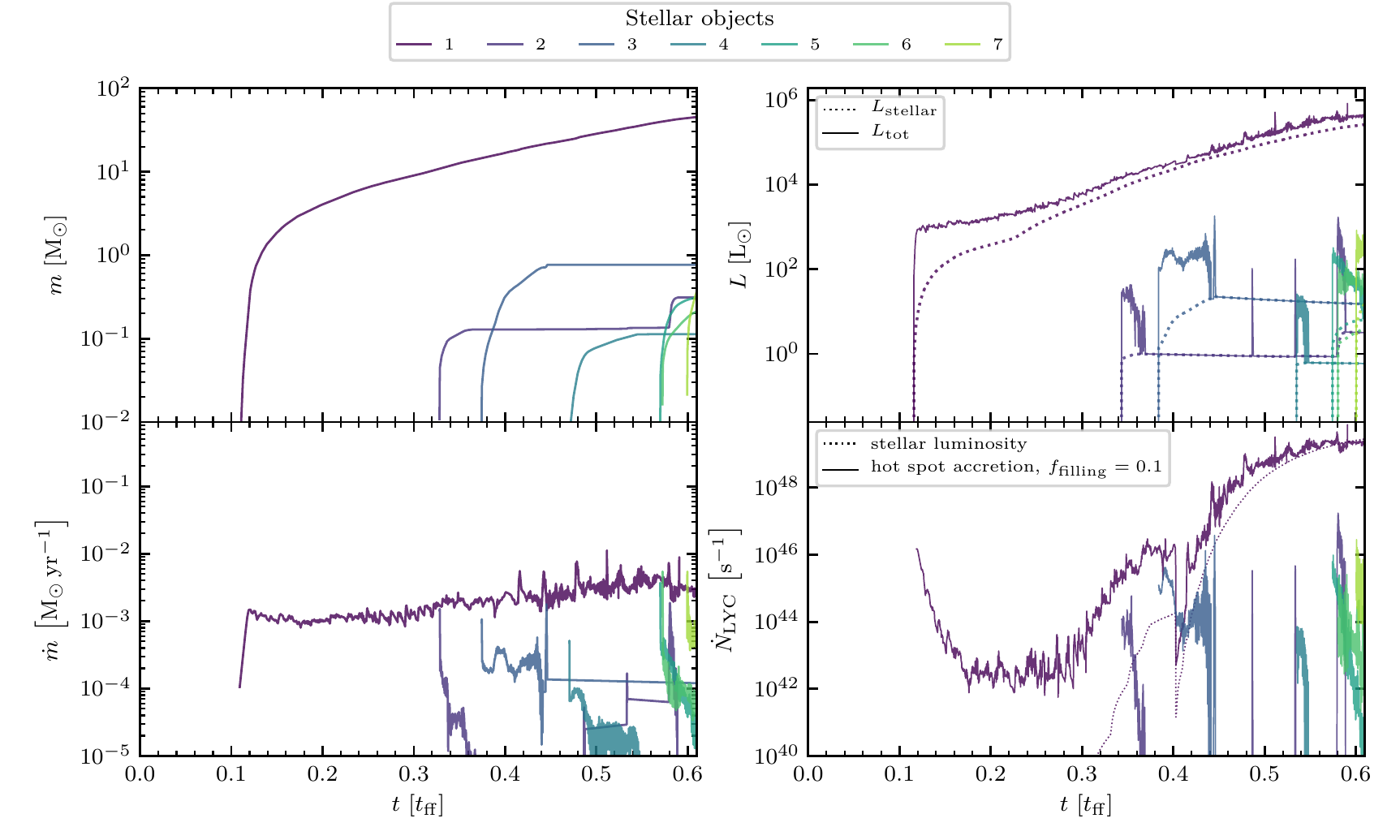}
    \caption{We show the mass, $m$, luminosity, $L$, accretion rate, $\dot{m}$, and rate of ionizing photons, $\dot{N}_\mathrm{LYC}$, vs time in this plot. In total we form 7 stellar objects represented by sink particles. The most massive star, labeled 1, forms early on embedded in the center (see Fig. \ref{fig: rosen_column}) It is the most massive star (see left upper panel) and the dominating source of radiation (see upper right panel) throughout time. \label{fig: sink_combined}}
\end{figure*}
\begin{figure*}
    \includegraphics[width=\textwidth]{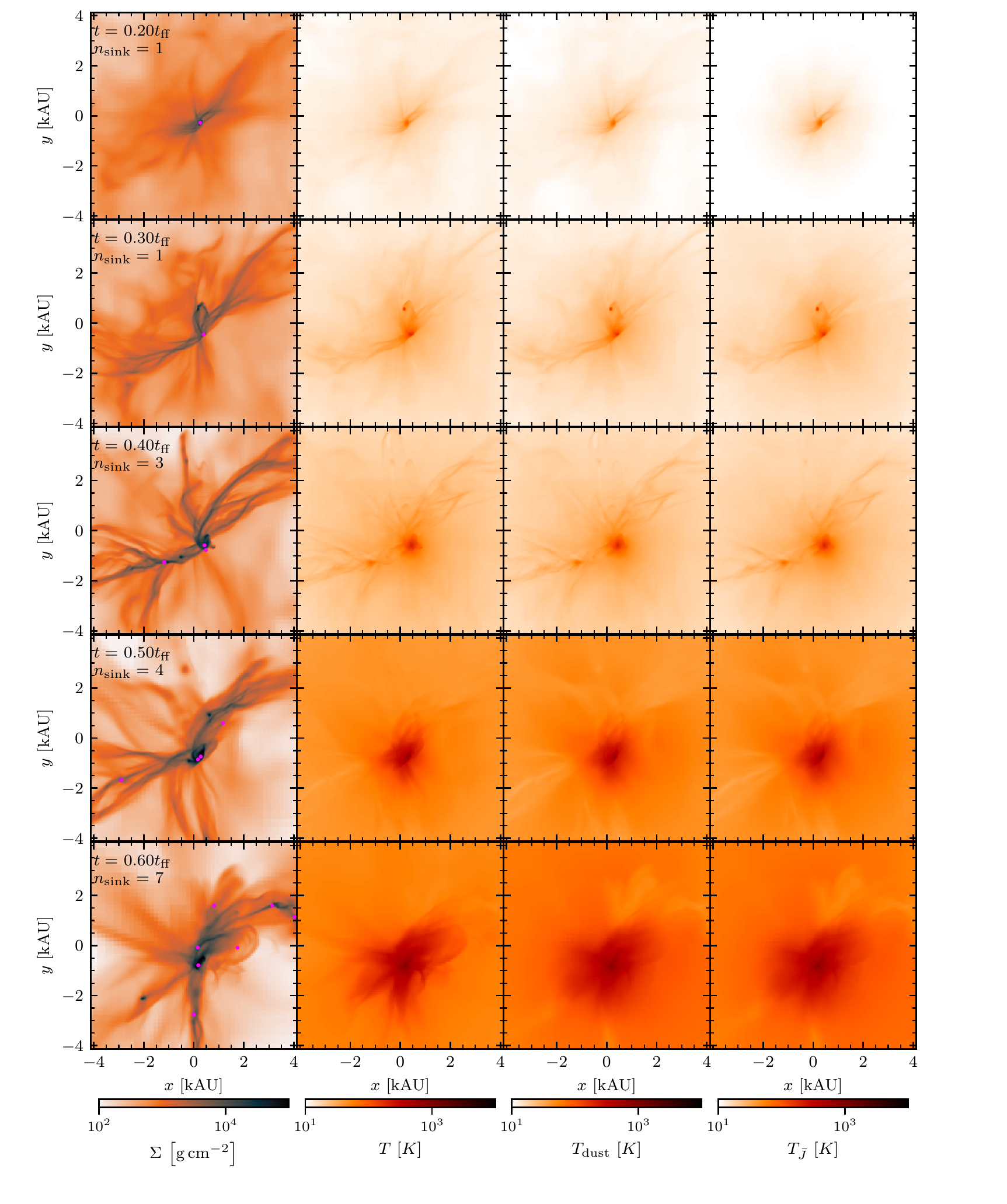}
    \caption{Projections along the $z$-coordinate axis of an ($8$ kAU)$^3$ cube around the origin for the turbulent star forming test (see \S\ref{s:star_forming_setup}). Organized in rows we show, from left to right, the column density and the density-weighted gas temperature, dust temperature and radiation temperature across different times (top to bottom). The time, $t$, is measured in free-fall times, $t_\mathrm{ff} = 42.3\, \mathrm{kyr}$. The number of formed sink particles present is labeled $n_\mathrm{sink}$. We can see stars forming in the central hub initially and later on forming in dens structures connected to the hub further out. The central hub is hotter than the ambient medium with gas and dust temperatures of around 100 K to 1000 K. At around $t=\,0.5\,t_\mathrm{ff}$ the luminosity output of the central hub increases, heating up the  surrounding material. Other stars than the primary do not cause visible changes in the radiation temperature as $T_{\bar{J}}$ is dominated by the most massive star.}\label{fig: rosen_column}
\end{figure*}

\begin{figure*}
    \includegraphics[width=\textwidth]{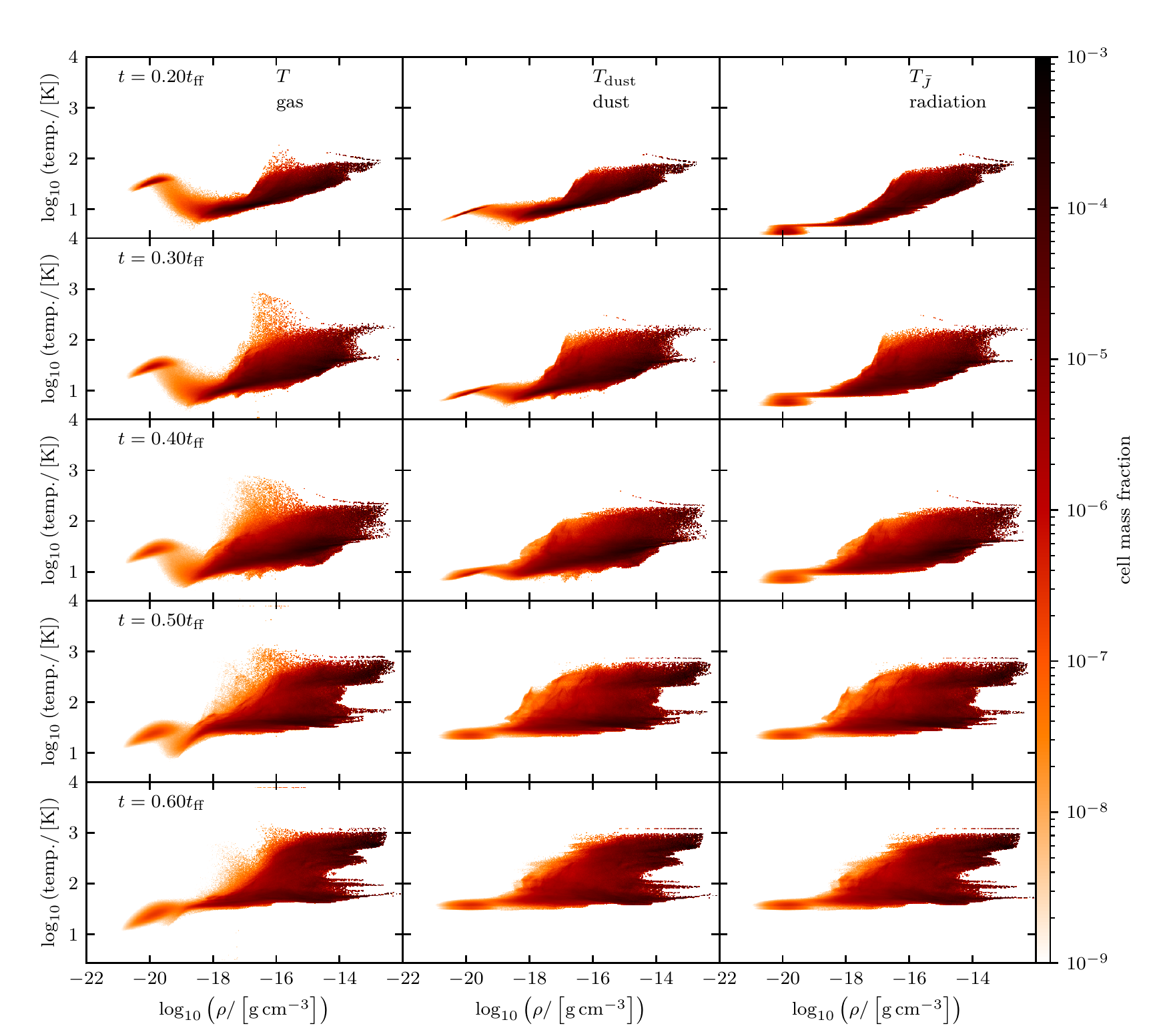}
    \caption{
    Phase diagrams of the gas, dust and radiation temperature (from left to right) for different times (from top to bottom).
    All cells are normalised by the total mass in the computational domain.
    The clump at $\rho\approx 10^{-20} \, \mathrm{g\, cm^{-3}}$ and $T\approx 10^2 \mathrm{K}$ represents the background medium which is slowly cooled by dust.
    The very dense part at around $\rho \approx10^{-14} \, \mathrm{g\, cm^{-3}}$ and $T\approx 10^2 \mathrm{K}$ shows the central hub getting hotter over time. For $t \le 0.4t_{\mathrm{ff}}$, we find $T > T_{\mathrm{dust}} > T_{\bar{J}}$ for some of the diffuse gas enabling dust to cool gas by radiating thermal energy away. Eventually, the global minimum radiation temperature increases and with it the gas and dust temperature follow.
    \label{fig:phase_combined}
    }
\end{figure*}
\begin{figure}
    \includegraphics[width=\columnwidth]{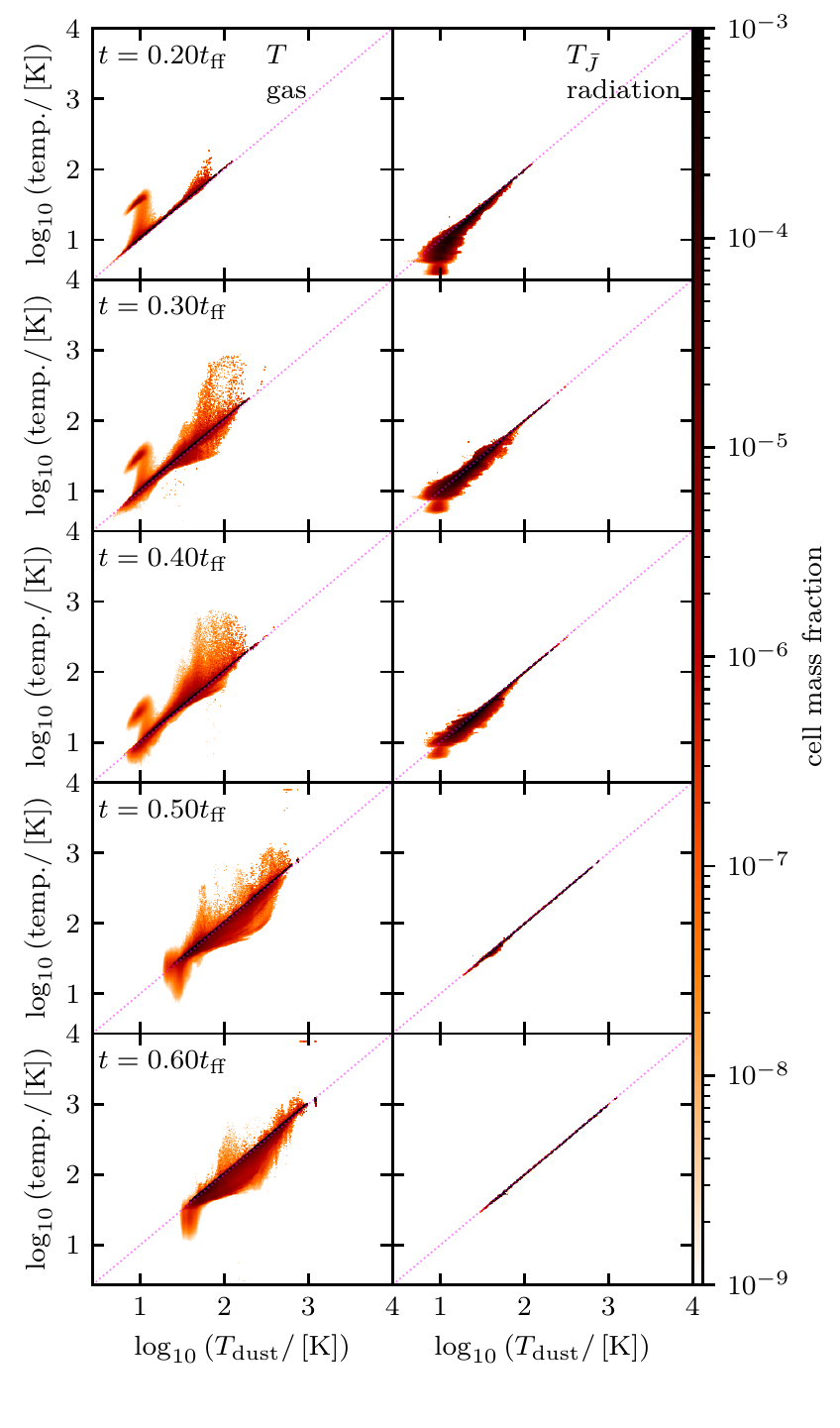}
    \caption{Phase diagrams of gas and radiation temperature vs dust temperature organized in columns. The phase diagrams are grouped into rows for different times. The dotted magenta line shows the one-to-one line. The cells are normalised to the current mass present in the entire computational domain. Most of gas and dust follow the one-to-one line with slight deviation (left column). The accretion shock heated gas is in a state able to be cooled by gas effectively ($T_\mathrm{dust} < T$). Dust is mostly hotter than the local radiation temperature (right column), which allows dust to cool radiatively.\label{fig:phase_trad_combined}}
\end{figure}
\begin{figure*}
    \includegraphics[width=\textwidth]{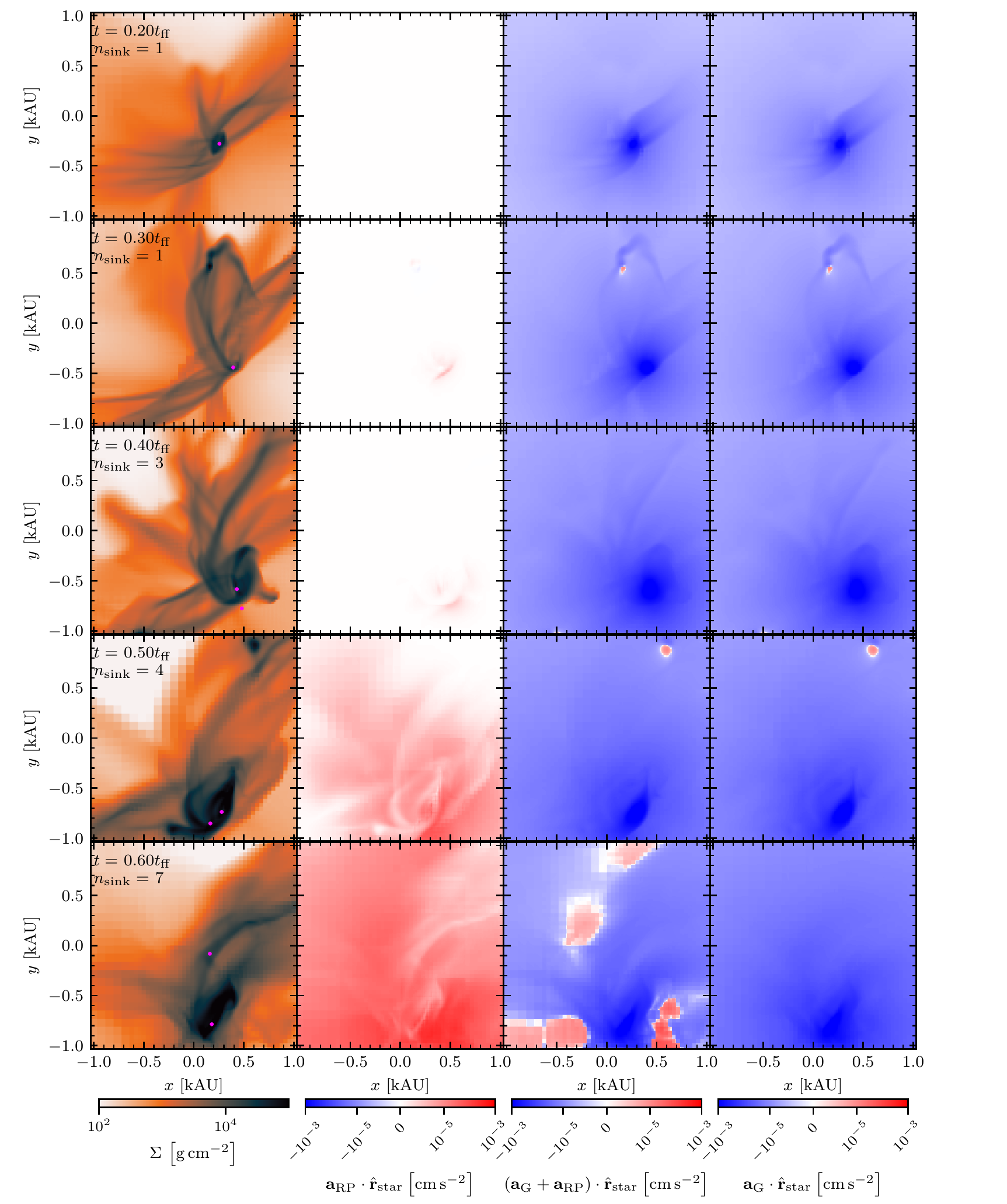}
    \caption{Projections along the $z$-axis of a ($2$ kAU)$^3$ cube around the origin. From left to right we show again the column density (for reference) and 3 different radial accelerations. All 3 acceleration vectors are density-weighted along the line of sight and projected on the line of sight towards the most massive star. Going from left to right we have the RP acceleration, the acceleration due to gravity and RP combined, and the gravitational acceleration. We can see that RP manages to oppose gravity in the diffuse gas close to the central star over time.\label{fig: rosen_radial}}
\end{figure*}

\begin{figure}
    \includegraphics[width=\columnwidth]{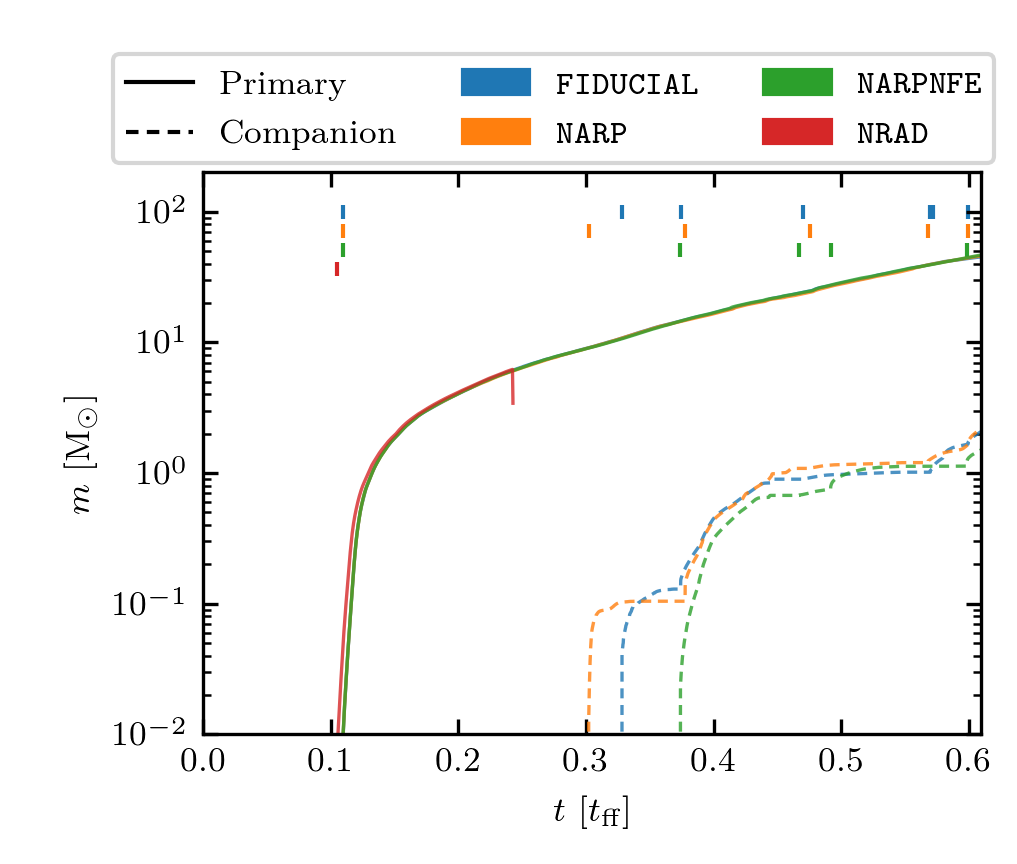}
    \caption{Mass vs. time of the central primary star and its companions for all performed runs. Tick marks at the top represent the formation of further sink particles. The primary mass does not change across different runs indicating that RP does not impact the accretion of the central star in our simulation. \label{fig: mass}}
\end{figure}

\begin{figure}
    \includegraphics[width=\columnwidth]{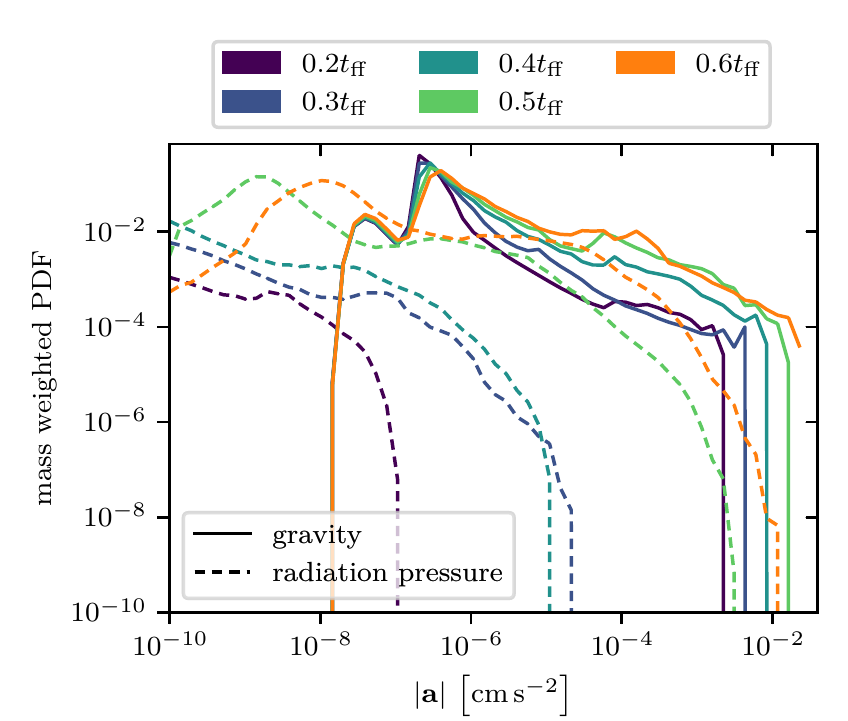}
    \caption{Mass weighted PDF of acceleration exerted as a consequence of gravity and radiation pressure across time. The impact of radiation pressure is increasing over time towards higher accelerations while the spectrum of gravity does not change by a lot.\label{fig: mw_accel}}
\end{figure}

\begin{figure}
    \includegraphics[width=\columnwidth]{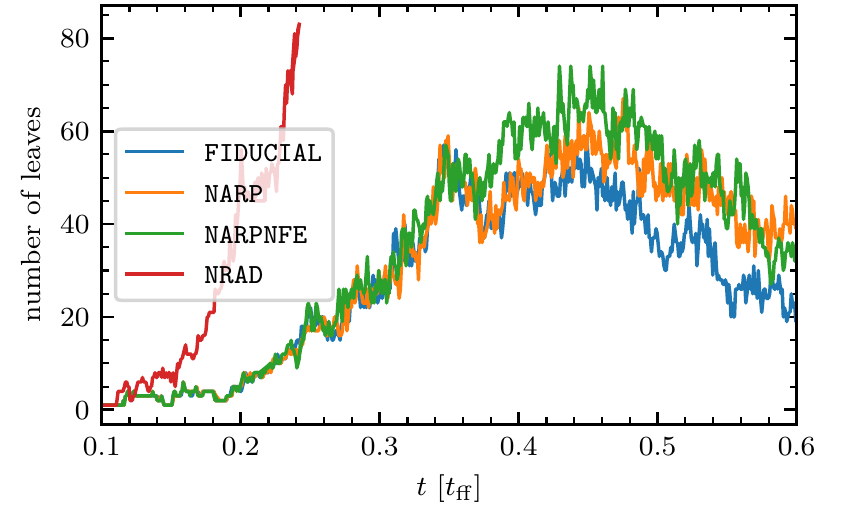}
    \caption{Number of leaves in a dendogram vs. time for all four runs. After 0.15 $t_\mathrm{ff}$ \texttt{NRAD} deviates from the runs \texttt{NARP}, \texttt{NARPNFE} and \texttt{FIDUCIAL} owing to its lack of both radiative heating from dust and stars, and RP. All other runs coincide until roughly 0.35 $t_\mathrm{ff}$ after which fragmentation is more suppressed the more physics is included. We can see that the addition of both RP and radiative heating suppresses fragmentation towards 0.6 $t_\mathrm{ff}$. \label{fig: dendro_cul}}
\end{figure}

In this section, we show results from a molecular clump undergoing gravitational collapse where we form stars self-consistently. The setup is a similar setup to the one presented by \cite{rosen2019}, except for the initial turbulent velocity field.

\subsection{Initial Conditions}
We model the evolution of a $150\, \mathrm{M}_\odot$ massive molecular clump with an initially seeded turbulent velocity field with a power spectrum, $P(k)\propto k^{-2}$, such that the clump is subvirial with a virial parameter of $\alpha = 0.14$. The density, $\rho$, follows a radial power-law profile of $\rho(r)\propto r^{-1.5}$ up to an outer cloud radius of $0.1 \, \mathrm{pc}$. The free fall time of the core is $t_\mathrm{ff} = 42.3 \mathrm{kyr}$. The cloud is pressure confined by a hot low density ambient medium of temperature $T_\mathrm{amb} = 2000 \, \mathrm{K}$ such that the density of the ambient medium is a factor of 100 lower than the density at the edge of the cloud. The gas temperature is set to $20 \, \mathrm{K}$ throughout the cloud. The entire setup is housed inside a $(0.4\, \mathrm{pc})^3$ computational domain with von Neumann boundary conditions, allowing for matter to flow in and out smoothly. We set the background infrared radiation field to $J_\mathrm{BG} = 1.205 \cdot 10^{-2} \, \mathrm{erg \, s^{-1} \, cm^{-2} \, sr^{-1}}$ which  corresponds to an equilibrium dust temperature of $T_{\mathrm{BG}} = 5 \, \mathrm{K}$.

We start with a base resolution of $(128)^3$ for the density and use 48 rays within \textsc{TreeRay}. We adaptively refine the grid to ensure that the Jeans length is refined by at least 8 grid cells following the conditions of \cite{truelove1997} to prevent artificial fragmentation. We allow for up to 5 additional levels of refinement such that the highest resolution of this simulation is $20\, \mathrm{AU}$. If the Jeans length is refined by 16 or more cells, we derefine. Condensations within our core arise self-consistently through gravitational collapse. Eventually, they may form sink particles.

\subsection{Sink Formation and Radiation from Protostars} \label{ss:sink_particles}

We only allow sink formation for cells with densities greater than $\rho_\mathrm{thresh} = 5.2\times 10^{-14} \, \mathrm{g\,cm^{-3}}$ which corresponds to 4 grid cells per Jeans-length  \citep{truelove1997} at the highest resolution, $20\,\mathrm{AU}$. We employ additional criteria for sink formation following \cite{federrath2011} and \cite{clarke2017}. These criteria are listed below. All volumes, $V_\mathrm{sink}$, which have a density greater than $\rho_\mathrm{thresh}$ qualify to form sink particles. We set $\rho_\mathrm{thresh}$ to be the Jeans density for a Jeans length corresponding to 4 grid cell sizes at highest refinement, fulfilling the conditions of \cite{truelove1997}. Additionally, we allow for the formation of a sink particle inside $V_\mathrm{sink}$ if and only if all of the following conditions are met:
\begin{enumerate}
  \item $V_\mathrm{sink}$ is on maximum refinement, \label{condition:1}
  \item $V_\mathrm{sink}$ is free of sink particles,
  \item gas inside $V_\mathrm{sink}$ is infalling,
  \item $V_\mathrm{sink}$ is located on a gravitational potential minimum,
  \item gas inside $V_\mathrm{sink}$ is gravitationally bound, \label{condition:4}
  \item gas inside $V_\mathrm{sink}$ will collapse within its own free fall time before it has the chance of being accreted by another sink particle (see \cite{clarke2017}, their eq. 4 and 5). \label{condition:6}
\end{enumerate}
Conditions \ref{condition:1} to \ref{condition:4} are taken from \cite{federrath2010} and condition \ref{condition:6} has been taken from \cite{clarke2017}. We find that \ref{condition:6} substantially reduces the number of sinks formed in accordance with \cite{clarke2017}.

We allow sink particles to model protoststars, where we use the implementation of \cite{klassen11} on GitHub\footnote{\url{https://github.com/mikhailklassen/protostellar_evolution}}. From the model we obtain an internal luminosity, $L_\mathrm{int}$, as well as an accretion luminosity, $L_\mathrm{acc}$. We assign each of these luminosities, $L_\mathrm{int}$ and $L_\mathrm{acc}$, a temperature, $T_\mathrm{int}$ and $T_\mathrm{acc}$, respectively, in the following way

\begin{eqnarray}
  T_\mathrm{int} &=& \left(\frac{L_\mathrm{int}}{ \sigma 4 \pi r_\mathrm{star}^2}\right)^{1/4} \, ,\label{eq:internalTemp}\\
  T_\mathrm{acc} &=& \left(\frac{L_\mathrm{acc}}{ f_\mathrm{filling}  \sigma 4\pi r_\mathrm{star}^2}\right)^{1/4}\, ,\label{eq:accTemp}
\end{eqnarray}

where $r_\mathrm{star}$ and $f_\mathrm{filling}$ are the stellar radius and the fraction of the area of the star upon which the star is accreting. We obtain $r_\mathrm{star}$ from the protostellar model and set $f_\mathrm{filling} = 0.1$ \citep{calvet1998}. Finally we split each luminosity, $L_\mathrm{X}$ (where $X = \mathrm{int}$ or $X=\mathrm{acc}$, based on its temperature, $T_\mathrm{X}$, to contribute towards ionizing radiation with a fraction, $\gamma(T_\mathrm{X})$, and its complement, $(1-\gamma(T_\mathrm{X}))$, to contribute towards non ionizing radiation. We compute the fraction $\gamma(T_\mathrm{X})$ by computing the ratio of luminosity emitted in the Lyman band and the bolometric luminosity at temperature $T_\mathrm{X}$. This is computed as such

\begin{eqnarray}
  \gamma(T_\mathrm{X}) = \frac{\int^\infty_{\nu_{\mathrm{LyC}}} \mathrm{d}\nu B_\nu(T_\mathrm{X})}{\int^\infty_{0} \mathrm{d}\nu B_\nu(T_\mathrm{X})}\, , \label{eq: bb_ratio}
\end{eqnarray}

where $\nu_\mathrm{LyC}$ marks the lower frequency of the Lyman continuum, and $B_\nu$ is Planck's law of black body radiation.

In total one can state the ionizing luminosity, $L_\mathrm{src,\, UV}$, and the non-ionizing luminosity,  $L_\mathrm{src,\, IR}$, of a sink particle  as:

\begin{eqnarray}
    L_\mathrm{src,\, UV} &=& \gamma(T_\mathrm{int}) L_\mathrm{int} + \gamma(T_\mathrm{acc}) L_\mathrm{acc}\, ,\\
    L_\mathrm{src,\, IR} &=& L_\mathrm{int} + L_\mathrm{acc} - L_\mathrm{src,\, UV}\, . \label{eq:LsrcIR}
\end{eqnarray}

We discuss the treatment of the non ionizing luminosity, $L_\mathrm{src,\, IR}$, in this work and treat the ionizing luminosity, $L_\mathrm{src,\, UV}$, as discussed by \cite{wunsch2021}.

In total, we include two bands of radiation. A non-ionizing band acting on dust with the novel scheme \textsc{TreeRay/RadPressure} and an ionizing band with the module \textsc{TreeRay/OnTheSpot} \citep{wunsch2021}. Both bands are emitted by sink particles, while the non-ionizing radiation may additionally be emitted by dust from everywhere inside the computational domain.

\subsection{Morphology across time}
The initial spherically symmetric configuration is quickly disturbed by the turbulent velocity field. Local high density regions are forming where streams of momentum, $\rho\mathbf{v}$, collide. In those regions the gas is heated through $p\mathrm{d}V$ work (adiabatic compression). The dust is heated by the gas through  collisional interactions, whilst trying to cool by radiating in the continuum infrared. As a result, the dust temperature is settling in between the gas temperature, $T_{\mathrm{gas}}$, and the radiation temperature, $T_{\bar{J}}$ (see eq. \ref{eq:radtemp}), as described in \S\ref{ss:heating_and_cooling_of_dust}. Fig. \ref{fig: rosen_all} shows the density, $\rho$, gas and dust temperatures, and the radiation temperature, $T_{\bar{J}}$, across different times for the entire computational domain. We can see that $T_{\bar{J}}$ is hottest at the central hub and decreases further out. This indicates that the central hub is dominating the IR luminosity output. As the simulation evolves, the radiation temperature increases globally, which can be seen most significantly comparing the times $0.4\, t_{\mathrm{ff}}$ and $0.5\, t_{\mathrm{ff}}$. The increased radiation temperature also heats the dust and therefore indirectly the gas at greater distances further out.

\subsection{Stellar population}

Fig. \ref{fig: sink_combined} shows an overview over the stellar population, showing the mass, accretion rate, luminosity and number of ionising photons vs. time. Early on, at around $0.1 \, t_\mathrm{ff}$ the first stellar object, labeled '1', forms in the central dense hub. Throughout the simulation '1' stays the most massive object and with that it is the most dominant source of stellar feedback. After its formation the simulation enters a quiescent phase, where no stars are formed for about another 0.2 $t_\mathrm{ff}$. This is followed by a phase of frequent star formation taking place outside of the central hub in the dense filamentary material accreting onto the central hub. 

Fig. \ref{fig: rosen_column} shows the column density, $\Sigma$, and the density-weighted temperatures of gas, dust, and radiation at 0.2 $t_\mathrm{ff}$, 0.3 $t_\mathrm{ff}$, 0.4 $t_\mathrm{ff}$, 0,5 $t_\mathrm{ff}$ and 0.6 $t_\mathrm{ff}$. One can see that the filamentary structure develops over time and feeds gas into the central hub. Stellar objects forming in the filamentary outskirts follow the general trend of the infalling gas and migrate to the central hub (see Fig. \ref{fig: rosen_column}). Inside the central hub their accretion rate is minuscule compared to the primary star embedded in the center. It is only a brief time during which the secondary sink particles accrete and after which their accretion drops significantly. Looking at their luminosity in the upper right panel of Fig. \ref{fig: sink_combined} one can see that the luminosity output of the secondary sink particles is dominated by their stellar luminosity due to the lack of accretion. Overall, the motion of the secondary sink particles is governed by the gravity of the central star and gas. The lower right panel of Fig. \ref{fig: sink_combined} shows the rates of photons emitted in the Lyman continuum, $\dot{N}_\mathrm{LYC}$, both from the stellar and hot spot accretion separately. The most massive star manages to form a very compact HII region between $0.4 t_\mathrm{ff}$ and $0.5 t_\mathrm{ff}$ (looking at the phase diagram, Fig. \ref{fig:phase_combined}). We show an exemplary spectral emission diagram in the appendix \ref{s:additional_figures}.

\subsection{The Role of Dust}
\subsubsection{Heating and Cooling}
With {\sc TreeRay/RadPressure} and our chemistry network (see \S\ref{ss:heating_and_cooling_of_dust}) we can follow the dust temperature evolution. Fig. \ref{fig:phase_combined} shows phase diagrams of gas, dust and radiation across different times.

We consider gas with $\rho<10^{-19} \, \mathrm{g\, cm^{-3}}$ to be part of the initial background medium. Throughout time the background medium is cooling down. Gas at densities $10^{-19} \, \mathrm{g\, cm^{-3}} < \rho < \, 10^{-16} \, \mathrm{g\, cm^{-3}}$ behaves mostly isothermal and is able to cool down over time via dust, since the radiation temperature is lower than the dust temperature. One can see this in Fig. \ref{fig:phase_combined} where $T_{\bar{J}}$ is lower than the dust temperature at $t=0.2\, t_{\mathrm{ff}}$ for the most diffuse gas. Over time $T_{\bar{J}}$ increases and hence the ability of dust to cool efficiently fades. The forming stars become more embedded as time progresses because of gravitational collapse, which forms compact structures. The average optical depth along all lines of sight at a given position increases. This causes the radiation temperature to increase due to contributions from hotter surrounding material. At later times one finds the luminosity output of the central object to be significant enough to influence $T_{\mathrm{dust}}$ even at greater distances, where the densities are lower. Here we find $T_{\bar{J}} \approx T_{\mathrm{dust}}>T$. We can also see the effects of accretion shock heating in the gas temperature at intermediate densities. Initially only a subtle characteristic can be found at $t=0.2 \, t_{\mathrm{ff}}$ at around  $\rho=10^{-16} \, \mathrm{g\, cm^{-3}}$, where the temperature spikes for a small fraction of the gas. At later times we find the feature of accretion shock heating to be more pronounced and shifted towards lower densities. This is because gas from the outer parts of the cloud starts hitting the central hub, which is at lower density due to the power-law profile. In addition, the kinetic energy with which the gas from the outside is accreting onto the central hub is increasing, resulting in more thermal energy upon impact. This results in the growing peak of accretion shock heating as seen in the evolution of the gas temperature for densities around $10^{-16}\,\mathrm{g\,cm^{-3}}$ in Fig. \ref{fig:phase_combined}. Because dust is coupled to the gas in this density regime (see Fig. \ref{fig: dtemp}), we expect dust to be heated by the accretion shock heated gas. This is further elaborated in the following paragraph.

Fig. \ref{fig:phase_trad_combined} shows phase diagrams for the gas and radiation temperature vs. dust temperature at different times. The dotted, magenta line indicates the one-to-one line. We can see that most of the gas is located around the one-to-one line in the left column at $t=0.2\,t_\mathrm{ff}$. Any deviations from the one-to-one line during $t=0.2\,t_\mathrm{ff}$ appear above the one-to-one line which allows gas to be cooled through dust, as $T > T_{\mathrm{dust}}$. In particular, this is the case for higher temperatures, most likely caused by accretion shock heating. For $t=0.3\,t_\mathrm{ff}$ and beyond, we find gas that may also be heated by dust ($T_\mathrm{dust} > T$). Looking at the right column for $T_{\bar{J}}$ vs. $T_\mathrm{dust}$, most deviations from the one-to-one line indicate that dust is preferably hotter than the radiation temperature and therefore capable of cooling radiatively until 0.4 $t_\mathrm{ff}$. This difference shrinks with increasing time, so that $T_\mathrm{dust} \approx T_{\bar{J}}$ indicating radiative cooling becomes less effective. The decrease in cooling efficiency allows both dust and gas to move upward along the one-to-one line reaching higher temperatures with time. The accretion shocked gas can be seen at high gas temperatures at around $10^2\, \mathrm{K}$ to $10^3\, \mathrm{K}$ while being far from the one-to-one line. Typically one finds this type of gas on the boundary of dense structures, giving the gas suitable conditions to be cooled by radiation. We expect that dust in this regime is also heated by the gas given that gas is up to ten times hotter in some instances (see $t=0.4\, t_\mathrm{ff}$ of Fig. \ref{fig:phase_trad_combined}). We find dust and radiation to follow the one-to-one line in the center of the core. Here dust can not cool as efficiently as on the boundary thus resulting in accumulation near the one-to-one line. This behaviour can be seen in Fig. \ref{fig:phase_trad_combined} in the right column at $t\ge0.5\, t_\mathrm{ff}$. Here dust is surrounded by optically thick material radiating at its own temperature making it unable to cool. Sitting at $T_\mathrm{gas}\approx 10^{4}\,\mathrm{K}$ and $T_\mathrm{dust} \approx 10^3 \mathrm{K}$ one finds few cells that represent ionized gas contained in a hypercompact HII region around the most massive star for $t \ge 0.5 \, t_\mathrm{ff}$. 
Besides that, we find radiation from the Lyman-Continuum to have no dynamical impact in the star forming setup for the simulated timescales. In our implementation, absorbed Lyman-Continuum radiation is immediately reprocessed by dust into IR radiation, thus contributing to heating the gas and dust which surrounds the young stars. Following the \textsc{TreeRay/OnTheSpot} method, ionizing radiation is locally ionising hydrogen that constantly recombines. Here, in particular, we reemit the  consumed ionizing radiation as IR radiation in an energy conserving manner, taking into account the heating by absorbed UV radiation in the chemical network. We note that in this way \textsc{TreeRay/OnTheSpot} faithfully recovers temperatures inside HII regions (see comparison with the code \textsc{mocassin} \citep{ercolano2003} as shown by \cite{haid18}).

\subsubsection{Additional runs}

To further investigate the effect of heating and cooling (see \S \ref{ss:heating_and_cooling_of_dust}) and RP on gas dynamics and fragmentation, we employ three additional runs. In the first run, we disable all radiation effectively turning \textsc{TreeRay/RadPressure} off. This run is called \texttt{NRAD}. Here, dust will always cool optically thin and without any background radiation, $\bar{J}=0$. The other two runs we label \texttt{NARP} and \texttt{NARPNFE}. For both runs, we do not apply accelerations from RP and in addition, we disable stellar feedback in the case of run \texttt{NARPNFE}. By comparing all runs including the original run (hereafter \texttt{FIDUCIAL}) to one another, we will benchmark the impact of \textsc{TreeRay/RadPressure} on the star forming setup in \S\ref{ss:radpressure} and \S\ref{ss:gas_dynamics_and_fragmentation}.

\subsubsection{Radiation Pressure} \label{ss:radpressure}

We find that gravity dominates over RP in the early stages of the \texttt{FIDUCIAL} run. Fig. \ref{fig: rosen_radial} shows a projection of the density along the line of sight and density-weighted acceleration due to RP, $\mathbf{a}_\mathrm{RP}$, gravity and RP, $\mathbf{a}_\mathrm{RP} + \mathbf{a}_\mathrm{G}$, and gravity, $\mathbf{a}_\mathrm{G}$. All accelerations shown are projected in the radial direction of the most massive star, where positive values of $\mathbf{a}_\mathrm{RP} \cdot \hat{\mathbf{r}}_\mathrm{star}$ and $\mathbf{a}_\mathrm{G} \cdot \hat{\mathbf{r}}_\mathrm{star}$ point away from the star. We can see that the region only close to the star is affected by radiation pressure and turns super-Eddington,  $\mathbf{a}_\mathrm{RP} \cdot \hat{\mathbf{r}}_\mathrm{star} > - \mathbf{a}_\mathrm{G} \cdot \hat{\mathbf{r}}_\mathrm{star}$, in the diffuse gas at later times. But still, RP does not manage to overcome gravity very close to the central star. From this, we conclude that RP will not influence the accretion of mass onto the central before $0.6 \, t_\mathrm{ff}$. 
The spherical fragment seen in the column density at 0.5 $t_\mathrm{ff}$ is collapsing. It shows red features in both, $\mathbf{a}_\mathrm{RP} + \mathbf{a}_\mathrm{G}  \cdot \hat{\mathbf{r}}_\mathrm{star}$ and $a_\mathrm{G} \cdot \hat{\mathbf{r}}_\mathrm{star}$, indicating that the blob is self-gravitating. However, this blob does not go on to form a sink particle because it is sheared apart before it can collapse (see \S\ref{ss:sink_particles}).

The mass of the principal star and all its companions is shown in Fig. \ref{fig: mass} for the four runs. Vertical tick marks on the top mark the formation of new sink particles for all simulations. Again, a direct comparison of the primary masses for the different runs as shown in Fig. \ref{fig: mass} underlines that RP is unable to impact the accretion rate onto the central star. The principal mass agrees in all three (four) runs (for the simulated time).

Fig. \ref{fig: mw_accel} shows the mass-weighted probability density function (PDF) of RP and gravitational acceleration at different times. As a function of time, the change in the PDF of the gravitational acceleration is minuscule, while the RP PDF changes significantly. The RP PDF shows higher accelerations with increasing luminosity output of the central hub. It is only after 0.5 $t_\mathrm{ff}$ that the maximum accelerations reached by RP are comparable to those caused by gravity within an order of magnitude. Yet, gravity dominates over RP in the high-acceleration regime. Given that these high accelerations are expected to occur close to the central hub, it is unlikely that RP from the central star has a substantial impact on accretion up to $0.6\,t_\mathrm{ff}$.

\subsubsection{Gas dynamics and Fragmentation}\label{ss:gas_dynamics_and_fragmentation}

Next, we investigate the effect of radiative heating of gas via dust. This requires the dust to be heated in the first place. In our method, dust may be heated by radiation originating from stars as well as by infrared radiation from the surrounding dust. Otherwise, the dust may cool indefinitely (run \texttt{NRAD}). 
In particular for run \texttt{NRAD}, one expects more fragmentation to occur as the overall temperature and the corresponding Jeans mass should be lower. We measure the degree of fragmentation in the density field using dendrograms\footnote{We use the following implementation:\url{https://github.com/dendrograms/astrodendro/}}. Dendrograms construct a tree that shows the structural hierarchy inside a given dataset. By counting the resulting leaves generated by a dendrogram tree, we can assess the degree of fragmentation. 

\begin{figure*}
    \includegraphics[width=\textwidth]{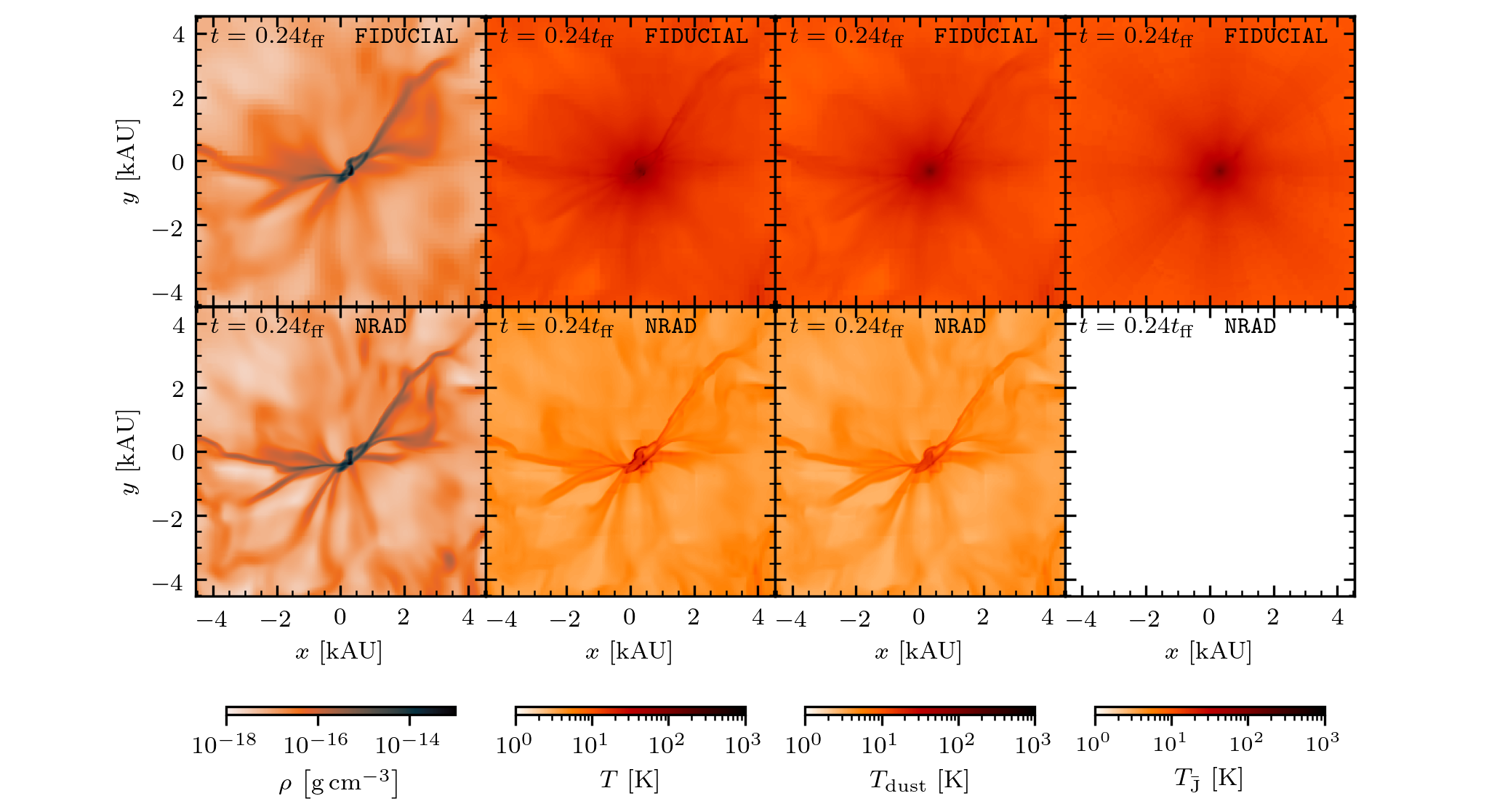}
    \caption{Slices through $z=0$ showing the density, gas temperature, dust temperature and radiation temperature from left to right at similar times. From top to bottom we show two different runs where we include all physics (\texttt{FIDUCIAL}) and no radiative transfer at all (\texttt{NRAD}). Comparing the densities one can see more fragmentation occurring in run \texttt{NRAD}. This is linked to optically thin cooling behaviour of dust ($T_{\bar{J}} = 0$ everywhere). We find more heating in the center for runs with stellar radiation (\texttt{FIDUCIAL} and \texttt{NARP} vs. \texttt{NARPNFE} although not shown here). Radiation pressure does not affect the early dynamics of the setup (\texttt{FIDUCIAL} vs. \texttt{NARP}). \label{fig: combined_4by4}}
\end{figure*}

Fig. \ref{fig: dendro_cul} shows the number of leaves as a function of time for each run. The run \texttt{NRAD} shows the most fragmentation as it cools the gas via optically thin dust emission standing in contrast to the other runs which do account for infrared radiation by dust (red line vs. others). Going a step further, the addition of radiation from stars reduces the number of leaves past 0.4 $t_\mathrm{ff}$ (compare the green and orange lines). After 0.45 $t_\mathrm{ff}$, RP provides an additional mechanism by which fragmentation is slightly suppressed. We show slices of the density and all three temperatures in Fig. \ref{fig: combined_4by4} for \texttt{FIDUCIAL} and \texttt{NRAD}. Here, we can see a major difference in $T$ and $T_\mathrm{dust}$ being lower for the run \texttt{NRAD} as \textsc{TreeRay/RadPressure} is turned off. This reduction in temperature is caused by the fact that dust is allowed to cool optically thin ($T_\mathrm{\bar{J}} = 0$). From this we conclude that RP and thermal feedback contribute towards reducing fragmentation in the star forming setup.

\subsection{Discussion: \textsc{TreeRay/RadPressure} and the Star Forming Setup} \label{ss:sf_discussion}

\begin{figure*}
    \includegraphics[width=\textwidth]{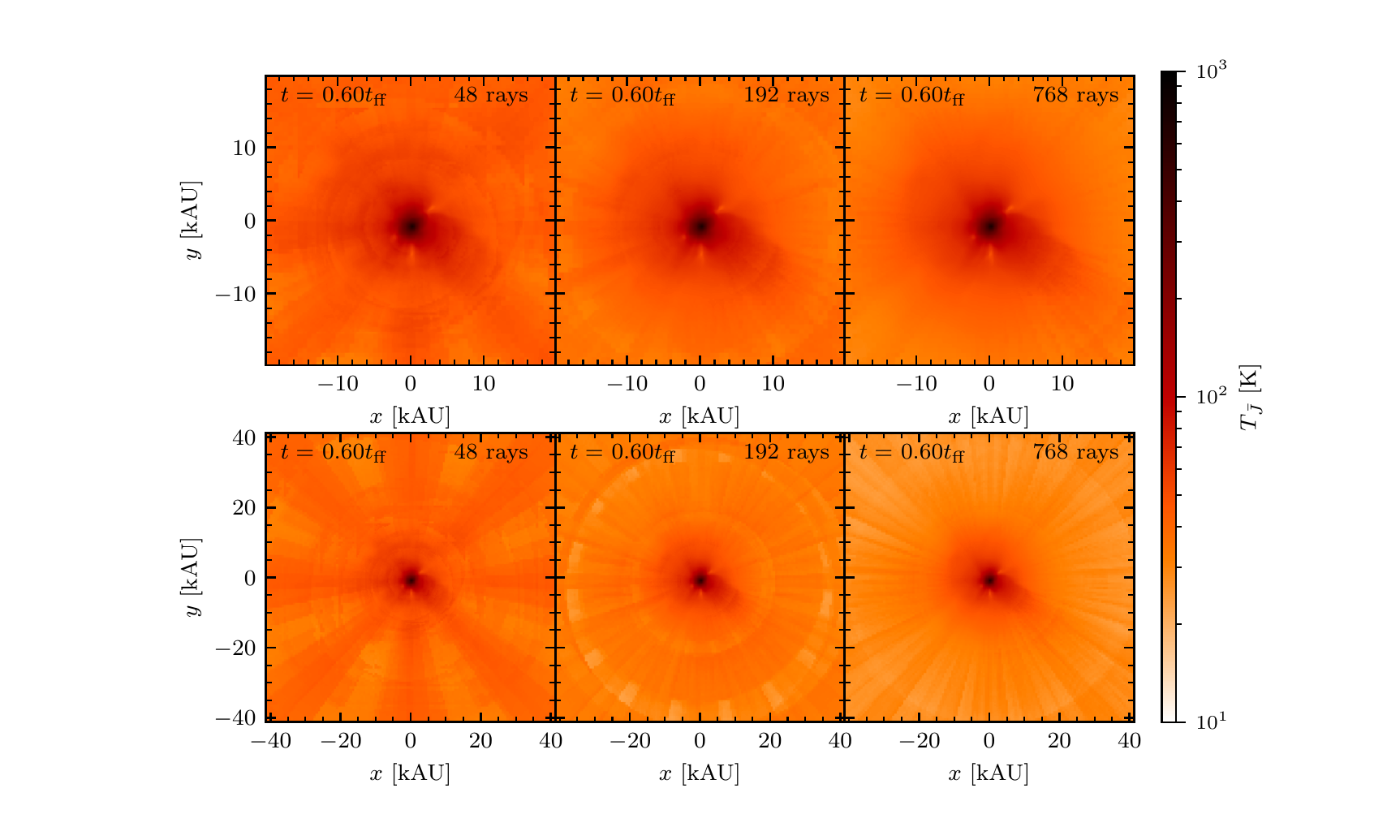}
    \caption{Slices through $z=0$ showing the radiation temperature, $T_{\bar{J}}$. From left to right, $T_{\bar{J}}$ is computed with 48, 192 and 768 rays, respectively. From top to bottom we vary the size of the shown region. Sink particles are not shown in this figure. By increasing the number of rays used, we tend towards an $\mathcal{O}(N^2)$ method as h-nodes are further opened to be comparable to the ray size. $T_{\bar{J}}$ changes mildly in the upper row showing that the scheme performs well scales of half a domain size. On scales of the full domain however, $T_{\bar{J}}$ shows ray artefacts that change with the number of rays used. We find the ratio of computational cost to be (1.00 : 4.39 : 15.98) for (48 rays : 192 rays : 768 rays). Additionally, we find a residual of $-15.4\pm14\%$ and $-3.77\pm5.31\%$ when comparing 48 rays and 192 rays with 768 rays, respectively, for the domain shown in the upper panel. In the lower panel, we find $-31.0\pm17.9\%$ and $-13.2\pm10.5\%$ for the analogous comparison. \label{fig: combined_trad}}
\end{figure*}

The star forming setup presents an application of \textsc{TreeRay/RadPressure} using a more realistic and ultimately more demanding setup other than the tests presented in \S\ref{s:benchmarks} and \S\ref{s:hii_region}. The setup is identical to the one presented in the work of \cite{rosen2019} except for its treatment of radiative transfer, the initial velocity seed and chemistry, including dust thermo-chemistry.

 On large scales, \textsc{TreeRay/RadPressure} shows artifacts that form a ray-like pattern in the radiation temperature. This pattern is concentric around the embedded source (see Fig. \ref{fig: rosen_all}). To investigate how \textsc{TreeRay/RadPressure} would perform against a perfectly accurate $\mathcal{O}(N^2)$-method we increase the number of rays sequentially. By doing so, h-nodes are opened further in comparison to the fiducial setup running with 48 rays, such that in the limit of a very high number of rays all nodes would be fully opened (i.e. all bottom nodes would be considered). We increase the number of rays and perform two additional restarts of the \texttt{FIDUCIAL} run (48 rays) at $t=0.6\,t_{\mathrm{ff}}$ using 192 and 768 rays, respectively. We find a relative cost increase of (1.00 : 4.39 : 15.98) for the algorithm with (48 rays : 192 rays : 768 rays) (i.e. close to the theoretical expectation). Fig. \ref{fig: combined_trad} shows $T_{\bar{J}}$ in a slice through $z=0$ for the three different ray resolutions. The upper row shows slices, which are half of the domain size in width, while the lower row shows the entire domain. \textsc{TreeRay/RadPressure} captures the temperature features at distances of 10~kAU to the central hub well with only 48 rays. On scales of the entire domain, one finds a star-like pattern surrounding the central hub. This pattern becomes more whispy as the number of rays increases, and it is expected to vanish in the perfect $\mathcal{O}(N^2)$ limit.
 
 The resulting pattern may originate from splitting optically thick labeled volumes onto rays linearly followed by a non-linear way of generating an angular size of the mapped optically thick volume (see Eq. \ref{eq: omegaThick}). In this way, the perceived total area may oscillate as one circles an optically thick clump at a fixed distance. The radiative intensity is the lowest if all material is contained in one ray and the greatest if all material is split among many rays resulting in a star-like pattern. The resulting oscillations are expected to be on the order of 26\% for the following reasons. A single optically thick control volume, $V_{\mathrm{control}}$, mapped to a single ray produces an area equivalent to $A_{\mathrm{control}} = V_{\mathrm{control}}^{2/3}$. If $V_{\mathrm{control}}$ is mapped to two rays equally instead, the equivalent area is $A_{\mathrm{two\,rays}} = (0.5 V_{\mathrm{control}})^{2/3} + (0.5 V_{\mathrm{control}})^{2/3} = 1.26 \times A_{\mathrm{control}}$, thus producing a maximum overestimate of $26\%$. For three rays involved, the overestimate reaches up to $44\%$. The previous estimates assume that the rays are not saturated,\ corresponding to the angular size of $V_{\mathrm{control}}$ being smaller than that of a ray. Therefore, the pattern in Fig. \ref{fig: combined_trad} only appears for large distances where rays are not saturated. To counteract this problem, we shift the nodes towards their center of emission during mapping to rays. It may also occur that shielding of a luminous sink particle will oscillate in a similar manner. These effects will be less pronounced if additional sources contribute to $T_\mathrm{rad}$ or dust provides additional shielding in the line of sight. Thus, the error estimate is a drastic overestimation of the real deviation.
 We find a mean residual of $-15.4\%\pm14\%$ and $-3.77\%\pm5.31\%$ when comparing 48 rays and 192 rays with 768 rays, respectively, for the domain shown in the upper panel of Fig. \ref{fig: combined_trad}. In the lower panel, we find $-31.0\%\pm17.9\%$ and $-13.2\%\pm10.5\%$ for the same comparison.
 We note that this star formation simulation, where all radiating sources are located very close to the center, is basically the worst-case setup for \textsc{TreeRay}. If other sources were present in the surroundings, those would dominate the radiation field locally, and hence the artifacts would be much smaller.
 Despite the presence of mild artifacts, we believe that 48 rays are the preferred setting given the significantly reduced computational cost (factor 4 compared to 192 rays).
 
 Currently, all rays have the same orientation leading to pronounced artifacts. A single orientation has been implemented because the ray intersection list is tabulated in the beginning of the simulation, and these tables are used every time when walking the tree to compute the contributions of tree nodes to ray segments. Different orientations require a number of these tables, which of course need to be communicated to, and stored on, all cores, which increases the memory footprint of the method and is hence not ideal for a pure MPI scheme. In a preliminary implementation, we find that introducing different \textsc{HEALPix} orientations mitigates the ray-like features while introducing some noise. However, the resulting physical properties of the gas and dust are better because the different orientations prevent the dust to locally heat up in a ray-like pattern and influence the dynamics and density distribution. This will be discussed in follow-up work.
 
 Given that \textsc{TreeRay/RadPressure} only tracks information of a single monopole on each node, the scheme can not distinguish a source to be embedded or not. This may cause over- or under-extinction of sources and anisotropic emission characteristics to be lost at large distances. This uncertainty in shielding may also contribute to the ray structures discussed above. The accuracy of \textsc{TreeRay/RadPressure} can be improved by introducing higher order terms containing information about the geometry within both h-nodes and the volumes of rays. We defer such improvements to later works.

\section{Conclusion}
This work presents a novel method to compute the radiative transfer of the infrared radiation by dust and stars on-the-fly in three-dimensional simulations of e.g. star formation. In particular, every cell in the computational domain and any number of present (sink) particles can be a source of radiation.

The general idea of the method can be transferred to compute radiative transfer on other macroscopic sources of radiation (e.g. sources that can not be considered as point sources like stars in a numerical simulation). We find that with a computational expense of $N\log (N)$, where $N$ is the number of grid cells (as for the original implementation in \cite{wunsch2021}), \textsc{TreeRay/RadPreesure} is applicable to solve large scale problems.

In a first step, for each grid cell, \textsc{TreeRay/RadPressure} maps contributions from dust and point sources (e.g. stars) onto rays along different directions, where each ray has an associated angular size similar to the shape of a cone. The rays span the surface of the unit sphere according to the HEALPix algorithm (typically we use 48 rays). Subsequently, each ray is integrated to compute a radiative intensity along its line of sight. With all rays integrated, a radiative flux and mean intensity are calculated. We take into account that the Planck-mean dust opacity is temperature-dependent. Finally, we compute the radiation pressure on dust by infrared radiation and its heating rate.

The novel approach presented in this work also couples the infrared radiative transfer to a chemical network (see \S\ref{ss:heating_and_cooling_of_dust}). In particular, the mean intensity of the infrared radiation serves as an input for the thermal equilibrium calculation of the dust temperature, i.e. the dust is heated by IR radiation. In this way, we are able to model the interplay of dust and gas in a self-consistent way. Given that a region is not deeply embedded, dust may cool gas through collisional interactions and radiate thermal energy away. In that case we recover an almost isothermal behavior. On the other hand, an embedded region behaves close to adiabatic if radiation can not escape efficiently. Our method allows to model radiative cooling in both embedded and exposed conditions.

The tests presented in \S\ref{s:benchmarks} show that we can reproduce the Beer-Lambert law and correct radiative energy profiles in optically thin and thick regimes. We are able to compute radiative transfer from macroscopic objects such as dense blobs as presented in \S\ref{ss:rp_test_opt_thick}. 
We verify the correct momentum transfer caused by radiation pressure onto dust and show that the method allows for the leakage through optically thin media as shown in \S\ref{ss:rp_bubble}.

Next, we connect the \textsc{TreeRay/RadPressure} module with the \textsc{TreeRay/OnTheSpot} module \citep{wunsch2021}, which treats the transport of ionizing radiation and the associated momentum input on gas and dust. In \S\ref{s:hii_region}, we show that the combined method faithfully models the expansion of an HII region in two different environments. We find that RP drives the expansion of an ultra-compact HII region over its limitation given by the pressure balance of the ionised medium and its surroundings.

In addition to the tests, we simulate massive star formation in a collapsing, turbulent prestellar core. The star forming setup highlights accretion shock heated gas on the boundary of filaments and the central hub. We find that dust is picking up thermal energy through collisional coupling with gas, resulting in dust temperatures slightly below 100 K. Under those circumstances, where dust is not fully surrounded by other hot dust, dust is able to radiatively cool. On the other hand, dust is not able to cool in the dense region close to the central star, where the local radiation temperature and dust temperature are equal. For times later than 0.5 $t_\mathrm{ff}$ gas may be warmer than dust due to accretion shock heating and in other instances, gas may be colder than dust due to insufficient dust-gas coupling where dust is radiatively heated. We find that the dust temperature is in agreement with the radiation temperature past 0.5 $t_\mathrm{ff}$. RP is minuscule compared to the effects of gravity in the early stage as the central star is not luminous enough to provide relevant feedback. The luminosity output of the central hub grows with the mass of the stellar population over time and with it the relative strength of RP compared to gravity. On the simulated timescales, however, RP does not manage to influence the dynamics of gas near the central star.  Eventually, gas may be blown away through stellar feedback in the later evolution of this setup, once the ram pressure generated by the infalling gas decreases.
    
Our method, \textsc{TreeRay/RadPressure}, is a first approach towards solving RT of extended sources and we plan on improving it further in future works, e.g. we plan to treat the effects of scattering on dust in future works. We aim at analysing the star forming setup in greater detail in a separate publication.

\section*{Acknowledgements}

We thank the referee J. Rosdahl for his valuable comments that helped to improve the quality of this work.
AK, SW, and SH gratefully acknowledge the European Research Council under the European Community's Framework Programme FP8 via the ERC Starting Grant RADFEEDBACK (project number 679852). SW and SH further thank the Deutsche Forschungsgemeinschaft (DFG) for funding through SFB~956 ''The conditions and impact of star formation'' (sub-project C5). DS acknowledges the DFG for funding through SFB~956 ''The conditions and impact of star formation'' (sub-project C6). RW acknowledges support by project 19-15008S of the Czech Science Foundation and by the institutional project RVO:67985815. FD gratefully acknowledges funding from the Grant Agency of the Czech Republic under grant number 20-21855S.
The software used in this work was in part developed by the DOE NNSA-ASC OASCR Flash Center at the University of Chicago.
The authors gratefully acknowledge the Gauss Centre for Supercomputing e.V. (www.gauss-centre.eu) for funding this project by providing computing time through the John von Neumann Institute for Computing (NIC) on the GCS Supercomputer JUWELS at \cite{JUWELS} (JSC).
We particularly thank the Regional Computing Center Cologne for providing the computational facilities for this project by hosting our supercomputing cluster "Odin".

\section*{Data availability}
The data underlying this article will be shared on reasonable request to the corresponding author.



\bibliographystyle{mnras}
\bibliography{bibfile} 

\begin{thebibliography}{}
\makeatletter
\relax
\def\mn@urlcharsother{\let\do\@makeother \do\$\do\&\do\#\do\^\do\_\do\%\do\~}
\def\mn@doi{\begingroup\mn@urlcharsother \@ifnextchar [ {\mn@doi@}
  {\mn@doi@[]}}
\def\mn@doi@[#1]#2{\def\@tempa{#1}\ifx\@tempa\@empty \href
  {http://dx.doi.org/#2} {doi:#2}\else \href {http://dx.doi.org/#2} {#1}\fi
  \endgroup}
\def\mn@eprint#1#2{\mn@eprint@#1:#2::\@nil}
\def\mn@eprint@arXiv#1{\href {http://arxiv.org/abs/#1} {{\tt arXiv:#1}}}
\def\mn@eprint@dblp#1{\href {http://dblp.uni-trier.de/rec/bibtex/#1.xml}
  {dblp:#1}}
\def\mn@eprint@#1:#2:#3:#4\@nil{\def\@tempa {#1}\def\@tempb {#2}\def\@tempc
  {#3}\ifx \@tempc \@empty \let \@tempc \@tempb \let \@tempb \@tempa \fi \ifx
  \@tempb \@empty \def\@tempb {arXiv}\fi \@ifundefined
  {mn@eprint@\@tempb}{\@tempb:\@tempc}{\expandafter \expandafter \csname
  mn@eprint@\@tempb\endcsname \expandafter{\@tempc}}}

\bibitem[\protect\citeauthoryear{{Abel} \& {Wandelt}}{{Abel} \&
  {Wandelt}}{2002}]{abel2002}
{Abel} T.,  {Wandelt} B.~D.,  2002, \mn@doi [\mnras]
  {10.1046/j.1365-8711.2002.05206.x}, \href
  {https://ui.adsabs.harvard.edu/abs/2002MNRAS.330L..53A} {330, L53}

\bibitem[\protect\citeauthoryear{{Altay} \& {Theuns}}{{Altay} \&
  {Theuns}}{2013}]{Altay2013}
{Altay} G.,  {Theuns} T.,  2013, \mn@doi [\mnras] {10.1093/mnras/stt1067},
  \href {https://ui.adsabs.harvard.edu/abs/2013MNRAS.434..748A} {434, 748}

\bibitem[\protect\citeauthoryear{{Baczynski}, {Glover}  \&
  {Klessen}}{{Baczynski} et~al.}{2015}]{fervent2015}
{Baczynski} C.,  {Glover} S.~C.~O.,   {Klessen} R.~S.,  2015, \mn@doi [\mnras]
  {10.1093/mnras/stv1906}, \href
  {https://ui.adsabs.harvard.edu/abs/2015MNRAS.454..380B} {454, 380}

\bibitem[\protect\citeauthoryear{{Bakes} \& {Tielens}}{{Bakes} \&
  {Tielens}}{1994}]{bakes1994}
{Bakes} E.~L.~O.,  {Tielens} A.~G.~G.~M.,  1994, \mn@doi [\apj]
  {10.1086/174188}, \href
  {https://ui.adsabs.harvard.edu/abs/1994ApJ...427..822B} {427, 822}

\bibitem[\protect\citeauthoryear{{Bisbas} et~al.,}{{Bisbas}
  et~al.}{2015}]{bisbas2015}
{Bisbas} T.~G.,  et~al., 2015, \mn@doi [\mnras] {10.1093/mnras/stv1659}, \href
  {https://ui.adsabs.harvard.edu/abs/2015MNRAS.453.1324B} {453, 1324}

\bibitem[\protect\citeauthoryear{Bouchut, Klingenberg  \& Waagan}{Bouchut
  et~al.}{2007}]{bouchut2007}
Bouchut F.,  Klingenberg C.,   Waagan K.,  2007, Numerische Mathematik, 108, 7

\bibitem[\protect\citeauthoryear{Bouchut, Klingenberg  \& Waagan}{Bouchut
  et~al.}{2010}]{bouchut2010}
Bouchut F.,  Klingenberg C.,   Waagan K.,  2010, Numerische Mathematik, 115,
  647

\bibitem[\protect\citeauthoryear{{Calvet} \& {Gullbring}}{{Calvet} \&
  {Gullbring}}{1998}]{calvet1998}
{Calvet} N.,  {Gullbring} E.,  1998, \mn@doi [\apj] {10.1086/306527}, \href
  {https://ui.adsabs.harvard.edu/abs/1998ApJ...509..802C} {509, 802}

\bibitem[\protect\citeauthoryear{{Clarke}, {Whitworth}, {Duarte-Cabral}  \&
  {Hubber}}{{Clarke} et~al.}{2017}]{clarke2017}
{Clarke} S.~D.,  {Whitworth} A.~P.,  {Duarte-Cabral} A.,   {Hubber} D.~A.,
  2017, \mn@doi [\mnras] {10.1093/mnras/stx637}, \href
  {https://ui.adsabs.harvard.edu/abs/2017MNRAS.468.2489C} {468, 2489}

\bibitem[\protect\citeauthoryear{{Dinnbier} \& {Walch}}{{Dinnbier} \&
  {Walch}}{2020}]{dinnbier2020}
{Dinnbier} F.,  {Walch} S.,  2020, \mn@doi [\mnras] {10.1093/mnras/staa2560},
  \href {https://ui.adsabs.harvard.edu/abs/2020MNRAS.499..748D} {499, 748}

\bibitem[\protect\citeauthoryear{{Draine}}{{Draine}}{2011}]{draine2011}
{Draine} B.~T.,  2011, \mn@doi [\apj] {10.1088/0004-637X/732/2/100}, \href
  {https://ui.adsabs.harvard.edu/abs/2011ApJ...732..100D} {732, 100}

\bibitem[\protect\citeauthoryear{{Ercolano}, {Barlow}, {Storey}  \&
  {Liu}}{{Ercolano} et~al.}{2003}]{ercolano2003}
{Ercolano} B.,  {Barlow} M.~J.,  {Storey} P.~J.,   {Liu} X.~W.,  2003, \mn@doi
  [\mnras] {10.1046/j.1365-8711.2003.06371.x}, \href
  {https://ui.adsabs.harvard.edu/abs/2003MNRAS.340.1136E} {340, 1136}

\bibitem[\protect\citeauthoryear{{Federrath}, {Banerjee}, {Clark}  \&
  {Klessen}}{{Federrath} et~al.}{2010}]{federrath2010}
{Federrath} C.,  {Banerjee} R.,  {Clark} P.~C.,   {Klessen} R.~S.,  2010,
  \mn@doi [\apj] {10.1088/0004-637X/713/1/269}, \href
  {https://ui.adsabs.harvard.edu/abs/2010ApJ...713..269F} {713, 269}

\bibitem[\protect\citeauthoryear{{Federrath}, {Banerjee}, {Seifried}, {Clark}
  \& {Klessen}}{{Federrath} et~al.}{2011}]{federrath2011}
{Federrath} C.,  {Banerjee} R.,  {Seifried} D.,  {Clark} P.~C.,   {Klessen}
  R.~S.,  2011, in {Alves} J.,  {Elmegreen} B.~G.,  {Girart} J.~M.,   {Trimble}
  V.,  eds,  IAU Symposium Vol. 270, Computational Star Formation. pp 425--428
  (\mn@eprint {arXiv} {1007.2504}), \mn@doi{10.1017/S1743921311000755}

\bibitem[\protect\citeauthoryear{Fryxell et~al.,}{Fryxell
  et~al.}{2000}]{fryxell2000flash}
Fryxell B.,  et~al., 2000, The Astrophysical Journal Supplement Series, 131,
  273

\bibitem[\protect\citeauthoryear{{Glover} \& {Mac Low}}{{Glover} \& {Mac
  Low}}{2007}]{glover2007}
{Glover} S. C.~O.,  {Mac Low} M.-M.,  2007, \mn@doi [\apjs] {10.1086/512238},
  \href {https://ui.adsabs.harvard.edu/abs/2007ApJS..169..239G} {169, 239}

\bibitem[\protect\citeauthoryear{{G{\'o}rski}, {Hivon}, {Banday}, {Wandelt},
  {Hansen}, {Reinecke}  \& {Bartelmann}}{{G{\'o}rski} et~al.}{2005}]{HEALPix}
{G{\'o}rski} K.~M.,  {Hivon} E.,  {Banday} A.~J.,  {Wandelt} B.~D.,  {Hansen}
  F.~K.,  {Reinecke} M.,   {Bartelmann} M.,  2005, \apj, \href
  {http://adsabs.harvard.edu/abs/2005ApJ...622..759G} {622, 759}

\bibitem[\protect\citeauthoryear{{Grond}, {Woods}, {Wadsley}  \&
  {Couchman}}{{Grond} et~al.}{2019}]{grond2019}
{Grond} J.~J.,  {Woods} R.~M.,  {Wadsley} J.~W.,   {Couchman} H.~M.~P.,  2019,
  \mn@doi [\mnras] {10.1093/mnras/stz525}, \href
  {https://ui.adsabs.harvard.edu/abs/2019MNRAS.485.3681G} {485, 3681}

\bibitem[\protect\citeauthoryear{{Haid}, {Walch}, {Seifried}, {W{\"u}nsch},
  {Dinnbier}  \& {Naab}}{{Haid} et~al.}{2018}]{haid18}
{Haid} S.,  {Walch} S.,  {Seifried} D.,  {W{\"u}nsch} R.,  {Dinnbier} F.,
  {Naab} T.,  2018, \mn@doi [\mnras] {10.1093/mnras/sty1315}, \href
  {https://ui.adsabs.harvard.edu/abs/2018MNRAS.478.4799H} {478, 4799}

\bibitem[\protect\citeauthoryear{{Haid}, {Walch}, {Seifried}, {W{\"u}nsch},
  {Dinnbier}  \& {Naab}}{{Haid} et~al.}{2019}]{haid19}
{Haid} S.,  {Walch} S.,  {Seifried} D.,  {W{\"u}nsch} R.,  {Dinnbier} F.,
  {Naab} T.,  2019, \mn@doi [\mnras] {10.1093/mnras/sty2938}, \href
  {https://ui.adsabs.harvard.edu/abs/2019MNRAS.482.4062H} {482, 4062}

\bibitem[\protect\citeauthoryear{{Hollenbach} \& {McKee}}{{Hollenbach} \&
  {McKee}}{1979}]{Hollenbach1979}
{Hollenbach} D.,  {McKee} C.~F.,  1979, \mn@doi [\apjs] {10.1086/190631}, \href
  {https://ui.adsabs.harvard.edu/abs/1979ApJS...41..555H} {41, 555}

\bibitem[\protect\citeauthoryear{{Hopkins}, {Quataert}  \& {Murray}}{{Hopkins}
  et~al.}{2011}]{hopkins2011}
{Hopkins} P.~F.,  {Quataert} E.,   {Murray} N.,  2011, \mn@doi [\mnras]
  {10.1111/j.1365-2966.2011.19306.x}, \href
  {https://ui.adsabs.harvard.edu/abs/2011MNRAS.417..950H} {417, 950}

\bibitem[\protect\citeauthoryear{{J\"{u}lich Supercomputing
  Centre}}{{J\"{u}lich Supercomputing Centre}}{2021}]{JUWELS}
{J\"{u}lich Supercomputing Centre} 2021, \mn@doi [Journal of large-scale
  research facilities] {10.17815/jlsrf-7-183}, 7

\bibitem[\protect\citeauthoryear{{Kannan}, {Vogelsberger}, {Marinacci},
  {McKinnon}, {Pakmor}  \& {Springel}}{{Kannan} et~al.}{2019}]{kannan2019}
{Kannan} R.,  {Vogelsberger} M.,  {Marinacci} F.,  {McKinnon} R.,  {Pakmor} R.,
    {Springel} V.,  2019, \mn@doi [\mnras] {10.1093/mnras/stz287}, \href
  {https://ui.adsabs.harvard.edu/abs/2019MNRAS.485..117K} {485, 117}

\bibitem[\protect\citeauthoryear{{Kessel-Deynet} \& {Burkert}}{{Kessel-Deynet}
  \& {Burkert}}{2000}]{kessel-deynet2000}
{Kessel-Deynet} O.,  {Burkert} A.,  2000, \mn@doi [\mnras]
  {10.1046/j.1365-8711.2000.03451.x}, \href
  {https://ui.adsabs.harvard.edu/abs/2000MNRAS.315..713K} {315, 713}

\bibitem[\protect\citeauthoryear{{Kim}, {Kim}  \& {Ostriker}}{{Kim}
  et~al.}{2018}]{kim2018}
{Kim} J.-G.,  {Kim} W.-T.,   {Ostriker} E.~C.,  2018, \mn@doi [\apj]
  {10.3847/1538-4357/aabe27}, \href
  {https://ui.adsabs.harvard.edu/abs/2018ApJ...859...68K} {859, 68}

\bibitem[\protect\citeauthoryear{{Klassen}, {Pudritz}  \& {Peters}}{{Klassen}
  et~al.}{2012}]{klassen11}
{Klassen} M.,  {Pudritz} R.~E.,   {Peters} T.,  2012, \mn@doi [\mnras]
  {10.1111/j.1365-2966.2012.20523.x}, \href
  {https://ui.adsabs.harvard.edu/abs/2012MNRAS.421.2861K} {421, 2861}

\bibitem[\protect\citeauthoryear{Krumholz \& Thompson}{Krumholz \&
  Thompson}{2012}]{krumholz2012direct}
Krumholz M.~R.,  Thompson T.~A.,  2012, The Astrophysical Journal, 760, 155

\bibitem[\protect\citeauthoryear{{Krumholz}, {Klein}, {McKee}  \&
  {Bolstad}}{{Krumholz} et~al.}{2007}]{krumholz2007}
{Krumholz} M.~R.,  {Klein} R.~I.,  {McKee} C.~F.,   {Bolstad} J.,  2007,
  \mn@doi [\apj] {10.1086/520791}, \href
  {https://ui.adsabs.harvard.edu/abs/2007ApJ...667..626K} {667, 626}

\bibitem[\protect\citeauthoryear{{Kuiper} \& {Hosokawa}}{{Kuiper} \&
  {Hosokawa}}{2018}]{kuiper2018}
{Kuiper} R.,  {Hosokawa} T.,  2018, \mn@doi [\aap]
  {10.1051/0004-6361/201832638}, \href
  {https://ui.adsabs.harvard.edu/abs/2018A&A...616A.101K} {616, A101}

\bibitem[\protect\citeauthoryear{{Kuiper}, {Yorke}  \& {Mignone}}{{Kuiper}
  et~al.}{2020}]{kuiper2020}
{Kuiper} R.,  {Yorke} H.~W.,   {Mignone} A.,  2020, arXiv e-prints, \href
  {https://ui.adsabs.harvard.edu/abs/2020arXiv200912374K} {p. arXiv:2009.12374}

\bibitem[\protect\citeauthoryear{{Lepp} \& {Shull}}{{Lepp} \&
  {Shull}}{1983}]{lepp1983}
{Lepp} S.,  {Shull} J.~M.,  1983, \mn@doi [\apj] {10.1086/161149}, \href
  {https://ui.adsabs.harvard.edu/abs/1983ApJ...270..578L} {270, 578}

\bibitem[\protect\citeauthoryear{{Levermore}}{{Levermore}}{1984}]{levermore1984}
{Levermore} C.~D.,  1984, \mn@doi [\jqsrt] {10.1016/0022-4073(84)90112-2},
  \href {https://ui.adsabs.harvard.edu/abs/1984JQSRT..31..149L} {31, 149}

\bibitem[\protect\citeauthoryear{Levermore \& Pomraning}{Levermore \&
  Pomraning}{1981}]{levermore1981flux}
Levermore C.,  Pomraning G.,  1981, The Astrophysical Journal, 248, 321

\bibitem[\protect\citeauthoryear{{Martin}, {Schwarz}  \& {Mandy}}{{Martin}
  et~al.}{1996}]{martin1996}
{Martin} P.~G.,  {Schwarz} D.~H.,   {Mandy} M.~E.,  1996, \mn@doi [\apj]
  {10.1086/177053}, \href
  {https://ui.adsabs.harvard.edu/abs/1996ApJ...461..265M} {461, 265}

\bibitem[\protect\citeauthoryear{{Matzner}}{{Matzner}}{2002}]{Matzner2002}
{Matzner} C.~D.,  2002, \mn@doi [\apj] {10.1086/338030}, \href
  {https://ui.adsabs.harvard.edu/abs/2002ApJ...566..302M} {566, 302}

\bibitem[\protect\citeauthoryear{{Menon}, {Federrath}, {Krumholz}, {Kuiper},
  {Wibking}  \& {Jung}}{{Menon} et~al.}{2022}]{menon2022}
{Menon} S.~H.,  {Federrath} C.,  {Krumholz} M.~R.,  {Kuiper} R.,  {Wibking}
  B.~D.,   {Jung} M.,  2022, \mn@doi [\mnras] {10.1093/mnras/stac485}, \href
  {https://ui.adsabs.harvard.edu/abs/2022MNRAS.512..401M} {512, 401}

\bibitem[\protect\citeauthoryear{Mihalas \& Mihalas}{Mihalas \&
  Mihalas}{2013}]{mihalas2013foundations}
Mihalas D.,  Mihalas B.~W.,  2013, Foundations of radiation hydrodynamics.
Courier Corporation

\bibitem[\protect\citeauthoryear{{Nelson} \& {Langer}}{{Nelson} \&
  {Langer}}{1997}]{nelson1997}
{Nelson} R.~P.,  {Langer} W.~D.,  1997, \mn@doi [\apj] {10.1086/304167}, \href
  {https://ui.adsabs.harvard.edu/abs/1997ApJ...482..796N} {482, 796}

\bibitem[\protect\citeauthoryear{{Raga}, {Cant{\'o}}  \&
  {Rodr{\'\i}guez}}{{Raga} et~al.}{2012}]{raga2012}
{Raga} A.~C.,  {Cant{\'o}} J.,   {Rodr{\'\i}guez} L.~F.,  2012, \mn@doi
  [\mnras] {10.1111/j.1745-3933.2011.01173.x}, \href
  {https://ui.adsabs.harvard.edu/abs/2012MNRAS.419L..39R} {419, L39}

\bibitem[\protect\citeauthoryear{Rosdahl \& Teyssier}{Rosdahl \&
  Teyssier}{2015}]{rosdahl2015scheme}
Rosdahl J.,  Teyssier R.,  2015, Monthly Notices of the Royal Astronomical
  Society, 449, 4380

\bibitem[\protect\citeauthoryear{{Rosen}, {Krumholz}, {Oishi}, {Lee}  \&
  {Klein}}{{Rosen} et~al.}{2017}]{rosen2017}
{Rosen} A.~L.,  {Krumholz} M.~R.,  {Oishi} J.~S.,  {Lee} A.~T.,   {Klein}
  R.~I.,  2017, \mn@doi [Journal of Computational Physics]
  {10.1016/j.jcp.2016.10.048}, \href
  {https://ui.adsabs.harvard.edu/abs/2017JCoPh.330..924R} {330, 924}

\bibitem[\protect\citeauthoryear{{Rosen}, {Li}, {Zhang}  \& {Burkhart}}{{Rosen}
  et~al.}{2019}]{rosen2019}
{Rosen} A.~L.,  {Li} P.~S.,  {Zhang} Q.,   {Burkhart} B.,  2019, \mn@doi [\apj]
  {10.3847/1538-4357/ab54c6}, \href
  {https://ui.adsabs.harvard.edu/abs/2019ApJ...887..108R} {887, 108}

\bibitem[\protect\citeauthoryear{Semenov, Henning, Helling, Ilgner  \&
  Sedlmayr}{Semenov et~al.}{2003}]{semenov2003rosseland}
Semenov D.,  Henning T.,  Helling C.,  Ilgner M.,   Sedlmayr E.,  2003,
  Astronomy \& Astrophysics, 410, 611

\bibitem[\protect\citeauthoryear{{Shapiro} \& {Kang}}{{Shapiro} \&
  {Kang}}{1987}]{shapiro1987}
{Shapiro} P.~R.,  {Kang} H.,  1987, \mn@doi [\apj] {10.1086/165350}, \href
  {https://ui.adsabs.harvard.edu/abs/1987ApJ...318...32S} {318, 32}

\bibitem[\protect\citeauthoryear{{Spitzer}}{{Spitzer}}{1978}]{spitzer1978}
{Spitzer} L.,  1978, {Physical processes in the interstellar medium}.
New York Wiley-Interscience, 1978.~333 p.

\bibitem[\protect\citeauthoryear{{Str{\"o}mgren}}{{Str{\"o}mgren}}{1939}]{stroemgren1939}
{Str{\"o}mgren} B.,  1939, \mn@doi [\apj] {10.1086/144074}, \href
  {https://ui.adsabs.harvard.edu/abs/1939ApJ....89..526S} {89, 526}

\bibitem[\protect\citeauthoryear{{Tan}, {Beltr{\'a}n}, {Caselli}, {Fontani},
  {Fuente}, {Krumholz}, {McKee}  \& {Stolte}}{{Tan}
  et~al.}{2014}]{massiveReview2014}
{Tan} J.~C.,  {Beltr{\'a}n} M.~T.,  {Caselli} P.,  {Fontani} F.,  {Fuente} A.,
  {Krumholz} M.~R.,  {McKee} C.~F.,   {Stolte} A.,  2014, in {Beuther} H.,
  {Klessen} R.~S.,  {Dullemond} C.~P.,   {Henning} T.,  eds, Protostars and
  Planets VI. p.~149 (\mn@eprint {arXiv} {1402.0919}),
  \mn@doi{10.2458/azu_uapress_9780816531240-ch007}

\bibitem[\protect\citeauthoryear{Thompson, Quataert  \& Murray}{Thompson
  et~al.}{2005}]{thompson2005radiation}
Thompson T.~A.,  Quataert E.,   Murray N.,  2005, The Astrophysical Journal,
  630, 167

\bibitem[\protect\citeauthoryear{{Truelove}, {Klein}, {McKee}, {Holliman},
  {Howell}  \& {Greenough}}{{Truelove} et~al.}{1997}]{truelove1997}
{Truelove} J.~K.,  {Klein} R.~I.,  {McKee} C.~F.,  {Holliman} John~H. I.,
  {Howell} L.~H.,   {Greenough} J.~A.,  1997, \mn@doi [\apjl] {10.1086/310975},
  \href {https://ui.adsabs.harvard.edu/abs/1997ApJ...489L.179T} {489, L179}

\bibitem[\protect\citeauthoryear{{Walch} et~al.,}{{Walch}
  et~al.}{2015}]{walch2015silcc}
{Walch} S.,  et~al., 2015, \mn@doi [\mnras] {10.1093/mnras/stv1975}, \href
  {https://ui.adsabs.harvard.edu/abs/2015MNRAS.454..238W} {454, 238}

\bibitem[\protect\citeauthoryear{{Weingartner} \& {Draine}}{{Weingartner} \&
  {Draine}}{2001}]{weingartner2001}
{Weingartner} J.~C.,  {Draine} B.~T.,  2001, \mn@doi [\apj] {10.1086/318651},
  \href {https://ui.adsabs.harvard.edu/abs/2001ApJ...548..296W} {548, 296}

\bibitem[\protect\citeauthoryear{{W{\"u}nsch}, {Walch}, {Dinnbier}  \&
  {Whitworth}}{{W{\"u}nsch} et~al.}{2018}]{2018wunsch}
{W{\"u}nsch} R.,  {Walch} S.,  {Dinnbier} F.,   {Whitworth} A.,  2018, \mn@doi
  [\mnras] {10.1093/mnras/sty015}, \href
  {http://adsabs.harvard.edu/abs/2018MNRAS.475.3393W} {475, 3393}

\bibitem[\protect\citeauthoryear{{W{\"u}nsch}, {Walch}, {Dinnbier}, {Seifried},
  {Haid}, {Klepitko}, {Whitworth}  \& {Palou{\v{s}}}}{{W{\"u}nsch}
  et~al.}{2021}]{wunsch2021}
{W{\"u}nsch} R.,  {Walch} S.,  {Dinnbier} F.,  {Seifried} D.,  {Haid} S.,
  {Klepitko} A.,  {Whitworth} A.~P.,   {Palou{\v{s}}} J.,  2021, arXiv
  e-prints, \href {https://ui.adsabs.harvard.edu/abs/2021arXiv210509644W} {p.
  arXiv:2105.09644}

\makeatother
\end{thebibliography}




\appendix

\section{information propagation upwards the tree}
\label{s:fillin_tree_from_bottom_to_top}
In this section we explain how quantities are propagated upwards the octtree. The crucial part is to store optically thin and thick contributions separately to maintain accuracy. In our implementation we propagate b-nodes and h-nodes differently in order to save memory on the biggest layer, made up of b-nodes. If the sub-node is a bottom node, we do the following:
\begin{eqnarray}
  \underline{L}_\mathrm{h-node,\, thick} = \sum^{\tau_{\mathrm{b-node}}\ge 1}_k \underline{L}_{\mathrm{b-node,\,thick}, k}, \label{eq: Lhnodethick} \\
  V_\mathrm{h-node,\, thick} = \sum^{\tau_{\mathrm{b-node}}\ge 1}_k V_{\mathrm{b-node}, k},  \label{eq: Vhnodethick}\\
    \widetilde{A}_\mathrm{h-node, \, thick} = \sum^{\tau_{\mathrm{b-node}}\ge 1}_k \widetilde{A}_{\mathrm{b-node}, k}, \label{eq: Ahnodethick} \\
  L_\mathrm{h-node,\, thin} = \sum^{\tau_{\mathrm{b-node,\,thick}}< 1}_k L_{\mathrm{b-node,\,thin}, k},  \label{eq: Lhnodethin} + S_{\mathrm{b-node}, \, k} \\
  V_\mathrm{h-node,\, thin} = \sum^{\tau_{\mathrm{b-node}}< 1}_k V_{\mathrm{b-node}, k},  \label{eq: Vhnodethin}\\
    \widetilde{A}_\mathrm{h-node, \, thin} = \sum^{\tau_{\mathrm{b-node}}< 1}_k \widetilde{A}_{\mathrm{b-node}, k}.\label{eq: Ahnodethin}
\end{eqnarray}
Otherwise, if the sub-nodes consist of h-nodes, we simply compute 
\begin{eqnarray}
  X_{\mathrm{h-node},\, Y} = \sum^{\mathrm{b-node}}_k X_{\mathrm{b-node},Y, k},  \label{eq: Xhnodethin}
\end{eqnarray}
where $X$ may be, $L$, $\widetilde{A}$, $V$ or $S$ and $Y$, thin or thick, respectively.  \\
Since a higher level node may contain optically thin and thick contributions at the same instance, we need to store them seperately. Eq. \ref{eq: Lhnodethick}, eq. \ref{eq: Vhnodethick} and eq. \ref{eq: Ahnodethick} trace optically thick quantities, while eq. \ref{eq: Lhnodethin}, \ref{eq: Vhnodethin}, and eq. \ref{eq: Ahnodethin} trace optically thin contributions.

In addition we compute the centre of luminosity (COL), $\mathbf{r}_\mathrm{COL}$, as seen from the geometric centre of the node, $\mathbf{r_\mathcal{O}}$, for higher level nodes. Again we distinguish if the sub-nodes are b-nodes or h-nodes. In the former case we compute
\begin{eqnarray}
  \xi_\mathrm{b-node}= L_\mathrm{b-node},
\end{eqnarray}
if the b-node is optically thin and
\begin{eqnarray}
  \xi_\mathrm{b-node}= L_\mathrm{b-node} \cdot (V_{\mathrm{b-node}})^{2/3},
\end{eqnarray}
otherwise. If the sub-nodes are h-nodes we use
\begin{eqnarray}
  \xi_\mathrm{h-node} = L_{\mathrm{h-node}, \mathrm{thin}} + \underline{L}_{\mathrm{h-node}, \, \mathrm{thick}} \cdot (V_{\mathrm{h-node}, \, \mathrm{thick}})^{2/3}\, .
\end{eqnarray}
Finally we compute:
\begin{eqnarray}
  \Xi_{\mathrm{h-node}} &=& \sum^{\mathrm{sub-nodes}}_k  \xi_{\mathrm{sub-node}, \, k}\, ,\\
    \mathbf{r_{\mathrm{h-node,\, COL}}} &=& \frac{1}{\Xi} \sum^{\mathrm{sub-nodes}}_k \xi_{\mathrm{sub-node}, k} \mathbf{r}_{\mathrm{sub-node}, \, k}, \label{eq: COL}
\end{eqnarray}
where $\mathbf{r}_{\mathrm{sub-node}, \, k}$ is the position of the sub-node relative to $\mathbf{r_\mathcal{O}}$.

\section{Limits of an optically thin and thick segment} \label{s: limits}
Here we want to show, that eq. \ref{eq: fthin} has the correct limits if the material inside segment $i$ becomes optically thin or thick. To do this we consider $\tau_{\mathrm{thin}, i} << 1$ and $\tau_{\mathrm{thin}, i} >> 1$. 

By using a Taylor expansion series one can show that all terms involving $\tau_{\mathrm{thin}, i}$ collapse to unity in the former case. We arrive at
\begin{eqnarray}
    \mathrm{eq.}\,(\ref{eq: fthin}) &\overset{\tau_{\mathrm{thin}, i} << 1}{\approx}& \frac{\hat{L}_{\mathrm{thin},i}}{4\pi R_i^2}\, . \label{eq:thinlimit}
\end{eqnarray}
Expression \ref{eq:thinlimit} is as expected, as would indeed see all luminosity content of segment $i$ to be radiating. This solution is similar to that of a point source.

In the latter case all terms involving $\tau_{\mathrm{thin}, i}$ can be approximated by $1/\tau_{\mathrm{thin}, i}$. The complete relation can be expressed in the following way
\begin{eqnarray}
    \mathrm{eq.}\,(\ref{eq: fthin}) &\overset{\tau_{\mathrm{thin}, i} >> 1}{\approx}& \frac{1}{\tau_{\mathrm{thin}, i}} \times \frac{\hat{L}_{\mathrm{thin},i}}{4\pi R_i^2}\, . \label{eq:opt_thick_thin_nodes}
\end{eqnarray}
Next we can introduce a factor of unity of the form $\frac{A_i}{A_i}$, where $A_i = \omega R_i^2$ is the area of segment $i$ measured in angular direction. Together with $1/\tau_{\mathrm{thin}, i}$ we can express this as
\begin{eqnarray}
    \frac{A_i}{A_i} \times \frac{1}{\tau_{\mathrm{thin}, i}} = \frac{A_i}{\hat{A}_{\mathrm{thin}, i}} \label{eq: inter_step}
\end{eqnarray}
by using the definition of eq. \ref{eq:tauthin}. Substituting eq. \ref{eq: inter_step} into eq. \ref{eq:opt_thick_thin_nodes} results in the following expression
\begin{eqnarray}
    \mathrm{eq.}\,(\ref{eq: fthin}) &\overset{\tau_{\mathrm{thin}, i} >> 1}{\approx}& \frac{\omega R_i^2}{4\pi R_i^2} \times \frac{\hat{L}_{\mathrm{thin}, i}}{\hat{A}_{\mathrm{thin}, i}} \, .    
\end{eqnarray}
The left factor yields $1/N_\mathrm{pix}$. The factor on the right hand side can be understood as optical depth weighted mean temperature by considering that each segment sums over the each node's quantities described by eq. \ref{eq: lthinbnode} and eq. \ref{eq: athinbnode}. Finally we arrive at
\begin{eqnarray}
    \mathrm{eq.}\,(\ref{eq: fthin}) &\overset{\tau_{\mathrm{thin}, i} >> 1}{\approx}& \frac{4\pi}{N_\mathrm{pix}} \times \frac{\sigma}{\pi} \, \langle T_\mathrm{dust}^4 \rangle_{\rho \mathrm{dV} \kappa} \, .   \label{eq:LthinFinalThickLimit} 
\end{eqnarray}
Note that eq. \ref{eq:LthinFinalThickLimit} is in agreement with eq. \ref{eq: fthick} which describes the flux seen from optically thick material contained within a ray's segment. The difference is that the optically thin material is stretched over the entire pixel and thus shining over the full ray's solid angle $\omega=\frac{4\pi}{N_\mathrm{pix}}$ while the optically thick material is not, but compact instead. This approximates the optically thick material to radiate from its surface.

\section{Radiation Pressure on Dust and gas by UV Radiation} \label{ss:uvrp}
For a given flux of ionizing photons per unit time and area, $\dot{N}_\mathrm{ph}$, with an average energy per photon, $\bar{E}$, we compute the momentum input per time caused by RP from UV radiation in the following way:
\begin{eqnarray}
    \dot{\mathbf{P}}_\mathrm{UV, \, dust} &=& \sigma_\mathrm{d} n \frac{\bar{E}_\mathrm{ph}}{c} \dot{N}_\mathrm{ph} \, \mathbf{u}_\mathrm{ph}\mathrm{d}V\, , \\
    \dot{\mathbf{P}}_\mathrm{UV, \, H} &=& \sigma_\mathrm{H} n_\mathrm{H} \frac{\bar{E}_\mathrm{ph}}{c} \dot{N}_\mathrm{ph} \, \mathbf{u}_\mathrm{ph} \mathrm{d}V\, , \\
    \dot{\mathbf{P}}_\mathrm{UV, \, H^+} &=& n^2 \alpha_\mathrm{B} \frac{\bar{E}_\mathrm{ph}}{c} \, \mathbf{u}_\mathrm{ph} \mathrm{d}V\, ,
\end{eqnarray}
where $\dot{\mathbf{P}}_\mathrm{UV, \, dust}$, $\dot{\mathbf{P}}_\mathrm{UV, \, H}$ and $\dot{\mathbf{P}}_\mathrm{UV, \, H^+}$ are the momentum deposition per unit time on dust, atomic hydrogen and ionized hydrogen, respectively. The quantities $\sigma_\mathrm{H}$, $\bar{E}_\mathrm{ph}$, $\dot{N}_\mathrm{ph}$ and $\mathbf{u}_\mathrm{ph}$ are the hydrogen cross section, the mean energy per photon, the incoming photon number per unit time per unit area and the normalised flux vector of the photons, respectively.

\section{Equations for heating and cooling of dust}\label{s:eq_heat_cool}
The gas-dust coupling is given by the following expression \cite{Hollenbach1979}:
\begin{eqnarray}
    \Gamma_\mathrm{dust-gas} = n n_\mathrm{dust} \sigma_\mathrm{dust} v_\mathrm{p} f 2 k_\mathrm{B} \left( T - T_\mathrm{dust} \right),
\end{eqnarray}
where $n$ and $n_\mathrm{dust}$ are the number density of gas and dust, $\sigma_\mathrm{dust}$ is the dust cross section and $v_\mathrm{p}$ is the thermal speed of protons for a given temperature $T$. $f$ corrects the coupling strength for different ionization fractions and temperatures.

We compute the heating by the interstellar radiation field with the following equation \citep{bakes1994}:
\begin{eqnarray}
    \Gamma_\mathrm{pe} = 10^{-24} \epsilon_\mathrm{pe} G_0 n_\mathrm{H} \, \mathrm{erg\, s^{-1}\, cm^{-3}},
\end{eqnarray}
where $G_0$ is the flux normalized to the Habing field for the solar neighborhood and $\epsilon_\mathrm{pe}$ is the photoelectric heating efficiency and $n_\mathrm{H}$ is the hydrogen number density. $\epsilon_\mathrm{pe}$ is given by eq. 43 of \cite{bakes1994}.

The H$_\mathrm{2}$ formation heating is calculated as such \citep{glover2007}
\begin{eqnarray}
    \Gamma_{\mathrm{H}_2} = 7.2 \times 10^{-12} \frac{R_{\mathrm{H}_2}}{1+n_\mathrm{cr}/n} \, \mathrm{erg\,s^{-1}\, cm^{-3}},
\end{eqnarray}
where $R_{\mathrm{H}_2}$ measures the rate of H$_2$ formation on dust grains and $n_\mathrm{cr}$ is the critical density given by
\begin{eqnarray}
    \frac{1}{n_\mathrm{cr}} = \frac{x_\mathrm{H}}{n_\mathrm{cr, H}} + \frac{x_\mathrm{H_2}}{n_\mathrm{cr,H_2}}.
\end{eqnarray}
$x_\mathrm{H}$ and $x_\mathrm{H_2}$ are the abundances of atomic and molecular hydrogen. As discussed by \citep{glover2007} we take the values for $x_\mathrm{H}$ from \cite{lepp1983} with the modification from \cite{martin1996} and the value for $x_\mathrm{H_2}$ from \cite{shapiro1987}.

\section{Additional figures: Dust Chemistry} \label{s: add_dust_chemistry}
\begin{figure}
    \includegraphics[width=\columnwidth]{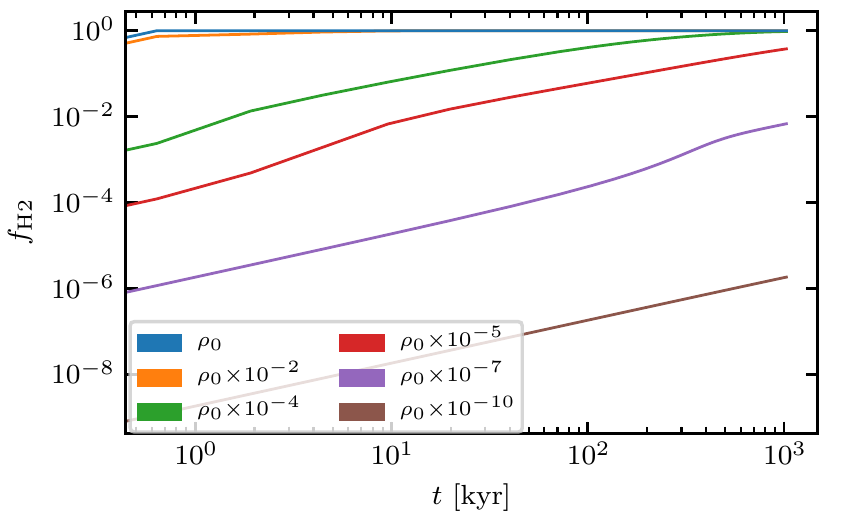}
    \caption{Molecular hydrogen fraction vs. time for dust chemistry setup in \S\ref{ss:dust_chemistry}. The different lines show different densities used for the dust chemistry setup. Higher densities form molecular hydrogen faster than lower ones.\label{fig: numd}}
\end{figure}

We show the corresponding molecular hydrogen fraction, $f_\mathrm{H2} = 2n_\mathrm{H2} / (n_\mathrm{H^+} + n_\mathrm{H} + 2n_\mathrm{H2})$, vs. time in Fig. \ref{fig: numd}. The formation timescale of H$_2$ is given by $T_\mathrm{H2} = 1 \mathrm{Gyr}/(n_\mathrm{H} / \mathrm{cm}^{-3})$ and for the density $\rho_0$ it is roughly 1 yr. We can see that both the blue and orange line of Fig. \ref{fig: numd} complete the formation of H$_2$ well within the simulated time period.

\section{Additional figures: Star Forming Setup} \label{s:additional_figures}
\begin{figure}
    \includegraphics[width=\columnwidth]{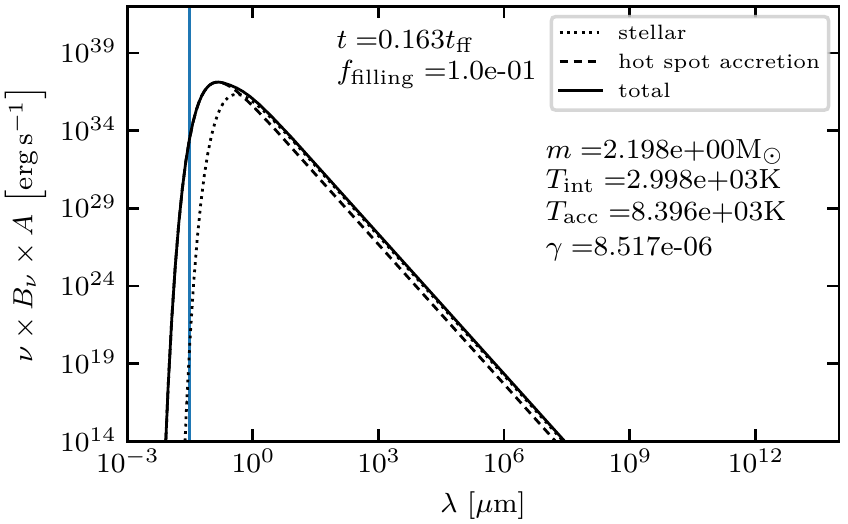}
    \caption{Stellar spectrum of the primary stellar particle at simulation time $t$. The y-axis is the resulting product of frequency, $\nu$, Planck's law, $B_\nu$, and the respective area from which the radiation is emitted, $A$. In the case of hot spot accretion we assume that the emission originates from $f_\mathrm{filling}$ times the area of the stellar object. The dotted line shows the purely stellar component and the dashed line shows the component due to hot spot accretion. The sum of both components is shown by the solid line. $T_\mathrm{int}$ and $T_\mathrm{acc}$ are calculated based on eq. \ref{eq:internalTemp} and eq. \ref{eq:accTemp}, respectively. $\gamma$ is the fraction of the power that is emitted in the ionising regime from both the internal and accretion luminosity combined (see eq. \ref{eq: bb_ratio}).\label{fig: sed_sink}}
\end{figure}

Fig. \ref{fig: sed_sink} shows the emitted energy rate at a given wavelength for a stellar particle modeled according to \S\ref{s:star_forming_setup}. Everything part of the Lyman-continuum is left of the blue vertical line and emitted in the ionizing band. Everything to the right of the blue line is emitted in the non-ionising band. Although the hot spot accretion is emitted from a smaller area it dominates in the bolometric luminosity output due to its higher temperature (see fig. \ref{fig: sink_combined} upper right panel). $\gamma$ is the fraction that is emitted in the ionising regime from the total luminosity output.

\section{Radiation Pressure acting on the interior of HII regions}
In this section we apply our scheme to the scenario of dusty HII regions and see if the scheme is producing expected outcomes. The work by \cite{draine2011} provides semi-analytical descriptions for such cases and we will use these as a reference.

\cite{draine2011} describe the interior of dusty HII regions in static equilibrium based on the following equation
\begin{eqnarray}
    n \sigma_d \frac{L_\mathrm{n} e^{-\tau} + L_\mathrm{i}\Phi(r)}{4\pi r^2 c} + \alpha_\mathrm{B} n^2 \frac{\langle h \nu \rangle_\mathrm{i}}{c} - \frac{\mathrm{d}}{\mathrm{d}r} (2nk_\mathrm{B}T)= 0\, , \label{eq:drain_hii}
\end{eqnarray}
where $n$, $\sigma_\mathrm{d}$, $L_\mathrm{n}$, $L_\mathrm{i}$, $\tau$ and $\Phi(r)$ are the proton density, dust opacity, neutral and ionizing luminosity, the optical depth and extinction of ionizing radiation, respectively. The quantity $\langle h \nu \rangle_i$ is the mean energy of a photon which is 18 eV in our case and $r$ measures the radial distance from the source of ionizing photons. Eq. \ref{eq:drain_hii} assumes a constant dust to gas ratio everywhere and takes into account the radiation pressure on dust by ionizing and non-ionizing radiation (first term) and radiation pressure on gas gas by ionizing radiation (second term). Radiation pressure is opposed by thermal pressure (third term) which governs the static equilibrium.

\begin{figure}
    \includegraphics[width=\columnwidth]{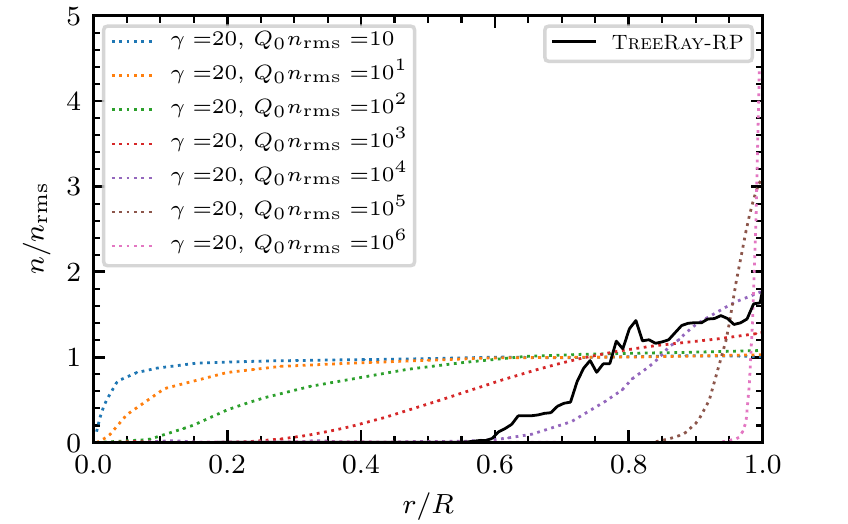}
    \caption{Radial profile showing the interior of an HII region taking into account radiation pressure. The plot shows the number density vs. radial distance. The solid line shows the curve obtained from our code and the dotted lines show semi-analytic models from \protect\cite{draine2011} (see their Fig. 2). \label{fig:draine_hii}}
\end{figure}

Fig. \ref{fig:draine_hii} shows both semi-analytic predictions based on the work of \cite{draine2011} in dotted lines and the result from our code in solid line. While our solution has not reached its steady state, it matches the semi-analytic prediction (in-between red and purple dotted lines; $10^3 < Q_0 n_\mathrm{rms} < 10^4$; for $\gamma=20$). The spikes are minor waves in the density field which is still adjusting for true static equilibrium. The different semi-analytic solutions are categorized based on the product of the number of ionising photons per second normalized to $10^{49} \, s^{-1}$ and the root mean square number density, which are labeled $Q_0$ and $n_\mathrm{rms}$, respectively. The latter is calculated in the following way according to the work of \cite{draine2011}
\begin{eqnarray}
    n_\mathrm{rms} = n_0 \left( \frac{3}{y_\mathrm{max}^3} \int_0^{y_\mathrm{max}} \frac{1}{u^2}y^2 \mathrm{d}y \right)^{1/2}\, ,
\end{eqnarray}
where $n_0$ is the characteristic density scale,
\begin{eqnarray}
    n_0 = \frac{4\pi \alpha_\mathrm{B}}{Q_0} \left( \frac{2ckT}{\alpha_\mathrm{B}\langle<h\nu\rangle}\right)^3\,.
\end{eqnarray}
The dimensionless quantities $u$ and $y$ refer to the density and length scale
\begin{eqnarray}
    u = \frac{n_0}{n} \\
    y = \frac{r}{\lambda_0}
\end{eqnarray}
with $\lambda_0$ being the characteristic length scale defined as
\begin{eqnarray}
    \lambda_0 = \frac{Q_0}{4\pi \alpha_\mathrm{B}} \left( \frac{\alpha_\mathrm{B}\langle h\nu\rangle}{2ckT}\right)^2\,.
\end{eqnarray}
Further the quantity $\gamma$ is calculated in the following way
\begin{eqnarray}
    \gamma = \left(\frac{2ckT}{\alpha_\mathrm{B} \langle h\nu \rangle} \right) \, \sigma_\mathrm{d} ,
\end{eqnarray}
where our value of $\gamma$ is roughly 15.  The quantity $\gamma$ refers to the dust opacity towards ionizing radiation.
Note that we do not employ a constant dust opacity for non ionizing radiation in our scheme, which is a significant difference to the assumptions made by \cite{draine2011}. Our dust opacity for ionizing radiation however is constant, namely $\sigma_\mathrm{d} = 1.5 \times 10^{-21} \, \mathrm{cm^2 \,H^{-1}}$.
Our setup compares to the setup $10^3 < Q_0 n_\mathrm{rms} < 10^4$ and features $\gamma\approx15$. We can see that our solution recovers features of both the purple and red dotted line.

\bsp	
\label{lastpage}
\end{document}